\documentclass[12pt,twoside]{article}

\usepackage{amsmath,amsfonts,amsthm,amssymb,amsopn,amstext,amscd}
\usepackage{enumerate}
\usepackage{graphicx}
\usepackage{float}
\usepackage{xcolor}
\usepackage{longtable}
\usepackage{wrapfig}
\usepackage{setspace} %For the space between lines https://es.overleaf.com/project/5ffe0099d7828d542ef16c86
\usepackage{enumitem}
\usepackage[round]{natbib}
\usepackage{hyperref}
\usepackage{dsfont}
\usepackage{soul}
\usepackage[normalem]{ulem} 
\usepackage{url}
\allowdisplaybreaks

\newcommand{\N}{\mathbb{N}}
\newcommand{\R}{\mathbb{R}}

\newcommand{\tp}{\boldsymbol{p}}
\newcommand{\talpha}{\boldsymbol{\alpha}}
\newcommand{\tomega}{\boldsymbol{\omega}}
\newcommand{\ind}[1]{\mathds{1}_{\{#1\}}}
\newcommand{\indi}[1]{\mathds{1}_{#1}}

\newcommand*\samethanks[1][\value{footnote}]{\footnotemark[#1]}

\newcounter{algorithm}
\newenvironment{alg}[1][\unskip]{%
\refstepcounter{algorithm}\par
\noindent
{\begin{center}\textsc{\texttt{Algorithm: #1}}\end{center}}
\vspace{-2ex}\ttfamily }

\begin{document}
\pagenumbering{arabic}
\singlespace
%\onehalfspace
%\doublespace

\title{Model choice and parameter inference in controlled branching processes}
\author{Miguel Gonz\'alez\thanks{All authors contributed equally to this work.}\ \footnote{Department of Mathematics, Faculty of Sciences and Instituto de Computaci\'on Cient\'ifica Avanzada, University of Extremadura, Badajoz, Spain. e-mail address: \url{mvelasco@unex.es}. ORCID: 0000-0001-7481-6561.}
\and Carmen Minuesa\samethanks[1]\  \footnote{Department of Mathematics, Faculty of Sciences, University of Extremadura, 06006, Badajoz, Spain. e-mail address: \url{cminuesaa@unex.es}. ORCID: 0000-0002-8858-3145.}
\footnote{Department of Mathematics, Faculty of Sciences, Autonomous University of Madrid, 28049, Madrid, Spain. e-mail address: \url{carmen.minuesa@uam.es}.}
\and In\'es del Puerto\samethanks[1]\  \footnote{Department of Mathematics, Faculty of Sciences and Instituto de Computaci\'on Cient\'ifica Avanzada, University of Extremadura, Badajoz, Spain. e-mail address: \url{idelpuerto@unex.es}. ORCID: 0000-0002-1034-2480.} }

\maketitle

%\author[label1]{Miguel Gonz\'alez}
%\ead{mvelasco@unex.es}
%\author[label2]{Carmen Minuesa}
%\ead{carmen.minuesa@uam.es}
%\author[label1]{In\'es M. del Puerto\corref{cor1}}
%\ead{idelpuerto@unex.es}
%\address[label1]{Department of Mathematics, Faculty of Sciences and Instituto de Computaci\'on Cient\'ifica Avanzada,
%University of Extremadura, Badajoz, Spain}
%\address[label2]{Department of Mathematics,
%Autonomous University of Madrid,
% Madrid, Spain.}

\begin{abstract}
Our purpose is to estimate the posterior distribution of the parameters of interest for controlled branching processes (CBPs) without prior knowledge of the maximum number of offspring that an individual can give birth to and without  explicit likelihood calculations. We  consider that  only the population sizes at each generation  and at least the number of progenitors of the last generation are observed, but the number of offspring produced by any individual at any generation is unknown. The proposed approach is two-fold. Firstly, to estimate the  maximum progeny per individual we  make use of an  approximate Bayesian computation (ABC) algorithm for model choice and based on sequential importance sampling  with the raw data. Secondly, given such an estimate and taking advantage of the simulated values of the previous stage, we approximate the posterior distribution of the main parameters of a CBP by  applying the rejection ABC algorithm  with an appropriate summary statistic and a post-processing adjustment. The accuracy of the proposed method is illustrated  by means of  simulated examples developed with the statistical software R. Moreover, we apply the methodology to two real datasets describing populations with logistic growth. To this end, different population growth models based on CBPs are proposed for the first time.
\end{abstract}


%%%%%%%%%%%%%%%%%%%%%%%%%%%%%%%%%%%%%%%%%%%%%%%%%%%%%%%%%%%%%%%%%%%%%%%%%%%%%%%%%%%%%%%%%%%%%%%%%%%%%
\section{Introduction}\label{sec:intro}
%%%%%%%%%%%%%%%%%%%%%%%%%%%%%%%%%%%%%%%%%%%%%%%%%%%%%%%%%%%%%%%%%%%%%%%%%%%%%%%%%%%%%%%%%%%%%%%%%%%%%

We focus our attention on inferential issues related to controlled branching processes. A controlled branching process is a discrete-time stochastic process that models populations developing in the following manner: the population begins with  a fixed number of individuals
or progenitors; each of them, independently of the others and
according to a common probability distribution, gives birth to offspring, and then ceases to participate in subsequent
reproduction processes. Thus, each individual lives for one unit of time and is replaced with a random number of offspring. Moreover, since  {by} several reasons of an environmental, social, or other nature the number of progenitors which take part in each generation {might} be controlled, a  random mechanism  is introduced in the model to determine the number of offspring  with reproductive capacity in each generation. Mathematically, a controlled branching process (CBP) is a process $\{Z_n\}_{n\in\N_0}$  defined recursively as
\begin{equation}\label{def:model}
Z_0=N,\quad Z_{n+1}=\sum_{j=1}^{\phi_n(Z_{n})}X_{nj},\quad n\in\N_0,
\end{equation}
where $\N_0=\N\cup\{0\}$, $N\in\N$, $\{X_{nj}:\ n\in\N_0;\ j\in\N\}$ and $\{\phi_n(k):n,k\in\N_0\}$ are independent families of non-negative integer valued random variables  and the empty sum in (\ref{def:model}) is considered to be 0. The random variables $X_{nj}$, $n\in\N_0$, $j\in\N$, are assumed to be independent and identically distributed (i.i.d.) with distribution $\tp=\{p_j=P(X_{01}=j): j\in \N_0\}$ and in terms of population dynamics they represent the number of offspring given by the $j$-th progenitor of the $n$-th generation.  Moreover, $\{\phi_n(k)\}_{k\in\N_0}$, for $n\in\N_0$, are independent stochastic processes with equal one-dimensional probability distributions. This property means that the control mechanism works in an independent manner in each generation, and once the population size at certain generation $n$, $Z_n$, is known, the probability distribution of the number of progenitors,  denoted by $\phi_n(Z_{n})$, is independent of the generation. Some particular cases collected in this general family of branching processes are the simplest model, the standard Bienaym\'e--Galton--Watson (BGW) process, by considering $\phi_n(k)=k$ a.s. for each $k$, or the branching processes with immigration, by setting $\phi_n(k)=k+Y_n$, where $\{Y_n\}_{n\in\N_0}$ is a class of i.i.d. random variables, among others. 
 
The recent monograph \cite{CBPs} provides an extensive description of its probabilistic theory. The behaviour of the long-time evolution of a CBP is determined by the parameters of the model associated to the offspring and control laws. Briefly, assuming that $m=E[X_{01}]$ and $\varepsilon(k)=E[\phi_{n}(k)]$, $k\in\N_0$, exist and are finite, and whenever the limit $\tau=\lim_{k\to\infty}k^{-1}\varepsilon(k)$ exists, the threshold parameter of this branching model is $\tau m$. The extinction occurs almost surely in subcritical populations, namely if $\tau m<1$, and different growth rates  on the non-extinction set are obtained depending on whether $\tau m=1$ (critical population) or $\tau m>1$ (supercritical population) with additional conditions.  In real situations, these parameters  are unknown. Until now, the  methodologies  proposed in the literature for the Bayesian inference on the offspring distribution have focused on the cases where either the support of the reproduction law is finite and known (see \cite{art-ABC})  or that the offspring law belongs to some one-dimensional parametric family (see \cite{art-ABC-summary}). A first paper in the context of the CBP that faces the problem of an unknown scenario on the offspring distribution could be \cite{chap-proceedings-2016}. The statistical procedures developed in this work did not include the estimation of the posterior distribution of the maximum number of offspring per progenitor given the sample of population sizes at each generation $\{Z_0,\ldots, Z_n\}$, but this quantity was set $1+\max_{1\leq k\leq n}Z_k$ as a primary approach.
%In the statistical procedures developed in this work, the sample considered was given by the population sizes at each generation, namely $\{Z_0,\ldots, Z_n\}$, and as a primary approach the maximum number of offspring per progenitor was set  $1+\max_{1\leq k\leq n}Z_k$. 
Within the class of other branching processes, this problem  has been only considered in the BGW process. In particular, from a probabilistic viewpoint, the asymptotic behaviour of the number of offspring of the most prolific individual in the $n$-th generation has been studied as an extreme value problem in \cite{RahimovYanev-1999} and \cite{Bertoin-2011}. From an inferential viewpoint, a particle Markov Chain Monte Carlo method was introduced to estimate the support of the offspring law  in \cite{drovandi-Pettit-al-2016}. However, the drawback of this approach is that its computational feasibility strongly depends on dealing with BGW processes with low values.

%\blue{\sout{One of the novelty of  this work is to propose  a  methodology to estimate the maximum offspring capacity per individual in the class of CBPs,  where this number is understood as the maximum progeny that an individual in the population can bear.}} 
The first aim of this work is to provide a methodology to estimate the maximum progeny that an individual in the population can bear (called maximum offspring capacity per individual) in the general class of CBPs and regardless  the magnitude of the observed samples. Having estimated the maximum offspring capacity per individual, we also make inference on the expected values  of offspring and control laws. To this end, we consider the maximum offspring capacity per individual as a model index and, for the first time, we tackle the problem of model choice and parameter estimation in a CBP. %\sout{The selection of an appropriate method depends on the dynamical structure of  the  data available.} 
We provide an algorithm based on approximate Bayesian computation (ABC) techniques to estimate both the maximum number of offspring that an individual is able to give birth to and the parameters of interest of the model.   The ABC methodology in the context of CBPs was already analysed and applied in \cite{art-ABC-summary} by assuming that the offspring distribution belongs to a parametric family. This means that the family of offspring distributions is known 
(for instance, geometric, Poisson or binomial distributions) and the only unknown elements are the parameters that determine them.  In this paper we drop this assumption and face the problem of making inference on the parameters of interest in a less informative scenario with respect to  the offspring distribution.

For our purpose, let us consider a CBP with an offspring distribution with an unknown support and control laws belonging to some known one-dimensional parametric family with unknown parameter. Let $\kappa=\sup\{j\in\N_0: p_j>0\}$ the maximum number of offspring per individual, denote   $\tp({\kappa})=\{p_j({\kappa})=P(X_{01}=j):\ j\in \N_0\}$ the offspring distribution when the maximum offspring capacity per individual is $\kappa$,  and  {let $\gamma$ be} the control parameter, with $\gamma\in \Gamma \subseteq\mathbb{R}$. {We recall} that in that case, the distribution of each control variable $\phi_n(k)$ {only} depends on $k$ and $\gamma$, and $E[\phi_n(k)]=\varepsilon(k,\gamma)$. Let us denote $m({\kappa})=\sum_{j=0}^\kappa j p_j(\kappa)$ and $\tau(\gamma)=\lim_{k\to\infty}k^{-1}\varepsilon(k,\gamma)$. We assume that {$m({\kappa})<\infty$} and $\tau(\gamma)$ exists for all $\gamma\in \Gamma$. Moreover we assume the existence of the inverse of  $\tau(\cdot)$.  Several preliminary simulation studies lead us to the conclusion that to approximate the posterior distributions of the parameters of interest reasonably well by making use of ABC methodology,  we have to assume that at least the population sizes at each generation and the number of progenitors in the last generation are observable (see \cite{art-ABC-summary}). Hence, let us  consider the observed sample $\widetilde{\mathcal{Z}}^{obs}_n =\{Z_0^{obs},\ldots,Z_n^{obs},\phi_{n-1}(Z_{n-1})^{obs}\}$.  Briefly, we will proceed as follows: firstly, we draw a sample from an estimate of the posterior  distribution of $\kappa$, denoted by $\pi(\kappa|\widetilde{\mathcal{Z}}^{obs}_n)$. Secondly, we generate a sample from an estimate of the posterior distribution of $(\tp(\widetilde\kappa_n),\gamma)$, where $\widetilde{\kappa}_n$ is a point estimate of $\kappa$. Next, from this sample we estimate the posterior distributions of $(\tp(\widetilde\kappa_n),\gamma)$, $m(\widetilde\kappa_n)$ and $\tau(\gamma)$ using kernel density estimation. We denote these posterior distributions by $\pi(\tp(\widetilde{\kappa}_n),\gamma|\widetilde{\kappa}_n,\widetilde{\mathcal{Z}}_n^{obs})$, $\pi(m \mid \widetilde{\kappa}_n,\widetilde{\mathcal{Z}}_n^{obs})$ and $\pi(\tau(\gamma)\mid \widetilde{\kappa}_n,\widetilde{\mathcal{Z}}_n^{obs})$, respectively.  

The performance of the proposed algorithm is firstly illustrated by two simulated examples. Next, the method is applied on two real datasets that show a logistic growth. To this end, we model the evolution of logistic growth of populations by CBPs, which represents another important novelty of this paper. These populations are characterised by the fact that  when their sizes are small enough, they grow with almost no restriction, but when the sizes increase, the limited resources of the environment lead to a control on the population sizes. As a consequence, there exists a maximum population size, usually called carrying capacity in an ecological context, that can be supported by the ecosystem. With the aim of describing mathematically these populations, we introduce CBPs with control distributions given by binomial distributions whose success probabilities mainly depend on the density of the population. We provide several models based on different success probability functions which are inspired in classical deterministic population growth models.

%\red{The performance of the proposed algorithm is firstly illustrated by two simulated examples. Next we apply it on two real datasets that show a logistic growth. To this end, we present another important novelty of this paper. This is the first time that the evolution of logistic growths of populations is modelled by  CBPs. These populations are developed freely when their sizes are small enough but when these grow up, they must be controlled due to the fact that the resources of the environment are limited and consequently there exists a maximum population size supported (usually called carrying capacity in an ecological context). We introduce CBPs with control distributions following binomial distributions whose success probabilities mainly depend on the density of the population. We provide several models based of different success probability functions, inspired in classical deterministic population growth models.
%\sout{The expected values of the populations size in a generation given the size of the previous one will be specified for different functions depending on the maximum population size supported bt the environment, the offspring means and the current size, inspired in classical deterministic models.}}

Apart from this introduction, the paper is organized as follows. In Section~\ref{sec:methods} we provide a detailed description of the ABC methodology for model choice and parameter estimation in the context of CBPs. Section~\ref{sec:example} gathers simulation studies to evaluate and illustrate the performance of the proposed ABC approach. In Section ~\ref{sec:example_real} we present the application of the proposed algorithm to two real datasets from populations that exhibit a logistic growth.  Additional information related to the examples are presented in the Appendix. In Section~\ref{sec:conclusion} we summarise the main contributions of this work.

\vspace{0.35cm}

%%%%%%%%%%%%%%%%%%%%%%%%%%%%%%%%%%%%%%%%%%%%%%%%%%%%%%%%%%%%%%%%%%%%%%%%%%%%%%%%%%%%%%%%%%%%%%%%%%%%%
\section{Methodology}\label{sec:methods}
%%%%%%%%%%%%%%%%%%%%%%%%%%%%%%%%%%%%%%%%%%%%%%%%%%%%%%%%%%%%%%%%%%%%%%%%%%%%%%%%%%%%%%%%%%%%%%%%%%%%%

In this section we describe the ABC approach  for model choice in the context of CBPs  and to estimate the posterior distribution of the main parameters of our model. ABC algorithms are a group of Monte Carlo algorithms used to find posterior distributions without requiring explicit knowledge of the likelihood function. These are very useful when the likelihood is  intractable or too costly to evaluate. The inference is mainly done with {samplings} from the model, and hence, their versatility in the framework of branching processes (see the monograph \cite{Sisson-2019} for further details).

In this context, the fact that the value of $\kappa$ is unknown and could be even infinite, increases the complexity of the problem of estimating the parameters of the CBP and requires to develop methodologies for model choice.  We assume $\kappa\geq 2$ to avoid trivial cases.  To implement the ABC methodology, we remark that even if our knowledge on the value of $\kappa$ is very poor, we usually have some information about an \emph{effective} upper bound for $\kappa$, denoted $K_{max}$, from the dynamics of the population that we model via the CBP. An example of this situation is the family of K-selected species (see \cite{Parri-1981}), which includes larger mammals such as elephants, horses, and primates, and whose species are relatively stable populations and produce relatively low numbers of offspring. %For example,  K-selected species  are species characterized by relatively stable populations and produce relatively low numbers of offspring. Among them are  larger mammals such as elephants, horses, and primates (see \red{\cite{Parri,1981}}). 
%For practical purposes and without loss of generality, %, as we will show in the simulated examples, 
%\blue{throughout this paper we consider offspring laws with finite support  and explore the case of offspring laws with infinite support in Subsection~\ref{subsec:example-infinite-sup}}.   Let us denote the upper bound \blue{of the support of the offspring distribution} by $K_{max}$.   Thus, $\kappa\in \{2,3,\dots,K_{max}\}$. 
For practical purposes and without loss of generality, throughout this paper we consider offspring laws with finite support. Thus, $\kappa\in \{2,3,\dots,K_{max}\}$. We can take the parameter $\kappa$ as a model index. We emphasise that as a consequence, for each value of $\kappa$ the parameter of interest in the corresponding model is $$(\tp(\kappa),\gamma)=(p_0(\kappa),\ldots,p_\kappa(\kappa),\gamma)\in \Delta_{\kappa}\times\R,$$
whose dimension depends on $\kappa$, and where $\Delta_{\kappa}$ is the $\kappa$-
standard simplex in $\R^{\kappa}$.

We recall that our final aim is to estimate the posterior $\pi(\tp(\kappa),\gamma\mid \widetilde{\kappa},\widetilde{\mathcal{Z}}_n^{obs})$, with $\widetilde{\kappa}$ a point estimate of $\kappa$, and to that end, we propose a two-fold procedure.

% We also emphasize that in this first stage using common summary statistics in this setting does not improve the results that we obtain when using the raw sample observed. This is due to the fact that the summary statistics proposed in the literature for CBPs are of low dimension and then, they are not suitable to identify large dimension parameters as it might happen in our model.

%We also introduce a convenient summary statistic to that end.
%In the second stage, once that we have obtained an estimate for $\hat{\kappa}_n$ and $\tilde{\kappa}_n$ based on the information of the sample $\{\kappa^{(1)},\ldots \kappa^{(N)}\}$ obtained in the first part,   we briefly describe the algorithms to approximate the distributions $\pi(\tp(\hat{\kappa}_n),\gamma|\widetilde{\mathcal{Z}}_n^{obs})$ and $\pi(\tp(\tilde{\kappa}_n),\gamma|\widetilde{\mathcal{Z}}_n^{obs})$. We also introduce a convenient summary statistic to that end.}

\subsection{First stage: estimation of $\pi(\kappa\mid \widetilde{\mathcal{Z}}_n^{obs})$}

In the first part, we estimate $\pi(\kappa\mid\widetilde{\mathcal{Z}}_n^{obs})$. We apply an ABC  {algorithm for} model choice based on sequential importance sampling, ABC SMC  {for} model choice, introduced in \cite{toni09} to draw a sample $\{(\kappa^{(1)},\tp(\kappa^{(1)})^{(1)},\gamma^{(1)}),\ldots, (\kappa^{(N)},\tp(\kappa^{(N)})^{(N)},\gamma^{(N)})\}$ from the joint posterior distribution of $(\kappa,\tp(\kappa),\gamma)$ given the observed sample $\widetilde{\mathcal{Z}}_n^{obs}$, denoted by $\pi(\kappa,\tp(\kappa),\gamma\mid \widetilde{\mathcal{Z}}_n^{obs})$. Next, using the information of the marginal sample $\{\kappa^{(1)},\ldots, \kappa^{(N)}\}$ we are able to estimate the distribution $\pi(\kappa\mid \widetilde{\mathcal{Z}}_n^{obs})$ and  propose the  {closer integer to} its posterior mean  as the Bayesian point estimator for the parameter $\kappa$. We refer to  {this estimator} as $\tilde{\kappa}_n$. Our choice is justified by the good asymptotic properties that this estimator usually exhibits even in the case of CBPs (see \cite{art-Dposterior}).
%Notice that from the sample obtained in the previous stage it is very easy to estimate these quantities by using the marginal sample $\{\kappa^{(1)},\ldots,\kappa^{(N)}\}$.

We now describe how to implement the ABC SMC  {algorithm for} model choice to draw samples from the posterior distribution $\pi(\kappa,\tp(\kappa),\gamma\mid \widetilde{\mathcal{Z}}_n^{obs})$. The algorithm  reaches the target distribution through a series of intermediate distributions sampling from appropriate proposal distributions and weighting the samples by importance weights. To that end,  we fix a number of $T$ iterations   and a decreasing sequence of tolerance levels $\epsilon_1 >\ldots > \epsilon_T$. In practice, the tolerance levels are selected as quantiles of the distances between the simulated and observed data (see the mathematical arguments  {for this choice} in \cite{Biau15}).

The first iteration consists in running the  tolerance-rejection ABC  {algorithm for} model choice. It starts by drawing a value $\kappa'$ from the prior distribution on the models, denoted  $\pi(\kappa)$. Assuming that we have no other knowledge than the lower and upper bounds of $\kappa$, we shall consider a uniform distribution on the points $2,\ldots, K_{max}$, denoted $U\{2,\ldots,K_{max}\}$, for the prior model distribution. %Next, we %sample a value of the parameter of interest from its prior distribution given $\kappa'$.
Using the fact that the reproduction and control laws are independent, we assume that the prior distribution for the model index $\kappa$, denoted  by $\pi(\tp(\kappa),\gamma\mid \kappa)$, satisfies
$$\pi(\tp(\kappa),\gamma\mid \kappa)=\pi(\tp(\kappa)\mid \kappa)\pi(\gamma),$$
where $\pi(\tp(\kappa)\mid \kappa)$ is the prior distribution of $\tp(\kappa)$ given the model index $\kappa$ and $\pi(\gamma)$ is a suitable prior for $\gamma$. Now, bearing in mind that the parameter $\tp(\kappa)$ is a probability distribution with support $\{0,\ldots,\kappa\}$, we propose a Dirichlet distribution with  a $(\kappa+1)$-dimensional parameter $\talpha_\kappa$, denoted $D(\kappa+1,\talpha_\kappa)$, as the distribution $\pi(\tp(\kappa)\mid \kappa)$. Let us also write $f(\widetilde{\mathcal{Z}}_n\mid \tp(\kappa), \gamma)$ {to refer to} the likelihood function given $\tp(\kappa)$ and $\gamma$, with $\widetilde{\mathcal{Z}}_n=\{Z_0,\ldots,Z_n,\phi_{n-1}(Z_{n-1})\}$. The next steps are the usual ones in  tolerance-rejection ABC algorithms. A sample $\widetilde{\mathcal{Z}}_n^{sim}=\{Z_0^{sim},\ldots,Z_n^{sim},\phi_{n-1}(Z_{n-1}^{sim})\}$ is generated by using the previously sampled parameters and accept them if the sample is close enough to the observed sample $\widetilde{\mathcal{Z}}_n^{obs}$ in terms of some \emph{distance} $\rho(\cdot,\cdot)$ and the tolerance level. In this stage, we compare directly the  raw  data without summary statistics. The jumps between the model indexes might lead to quite different dimensions of the prior distributions $\pi(\tp(\kappa)|\kappa)$, for each $\kappa$, and consequently,  {finding a low-dimensional summary statistic to identify parameters  of a large dimension is quite hard}   (see the discussion in \cite{Nott-2019}).

It is worth to mention that in order to quantify the disparities  between the simulated and the observed data we can use many different functions. However, based on the results of previous studies (see \cite{art-ABC-summary}), a good discrepancy measure in the CBP setting should satisfy the non-negative property, the identity of indiscernible and the symmetry, but it should also compare the simulated and observed data in relative terms to avoid any issue due to the magnitude of each coordinate. For these reasons, we propose the following {function}:
$$\rho({\bf{x}},{\bf{y}}) = d_e\left(\frac{{\bf{x}}}{{\bf{y}}},\frac{{\bf{y}}}{{\bf{x}}}\right), \mbox{ with } {\bf{x}}=(x_1,\ldots,x_L), {\bf{y}}=(y_1,\ldots,y_L)\in\mathbb{R}_+^L,$$
where $\frac{\bf{x}}{\bf{y}}=(\frac{x_1}{y_1},\ldots,\frac{x_L}{y_L})$, $\frac{\bf{y}}{\bf{x}}=(\frac{y_1}{x_1},\ldots,\frac{y_L}{x_L})$, $\mathbb{R}_+=(0,\infty)$ and  $d_e$ is the \emph{Euclidean distance}.

We can now describe the first iteration of the   ABC SMC algorithm {for} model choice {on} {$\kappa$} as follows:

\vspace{2ex}

\begin{alg}[ABC SMC algorithm {for} model choice {on} {$\kappa$}]\label{alg:SMC1}
\begin{enumerate}
  \item [ ] Specify a decreasing sequence of tolerance levels $\epsilon_1 > \ldots > \epsilon_T>0$ for $T$ iterations.
  \item [ ] For $i=1$ to $N$, do
  \begin{enumerate}
    \item [ ] Repeat
    \begin{enumerate}
    \item [ ] Generate ${\kappa}'$  from  $U\{2,\ldots,K_{max}\}$
    \item [ ] Generate $\tp({\kappa}')$ from $D({{\kappa}'+1},\talpha_{{\kappa}'})$ and $\gamma'$ from $\pi(\gamma)$
.
    \item [ ] Generate $\widetilde{\mathcal{Z}}_n^{sim}$ from $f(\widetilde{\mathcal{Z}}_n\mid \tp(\kappa'), \gamma')$
    \end{enumerate}
    \item [ ] Until $\rho(\widetilde{\mathcal{Z}}_n^{sim}, \widetilde{\mathcal{Z}}_n^{obs})\leq\epsilon_1$.
    \item [ ] Set  $({\kappa}_1^{(i)},\tp({\kappa}_1^{(i)})^{(i)}, \gamma_1^{(i)})=({\kappa}', \tp({\kappa}'), \gamma')$.
    \item [ ] Set $\omega_1^{(i)}=1/N$.
  \end{enumerate}
  \item [ ] End for
\end{enumerate}
\end{alg}

To run the following iterations, the idea is to draw the parameters from  proposal distributions that are closer to the  target distributions so that we can reduce the variance of the final sample. For each iteration $t$, $t=2,\ldots, T$, we have to specify a joint proposal distribution for each $\tp({\kappa}^*)$ and $\gamma^*$, denoted by $q_t(\tp({\kappa}), \gamma \mid \tp({\kappa}^*), \gamma^*)$. However, in real applications finding a joint distribution that leads to a good performance of the ABC SMC algorithm represents a challenge.
%This is one of the challenge to be fixed  in real applications in order to obtain a good performance of the ABC SMC algorithm.

%To that end, it is important to highlight that despite the independence between offspring and control distributions,  once the sample is given, their posterior distributions are \red{usually} highly correlated, as shown empirically in the \red{second} simulated \red{example} \sout{examples} in Section~\ref{sec:example} (see \red{\sout{Figures~\ref{fig:c1mgama-second-step}-\ref{fig:c4mgama-second-step} or}}  Figure~\ref{fig:mgama-second-step}, left). 
To that end, it is important to highlight that despite the independence between offspring and control distributions,  once the sample is given, their posterior distributions are usually highly correlated, as shown empirically in the second simulated example  in Section~\ref{sec:example} (see  Figure~\ref{fig:mgama-second-step}, left). Indeed, the outputs $({\kappa}', \tp({\kappa}'), \gamma')$ of each iteration of the algorithm satisfy $\tau(\gamma')m(\kappa')\approx\tau m$, where recall  $m(\kappa)=\sum_{j=0}^\kappa j p_j(\kappa)$, $\tau(\gamma)=\lim_{k\to\infty}k^{-1}\varepsilon(k,\gamma)$, with $\varepsilon(k,\gamma)=E[\phi_n(k)]$, and $\tau m$ represents the true value of the threshold parameter. Thus, the use of component-wise perturbation proposals might lead to an inappropriate structure of the true posterior. Taking into account the relationship described above, we suggest the following proposal distribution: 
\begin{equation}
\label{eqpro}
q_t(\tp({\kappa}), \gamma \mid \tp({\kappa}^*), \gamma^*)=q_t(\tp({\kappa})\mid   \tp({\kappa}^*), \gamma^*)q_t(\gamma\mid \tp({\kappa}), \tp({\kappa}^*), \gamma^*).
\end{equation}
We set $q_t(\tp({\kappa})\mid   \tp({\kappa}^*), \gamma^*)$ to be a Dirichlet distribution with mean vector $\tp(\kappa^*)$ and variance controlled by a single tuning parameter $a>0$, i.e., a Dirichlet distribution of order $\kappa^*+1$ and parameter $a \tp(\kappa^*)$, $D(\kappa^*+1, a \tp(\kappa^*))$. Given a value $\tp({\kappa})$ from $q_t(\tp({\kappa})\mid   \tp({\kappa}^*), \gamma^*)$, we fix
$q_t(\gamma\mid \tp({\kappa}), \tp({\kappa}^*), \gamma^*)$  as the distribution of the variable $\tau^{-1}(U^*/m(\kappa))$, where $\tau^{-1}(\cdot)$ is the inverse of the function $\tau(\cdot)$, the random variable $U^*$ follows a normal distribution with mean $\tau(\gamma^*)m(\kappa^*)$ and some variance $\sigma_t^2$, $N( \tau(\gamma^*)m(\kappa^*), \sigma_t^2)$, and $m(\kappa)$ is the offspring mean of the distribution $\tp({\kappa})$.  Notice that we keep the variability of the proposal distribution fixed when the value $\tp({\kappa}^*)$ is perturbed, however an  adaptive dispersion is chosen to  perturb the control parameter $\gamma$. In particular, $\sigma_t^2$ is twice the weighted empirical variance of selected $\gamma$'s in the $t-1$ iteration (see \cite{filippi2013} for further discussion on optimality of proposals for ABC SMC).

For a step-by-step description of the remaining iterations of the algorithm in the first phase, let us write $\mathcal{P}_t=\{({\kappa}_t^{(1)},\tp({\kappa}_t^{(1)})^{(1)},\gamma_t^{(1)})\ldots, ({\kappa}_t^{(1)},\tp({\kappa}_t^{(N)})^{(N)},\gamma_t^{(N)})\}$, for the output of the $t$ stage of the algorithm. Moreover, let us denote $\mathcal{P}_t(\kappa)$ the family defined by the elements of $\mathcal{P}_t$ %the matrix whose rows corresponds to the vectors of the family $\mathcal{P}_t$
such that ${\kappa}_t^{(j)}={\kappa}$, $j=1,\ldots,N$. We also write $\indi{A}$ to refer to the indicator function of the set $A$ and $\tomega_t=(\omega_t^{(1)},\ldots, \omega_t^{(N)})$ to refer to the vector of weights in the iteration $t$.
%For a {step-by-step} description of \blue{the remaining iterations of the algorithm in the first phase}, let us write $\mathcal{K}_t=\{{\kappa}_t^{(1)},\ldots,{\kappa}_t^{(N)}\}$, $\mathcal{P}(\mathcal{K}_t)=\{\tp({\kappa}_t^{(1)})^{(1)},\ldots, \tp({\kappa}_t^{(N)})^{(N)}\}$, and $\Gamma_t=\{\gamma_t^{(1)}, \ldots, \gamma_t^{(N)}\}$ for the output of the $t$ stage of the algorithm. Moreover, let {us} denote $\mathcal{P}_t(\kappa)$ the matrix whose rows corresponds to the distributions of the family $\mathcal{P}(\mathcal{K}_t)$ such that ${\kappa}_t^{(j)}={\kappa}$, $j=1,\ldots,N$, and $\Gamma_t({\kappa})$ the vector  whose coordinates are the elements of $\Gamma_t$ associated with the rows of $\mathcal{P}_t(\kappa)$. We also write $\indi{A}$ to refer to the indicator function of the set $A$ and $\tomega_t=(\omega_t^{(1)},\ldots, \omega_t^{(N)})$ to refer to the vector of weights in the iteration $t$.

\begin{alg}[Continuation of the  ABC SMC algorithm for model choice on $\kappa$]\label{alg:SMC1}
\begin{enumerate}
  %  \item [ ] $\mathbf{\sum_1}=2\ \text{Cov}[\theta_1,\gamma_1]$ (twice the sample covariance matrix).
  \item [ ] For $t=2$ to $T$, do
  \begin{enumerate}
  \item [ ] For $i=1$ to $N$, do
  \begin{enumerate}
    \item [ ] Repeat
    \begin{enumerate}
    \item [ ] Generate ${\kappa}^*$  from  $U\{2,\ldots,K_{max}\}$.
    %\item [ ] Generate {$\tp({\kappa}^*)$ from $\mathcal{P}_{t-1}(\kappa^*)$} and {$\gamma^*$ from $\Gamma_{t-1}(\kappa^*)$} with the \linebreak corresponding weights $\tomega_{t-1}$.
    \item [ ] Generate $({\kappa}^*,\tp({\kappa}^*),\gamma^*)$ from $\mathcal{P}_{t-1}(\kappa^*)$ with the corresponding \linebreak weights $\tomega_{t-1}$.
    \item [ ] Sample $(\tp({\kappa}^{*})^{**}, \gamma^{**})$ from {$q_t(\tp({\kappa}), \gamma \mid \tp({\kappa}^*), \gamma^*)$} described in (\ref{eqpro}).
    \item [ ] Sample $\widetilde{\mathcal{Z}}_n^{sim}$ from $f(\widetilde{\mathcal{Z}}_n\mid \tp({\kappa^*})^{**}, \gamma^{**})$.
    \end{enumerate}
    \item [ ] Until  $\rho(\widetilde{\mathcal{Z}}_n^{sim},\widetilde{\mathcal{Z}}_n^{obs})\leq\epsilon_t$.
    \item [ ] Set $({\kappa}_t^{(i)},{\tp(\kappa_t^{(i)})^{(i)}}, \gamma_t^{(i)})=({\kappa}^{*}, \tp({\kappa}^{*})^{**}, \gamma^{**})$.
    \item [ ] Set
    $${\omega_t^{(i)}=\frac{\pi({\tp(\kappa_t^{(i)})^{(i)}}\mid \kappa_t^{(i)})\pi({\gamma_t^{(i)}})}{\displaystyle\sum_{j=1}^{N} \omega_{t-1}^{(j)}q_{t}(\tp(\kappa_t^{(i)})^{(i)},\gamma_t^{(i)}|\tp(\kappa_{t-1}^{(j)})^{(j)}, \gamma_{t-1}^{(j)})\ind{\kappa_t^{(i)}=\kappa_{t-1}^{(j)}}}}.$$
  \end{enumerate}
  \item [ ] End for
  \item [ ] {For every $k=2,\ldots,K_{max}$, normalise the weights}.
  \end{enumerate}
  \item [ ] End for
\end{enumerate}
\end{alg}

\subsection{Second stage: estimation of $\pi(\tp(\widetilde{\kappa}_n),\gamma\mid\widetilde{\kappa}_n,\widetilde{\mathcal{Z}}_n^{obs})$}\label{subsec:second-part}

Having obtained the estimate for $\kappa$, denoted $\widetilde{\kappa}_n$, given by the  closest integer to the mean of the sample $\{\kappa^{(1)},\ldots \kappa^{(N)}\}$ drawn in the first stage,   we now describe how to draw a sample from the ABC approximation of the distribution $\pi(\tp({\widetilde{\kappa}_n}),\gamma|\tilde{\kappa}_n,\widetilde{\mathcal{Z}}_n^{obs})$.

Besides the approximation of the marginal posterior distribution of the model index, the output of the first stage provides a sample from the ABC estimate of the marginal posterior distributions of parameters, i.e. $\pi(\tp(\kappa),\gamma|\kappa,\widetilde{\mathcal{Z}}_n^{obs})$,  for $\kappa=2,\ldots, K_{max}$. Although  the ABC methodology for the inference on $\kappa$ works quite well without the use of summary statistics as pointed out before, its use does  improve the output of the ABC algorithm when the aim is to make inference on the parameters once the model index is known (see \cite{art-ABC-summary}). Thus, our proposal is to proceed as follows:  when the first stage is implemented all the generated parameter values together with their data sets in the last iteration  are stored. Consequently, they can be used  to run an ABC algorithm to estimate the posterior distribution of the parameters of the model given $\kappa=\tilde{\kappa}_n$ without having to generate new data. Let us denote  as $\{\widetilde{\mathcal{Z}}_n^{(1)},\ldots, \widetilde{\mathcal{Z}}_n^{(N)}\}$ the simulated data corresponding to the sample $\{(\kappa^{(1)},\tp(\kappa^{(1)})^{(1)},\gamma^{(1)}),\ldots, (\kappa^{(N)},\tp(\kappa^{(N)})^{(N)},\gamma^{(N)})\}$, and let $\{\kappa^{(i_1)},\ldots, \kappa^{(i_L)}\}$ be all the elements of the sample $\{\kappa^{(1)},\ldots, \kappa^{(N)}\}$ such that $\kappa^{(i_l)}=\tilde{\kappa}_n$, for $l=1,\ldots,L$. Next, we use of the simulated marginal values
$$\{(\tp(\kappa^{(i_1)})^{(i_1)},\gamma^{(i_1)},\widetilde{\mathcal{Z}}_n^{(i_1)}),\ldots, (\tp(\kappa^{(i_L)})^{(i_L)},\gamma^{(i_L)},\widetilde{\mathcal{Z}}_n^{(i_L)})\},$$
to check the rejection condition in the  tolerance-rejection ABC algorithm considering a suitable summary statistic. We use the following  summary statistic
\begin{align}\label{eq:summary-statistic}
\mathcal{S}(\widetilde{\mathcal{Z}}_n)=\left(\sum_{i=1}^n Z_i,\frac{\sum_{i=1}^n Z_i}{\sum_{i=0}^{n-1} Z_i},\frac{\phi_{n-1}(Z_{n-1})}{Z_{n-1}}, \frac{Z_{n}}{\phi_{n-1}(Z_{n-1})}\right).
\end{align}
This statistic results from adding a fourth coordinate to the one in \cite{art-ABC-summary}. The properties of the model (see \cite{CBPs}) ensure that in a general setting, as $n\to\infty$,
$$\frac{\sum_ {i=1}^n Z_i}{\sum_{i=0}^{n-1} Z_i}\to\tau m,\quad  \frac{\phi_{n-1}(Z_{n-1})}{Z_{n-1}}\to \tau,\quad \frac{Z_{n}}{\phi_{n-1}(Z_{n-1})}\to m,$$
almost surely on $\{Z_n \to \infty\}$, regardless of whether we consider parametric frameworks for the offspring or control distributions. %where recall {$m$ is the offspring mean}, $\tau=\lim_{k\to\infty}k^{-1}\varepsilon(k)$, with $\varepsilon(k)=E[\phi_0(k)]$, whenever the limit exists. 
Consequently, the third and the new coordinate enable us to identify each factor of the threshold parameter. Our simulation results show that the four dimensional summary statistic proposed improves the results compared to previous summary statistics. More details about the efficiency of  adding a new coordinate to the summary statistic can be found in \cite{Joyce08}.

Finally, we apply a post-processing method based on a local linear regression on the output sample. The outputs $\tp(\kappa^{(i_j)})^{(i_j)}$ are ($\tilde{\kappa}_n$+1)-dimensional vectors whose coordinates sum one, but, after regression, some of them could be negative. Such outputs must be removed from the sample (see \cite{art-ABC-summary} for details on both methods).

\vspace{0.15cm}

%%%%%%%%%%%%%%%%%%%%%%%%%%%%%%%%%%%%%%%%%%%%%%%%%%%%%%%%%%%%%%%%%%%%%%%%%%%%%%%%%%%%%%%%%%%%%%%%%%%%
\section{Simulated examples}\label{sec:example}
%%%%%%%%%%%%%%%%%%%%%%%%%%%%%%%%%%%%%%%%%%%%%%%%%%%%%%%%%%%%%%%%%%%%%%%%%%%%%%%%%%%%%%%%%%%%%%%%%%%%
Our methodology is illustrated via several simulated examples. First, we show how well the methodology works in situations as described above, where the reproduction law has finite support. More precisely, we fix the value of the threshold parameter $\tau m$ and consider different CBPs where we vary the support of the offspring distribution, the mean of the offspring distribution $m$, and the control parameter $\gamma$ in such a way that the value of $\tau m$ remains constant for all of the cases. Second, we return to the previous simulated study  in \cite{art-ABC-summary}. Our aim is to estimate the posterior distribution of the parameters of interest without assuming a parametric offspring distribution. The true offspring distribution in this scenario has an infinite support, but we show our methodology is also useful in this context if the main aim is to approximate the posterior distributions of stable parameters, namely, the offspring mean and control parameter. % The second example  %different finite supports and types of skewness  and different control parameters but with the same threshold parameter.

\subsection{Example 1}\label{subsec:example-finite-sup}

We begin our simulation study focusing on offspring distributions with finite support. We show the suitability of the methodology in this framework by considering reproduction laws with different supports and means, and also various control laws with different parameters, keeping the same threshold parameter.

To that end, we explore four different models/cases of CBPs  where the initial number of individuals is $Z_0=1$ and the control  variables $\phi_n(j)$ follow binomial distributions with parameters $\xi(j)$ and $\gamma$, where $\xi(j)= j+\lfloor\log (j)\rfloor$, for each $j\in\N$, $\xi(0)=0$, and $\lfloor x \rfloor$ denotes the integer part of a number $x$. We observe that these control laws are a mixture of a deterministic component and a random one. The introduction of these control functions can be explained in an ecological context, where, first,  we allow the introduction of new individuals in the ecosystem as described by the deterministic function $\xi(\cdot)$, and next, the binomial control models situations as the emigration or death of individuals due to their hunt by predators. Here, $\gamma$ represents the probability that an individual does not participate in the subsequent reproduction process as it is no longer present in the ecosystem. We note that for these CBPs $\varepsilon(j,\gamma)=\gamma \xi(j)$ and $\tau=\gamma$. For our purpose, we vary the value of the control parameter $\gamma$ across the four cases. For the reproduction law we chose binomial distributions with different sizes, $\kappa$, and probabilities of success, $\varrho$, in such a way that the four models satisfy $\tau m=2.88$, i.e. the CBPs are supercritical. The values of the parameters are gathered in Table \ref{tab:binomialcases}.
We emphasise that our choice of the parameters enables us to compare the results obtained by the methodology proposed when examining different finite supports and types of skewness of the offspring distribution (see Table \ref{tab:binomialprob}).

%The aim is to illustrate how  the methodology works for situations in which the offspring probabilities have different finite supports and types of skewness (see Table \ref{tab:binomialprob}) and different control parameters but with the same threshold parameter. 
\begin{table}[H]
    \centering
    \begin{tabular}{|c|c|c|c|c|c|}
    \cline{2-6}\cline{2-6}
  \multicolumn{1}{c|}{ }  &  {$\kappa$} &   {$\varrho$} &$\gamma$ & $m$& $\tau m$\\  \hline
   Case 1      & 4&0.9&0.8&3.6&2.88 \\
    Case 2     & 10&0.36&0.8&3.6&2.88 \\
    Case 3      & 7&0.8&0.5&5.6&2.88  \\
    Case 4     & 10&0.8&0.36&8&2.88\\
    \hline\hline
    \end{tabular}
    \caption{ {Value of the parameters of each CBP.}}%Offspring, control and threshold parameters.}
    \label{tab:binomialcases}
\end{table}

\begin{table}[H]
    \centering
    \begin{tabular}{|c|c|c|c|c|c|c|c|c|c|c|c|}
    \cline{2-12}\cline{2-12}
  % \multicolumn{1}{c}{ }& \multicolumn{11}{c}{}\\
  \multicolumn{1}{c|}{ }  & 0 & 1 & 2 & 3 & 4 & 5 & 6 & 7  & 8 & 9 & 10\\  \hline
%   & $P_0(\kappa)$& $P_1(\kappa)$&$P_2(\kappa)$ & $P_3(\kappa)$& $P_4(\kappa)$&$P_5(\kappa)$&$P_6(\kappa)$&$P_7(\kappa)$&$P_8(\kappa)$&$P_9(\kappa)$&$P_{10}(\kappa)$\\  \hline
   Case 1      &  0.0001& 0.004& 0.052& 0.344& 1.000& &&&&& \\
    Case 2     &  0.012& 0.076& 0.241& 0.487& 0.729& 0.893& 0.970& 0.994& 0.999& 1.000& 1.000 \\
    Case 3&     0.000& 0.000& 0.005& 0.033& 0.148& 0.423& 0.790& 1.000&&& \\
    Case 4     & 0.000& 0.000& 0.001& 0.001& 0.006& 0.033& 0.121& 0.322& 0.624& 0.893& 1.000\\
    \hline\hline
    \end{tabular}
    \caption{ {Values of the cumulative distribution function associated with each offspring distribution.}}%Cumulative offspring distribution functions.}
    \label{tab:binomialprob}
\end{table}

\begin{figure}[H]
\centering\includegraphics[width=0.38\textwidth]{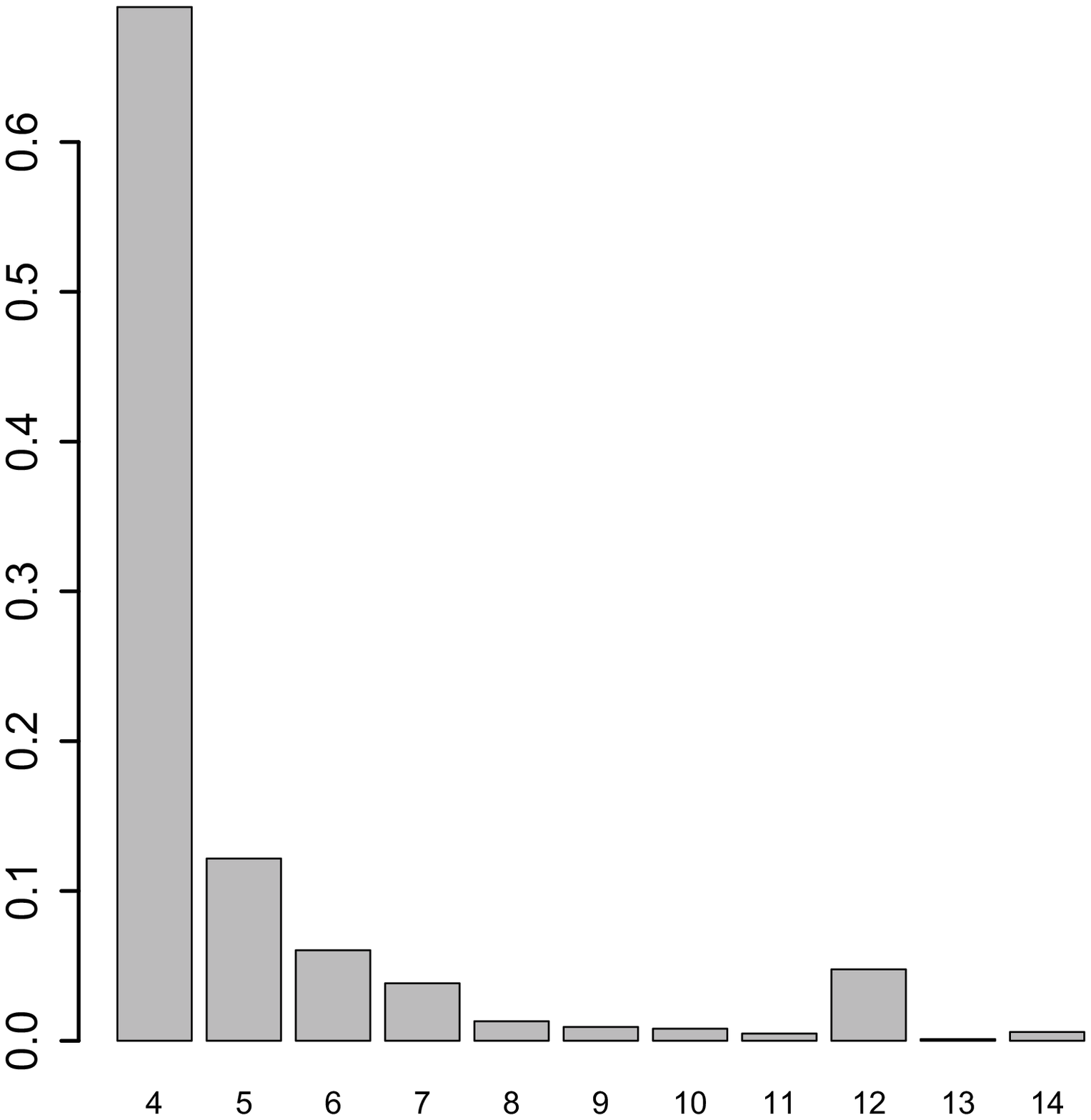}
\hspace*{0.1\textwidth}
\includegraphics[width=0.38\textwidth]{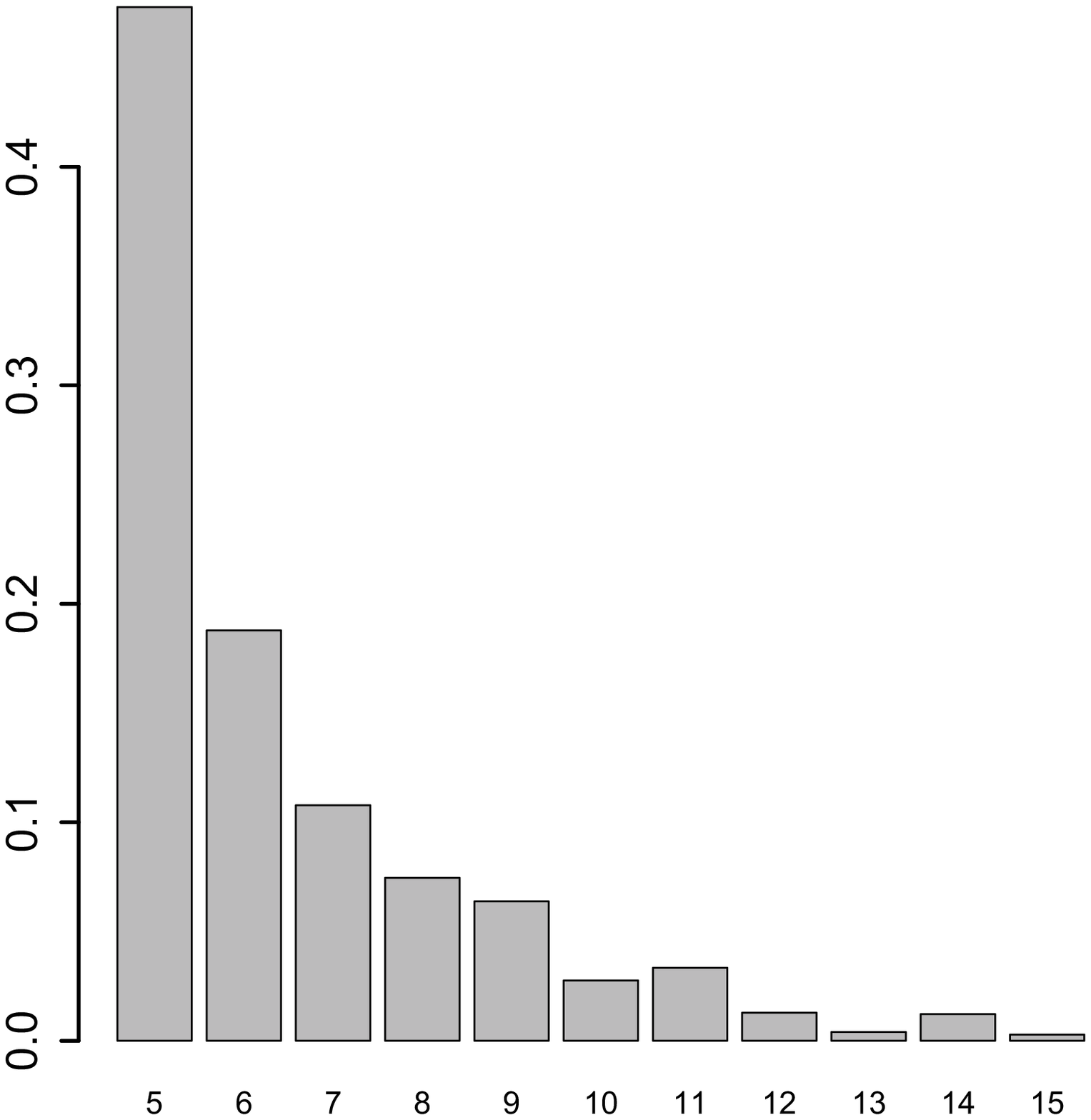} 
\caption{ Estimate of  the posterior  of $\kappa$ obtained in the first step of the ABC SMC algorithm for model choice in Example 1. Left: Case 1. Right: Case 2.}\label{fig:particular cases12}
\end{figure}

\begin{figure}[H]
\centering\includegraphics[width=0.38\textwidth]{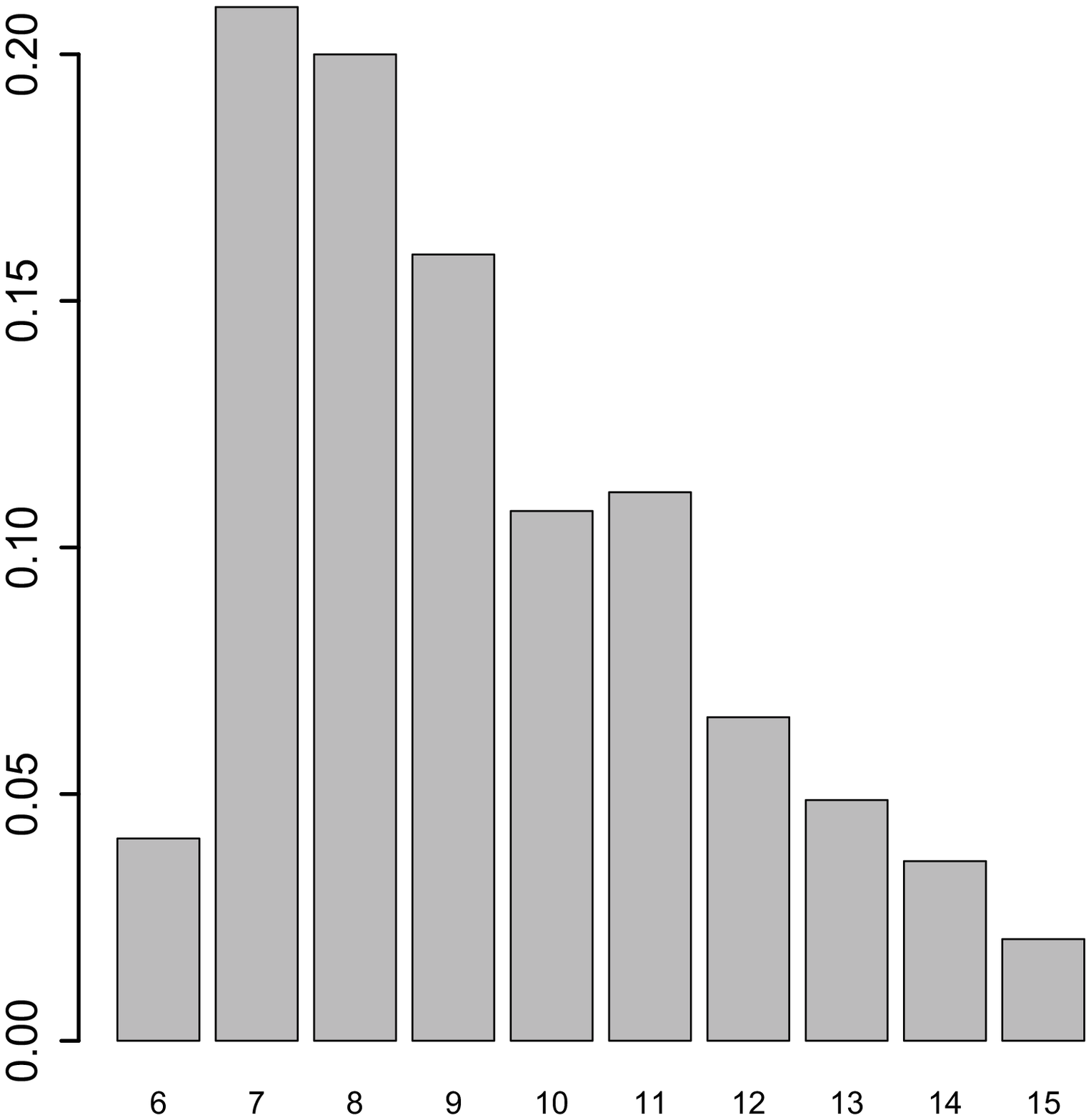}\hspace{0.1\textwidth}
\includegraphics[width=0.38\textwidth]{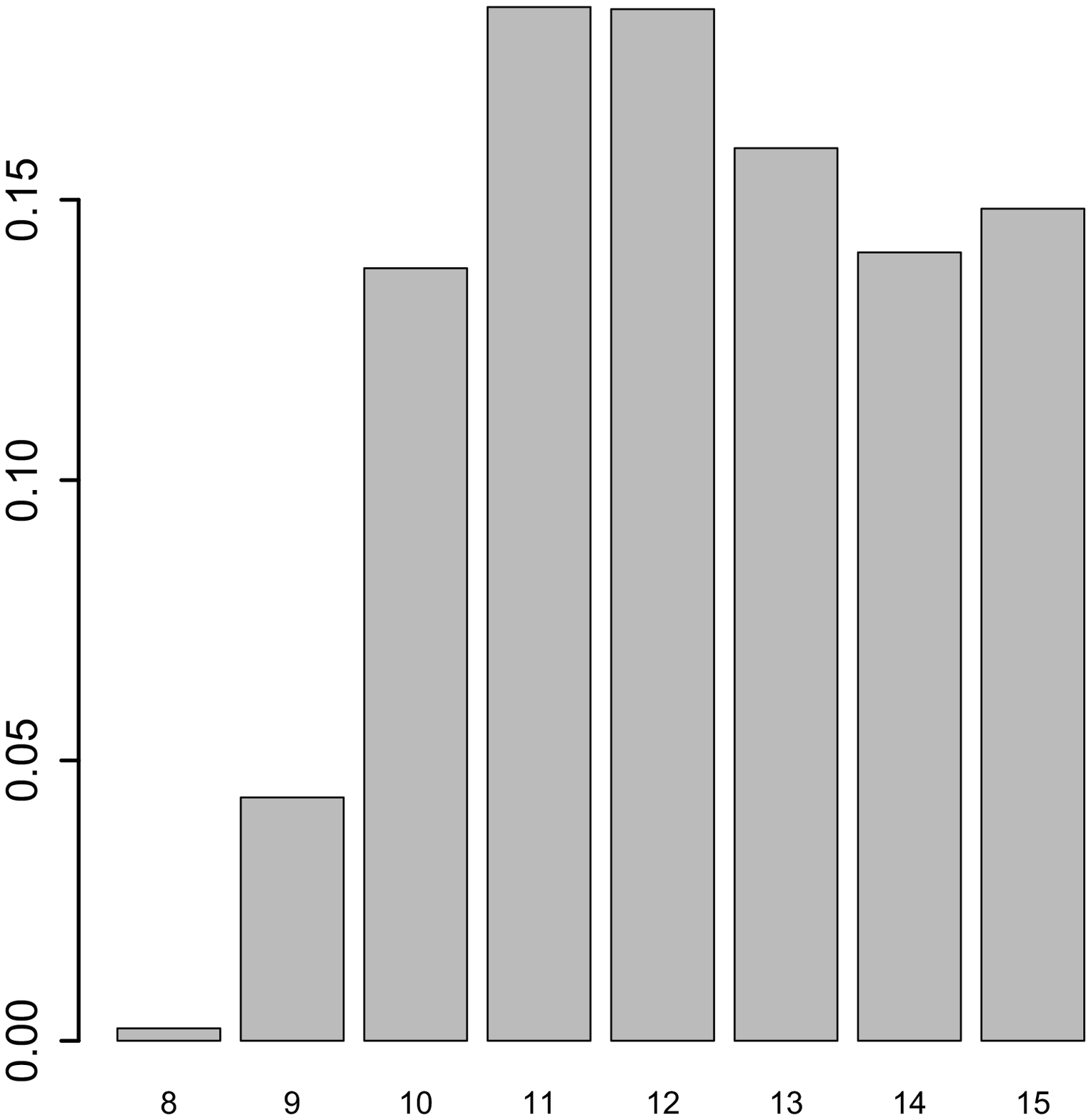}
\caption{Estimate of  the posterior of $\kappa$ obtained in the first step of the ABC SMC algorithm for model choice in Example 1. Left: Case 3. Right: Case 4.}\label{fig:particular cases34}
\end{figure}

For each case described above we simulated the first 10 generations of a CBP and we ran the ABC SMC algorithm for model choice with each of the corresponding samples as observed data (see Table~\ref{tab:sup:sim-data-ex2}  in Appendix for details on the samples). For that purpose, we assumed that our only knowledge on the offspring distribution is an upper bound for $\kappa$, and the fact that the control laws for a population size $j$ are binomial distributions with parameters $\xi(j)$ and $\gamma$, with $\gamma\in (0,1)$ unknown. To run the ABC SMC algorithm for model choice we fixed $T=3$ iterations, an upper bound $K_{max}=15$,  $\talpha_\kappa=(1,\ldots, 1)$, where the prior for $\gamma$ is a beta distribution with both parameters equal to 1, and the tuning parameter is $a=30$. The choice of the value of $a$ was justified by the results of several simulated experiments to avoid that the proposal distribution becomes a Dirac measure at the point where it is perturbed. We simulated pools of  $4\cdot 10^5$, $2\cdot 10^6$, $20\cdot 10^6$ of non-extinct CBPs at the corresponding iterations and fixed as the tolerance levels $\epsilon_1$ , $\epsilon_2$ and $\epsilon_3$ the quantiles of orders 0.0125, 0.0025, and 0.00025, respectively, of the
sample of the distances between the paths of the simulated and observed processes. As a result, for each sample path observed we obtained a sample of size 5000 of the corresponding posterior distribution of $\kappa$. The barplots of these samples are given in Figure \ref{fig:particular cases12} in the Cases 1 and 2, and in Figure \ref{fig:particular cases34} in the Cases 3 and 4. %, whereas the point estimates of the maximum offspring capacity for each case are summarized in Table \ref{tab:particular cases} \blue{Do we keep Table 3?}.  
In the Case 1, with $\kappa=4$, the distribution is concentrated around 4 and the point estimate is $\tilde{\kappa}_n=5$ due to the tail of the distribution. In the Case 2, with $\kappa=10$, $\tilde{\kappa}_n=7$ while the posterior distribution is right-skewed too. The posterior distribution in the Case 3, with $\kappa=7$, has a similar shape, but with support $\{6,\ldots,15\}$, and $\tilde{\kappa}_n=9$. Finally, in the Case 4, with $\kappa=10$, the posterior distribution is more symmetric than in the previous cases and the point estimate is $\tilde{\kappa}_n=12$. Taking into account the cumulative distribution function associated with each of the offspring distributions (see Table \ref{tab:binomialprob}), the proposed estimate of $\tilde{\kappa}_n$ in each case is quite reasonable.  %\red{\sout{Moreover, to illustrate the variability associated with the possible results,  we repeated the algorithm  100 times for each model/case. We show the results at the end of this example in Table~\ref{tab:freq} and Figures~\ref{fig:estimate1}-\ref{fig:estimate4}, and they indicate that the estimates obtained for the particular examples developed are among the most frequent (see Table \ref{tab:freq}).}}

%\begin{table}[H]
%    \centering
%    \begin{tabular}{|c|c|c|c|c|}
%    \cline{2-5}\cline{2-5}
%   \multicolumn{1}{c|}{ } & Case 1& Case 2&Case 3 & Case 4\\  \hline
%  $\widetilde{\kappa}_n$& 5& 7 & 9 & 12\\
%  \hline
%  $\kappa$ & 4 & 10 & 7 & 10\\
%  \hline\hline
%    \end{tabular}
%    \caption{Point estimates and true value of maximum number of offspring per individual.}
%    \label{tab:particular cases}
%\end{table}

%Following with the development of the particular examples, next, 
\begin{figure}[H]
\centering\includegraphics[width=0.32\textwidth]{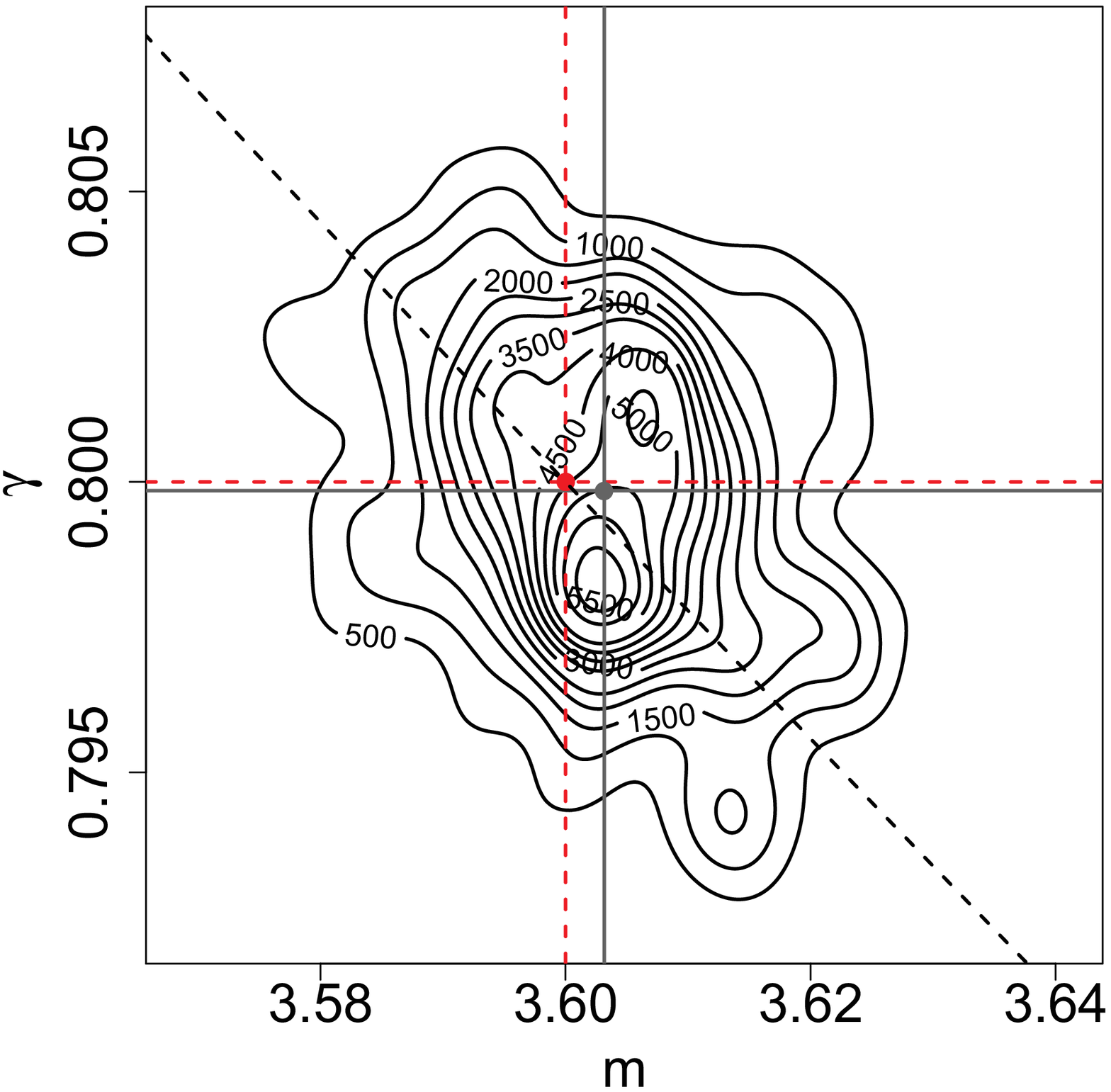}
\includegraphics[width=0.32\textwidth]{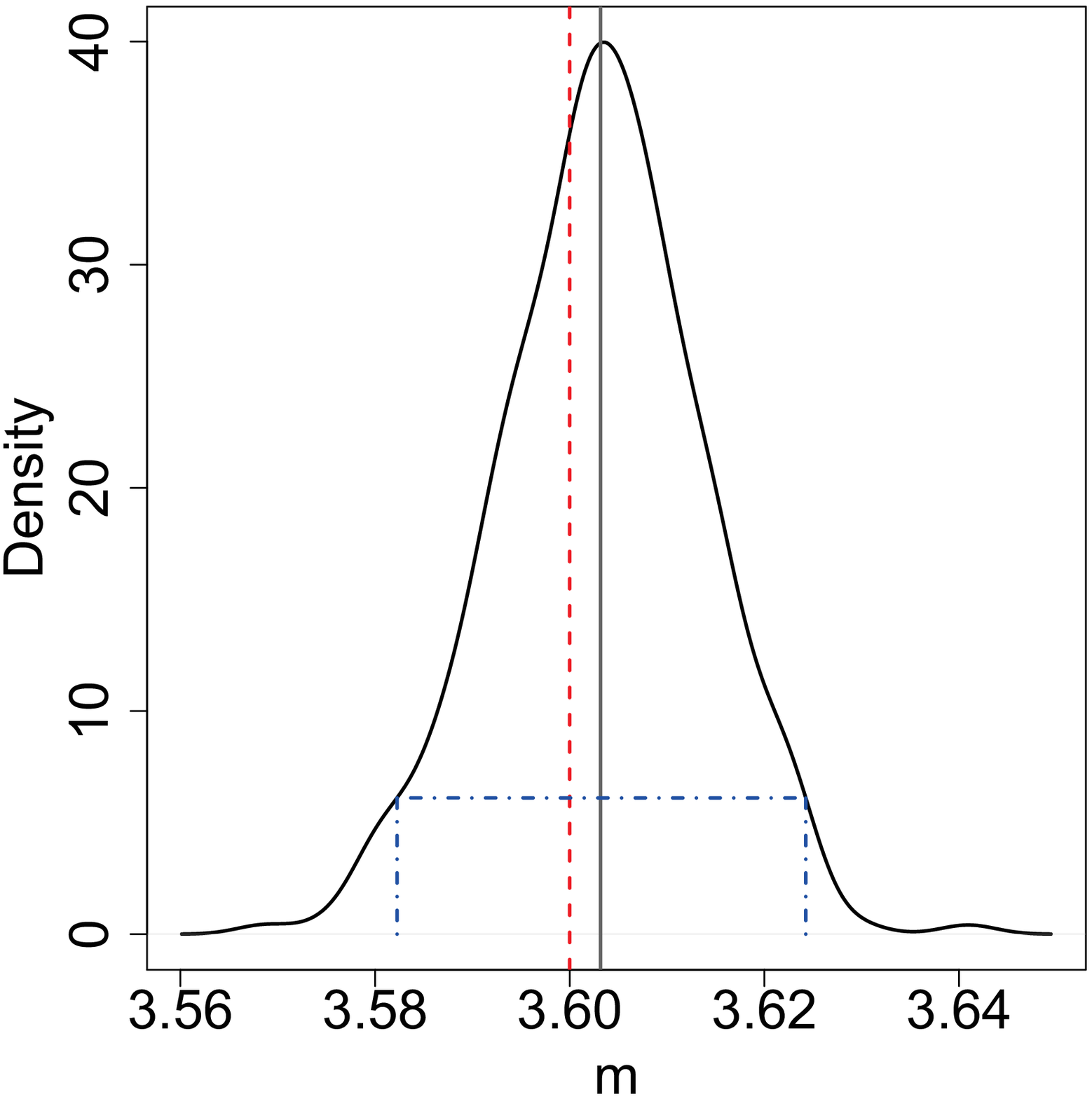}
\includegraphics[width=0.32\textwidth]{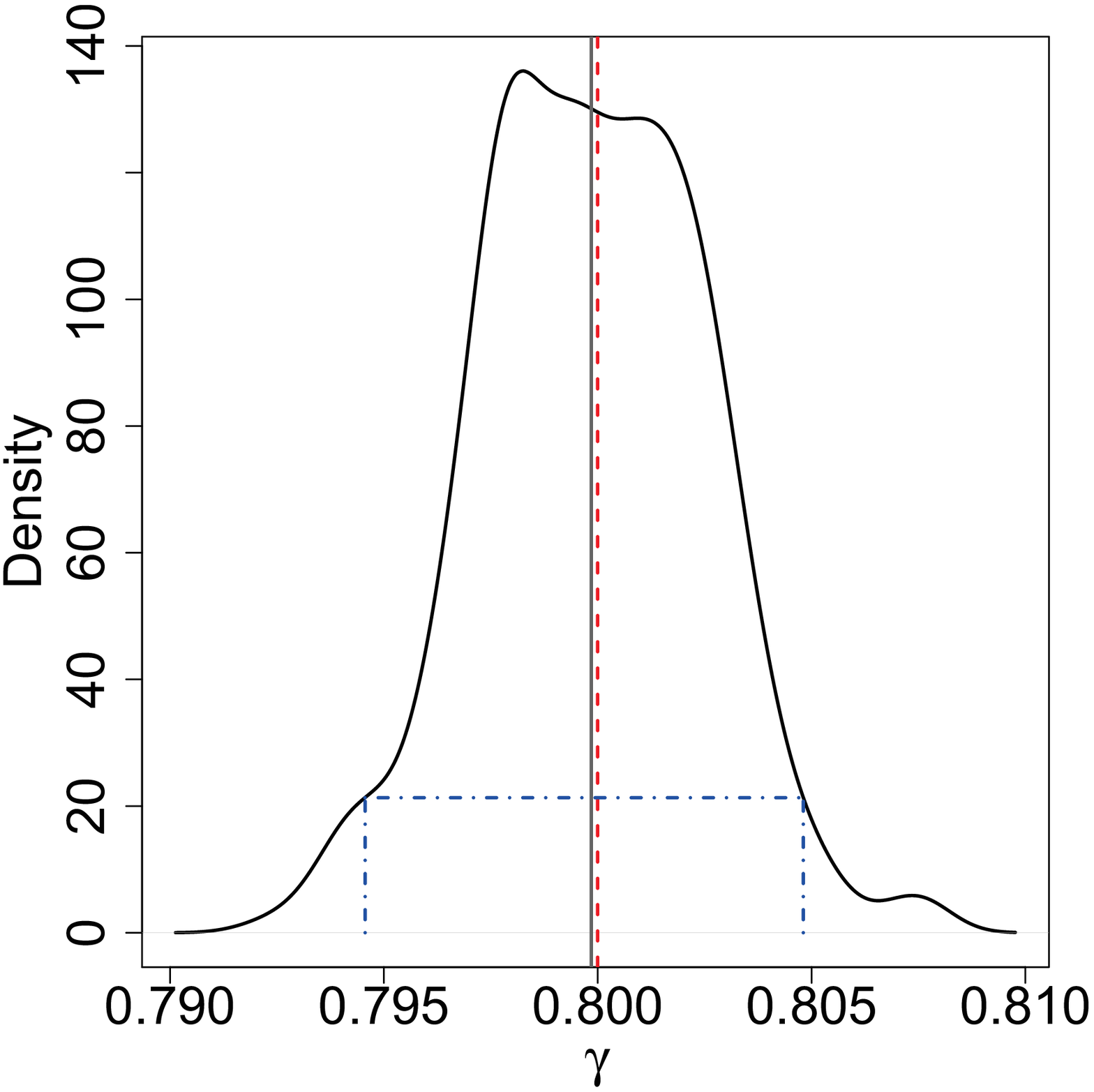}
\caption{Case 1. Estimates of the posterior distributions via the ABC algorithm with the local linear regression adjustment with $\widetilde{\kappa}_n=5$. Left: Contour plot of the estimates of the joint density $\pi(m,\gamma\mid\widetilde{\kappa}_n,\widetilde{\mathcal{Z}}_n^{obs})$, together with the curve $\tau m = 2.88$. The red point corresponds to the true values of the parameters and the grey point corresponds to the sample means. Centre: Estimate of  $\pi(m\mid\widetilde{\kappa}_n,\widetilde{\mathcal{Z}}_n^{obs})$. Right: Estimate of  $\pi(\gamma\mid\widetilde{\kappa}_n,\widetilde{\mathcal{Z}}_n^{obs})$.  Red dashed vertical lines represent the true value of the parameter, grey solid vertical lines are the sample means, and blue dashed-dotted vertical lines correspond to 95\%  HPD intervals.}\label{fig:c1mgama-second-step}
\end{figure}

\begin{figure}[H]
\centering\includegraphics[width=0.32\textwidth]{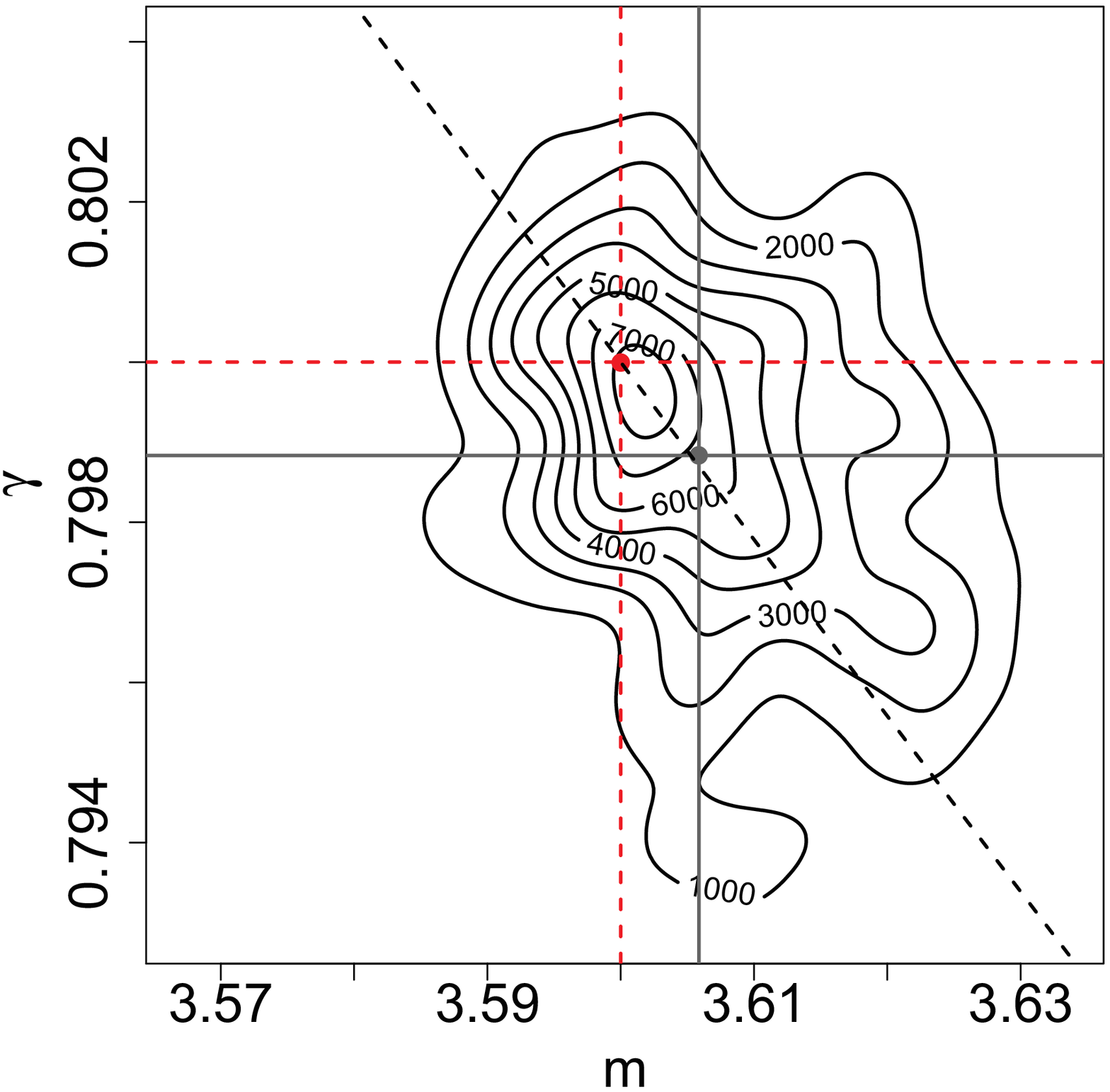}
\includegraphics[width=0.32\textwidth]{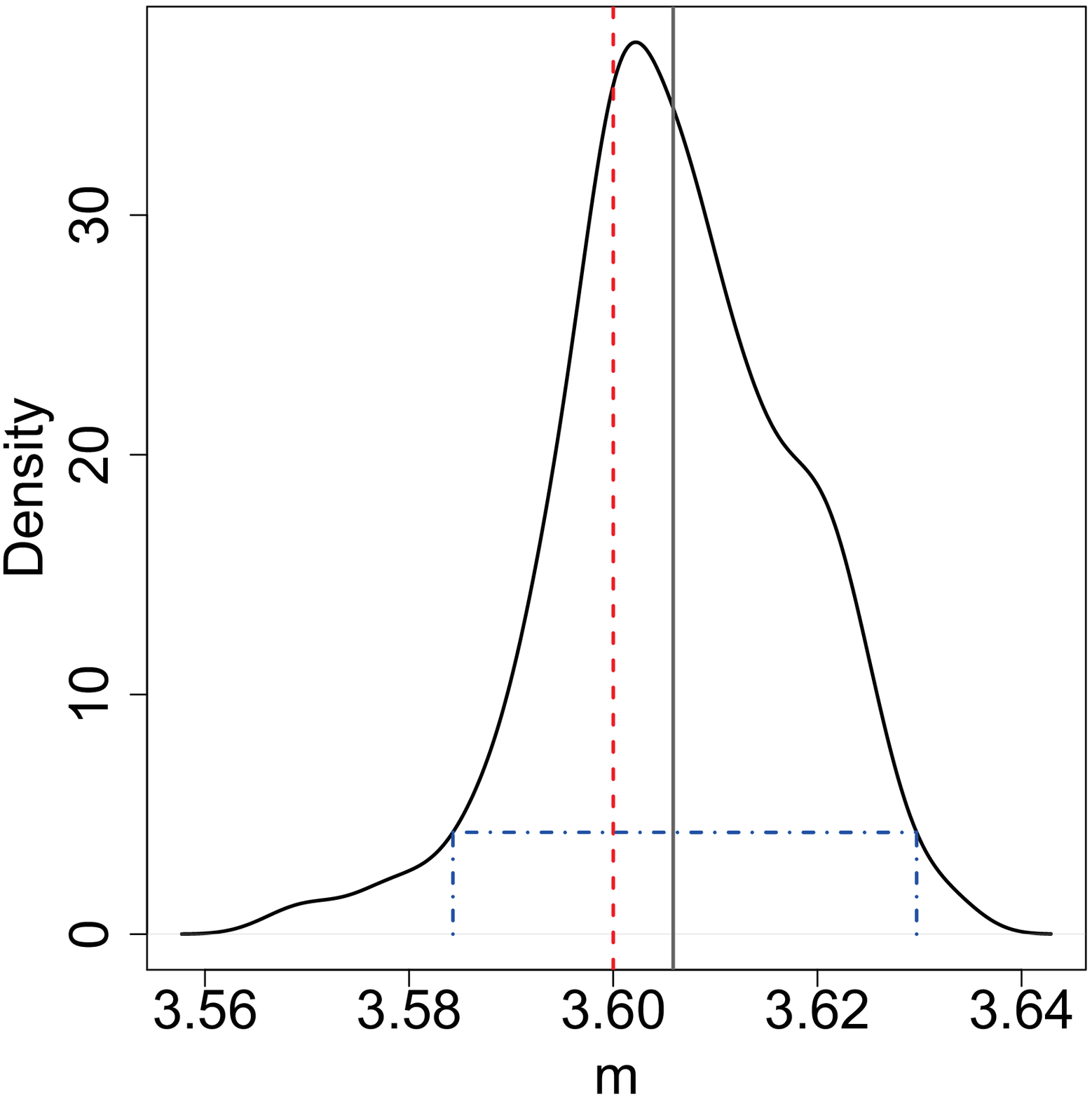}
\includegraphics[width=0.32\textwidth]{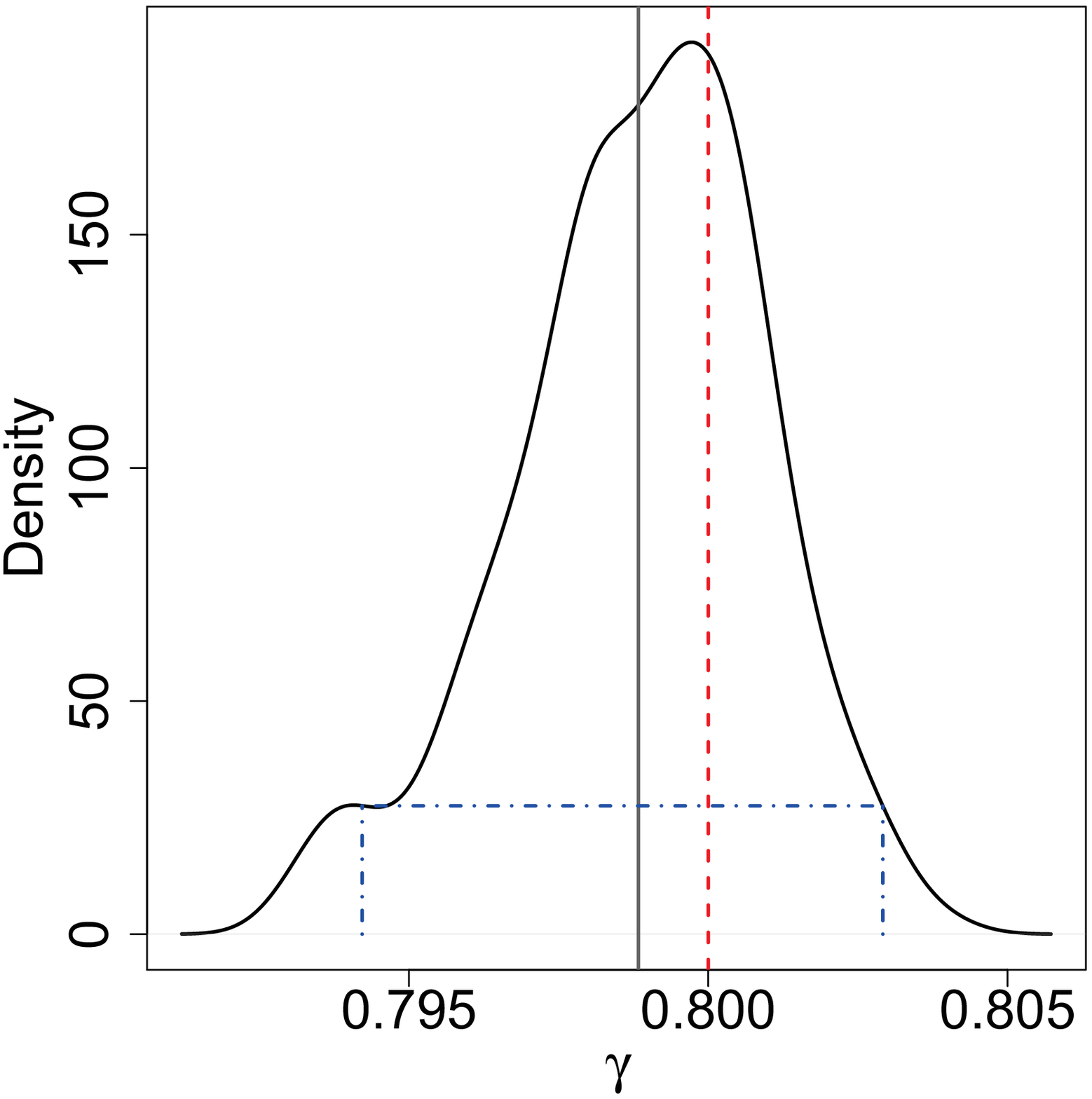}
\caption{Case 2. Estimates of the posterior distributions via the ABC algorithm with the local linear regression adjustment with $\widetilde{\kappa}_n=7$. Left: Contour plot of the estimates of the joint density $\pi(m,\gamma\mid\widetilde{\kappa}_n,\widetilde{\mathcal{Z}}_n^{obs})$, together with the curve $\tau m = 2.88$. The red point corresponds to the true values of the parameters and the grey point corresponds to the sample means. Centre: Estimate of  $\pi(m\mid\widetilde{\kappa}_n,\widetilde{\mathcal{Z}}_n^{obs})$. Right: Estimate of  $\pi(\gamma\mid\widetilde{\kappa}_n,\widetilde{\mathcal{Z}}_n^{obs})$.  Red dashed vertical lines represent the true value of the parameter, grey solid vertical lines are the sample means, and blue dashed-dotted vertical lines correspond to 95\%  HPD intervals.}\label{fig:c2mgama-second-step}
\end{figure}

\begin{figure}[H]
\centering\includegraphics[width=0.32\textwidth]{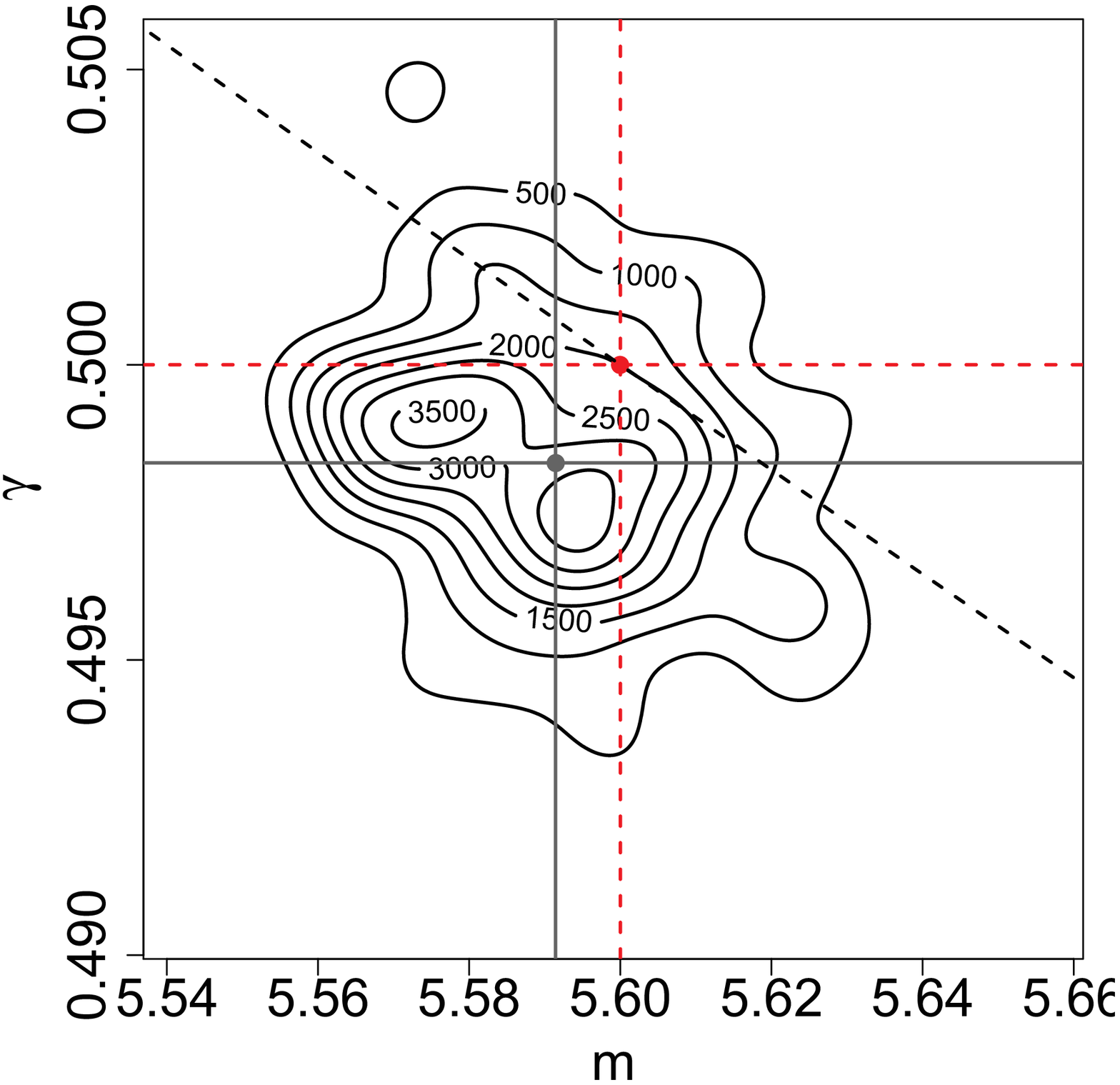}
\includegraphics[width=0.32\textwidth]{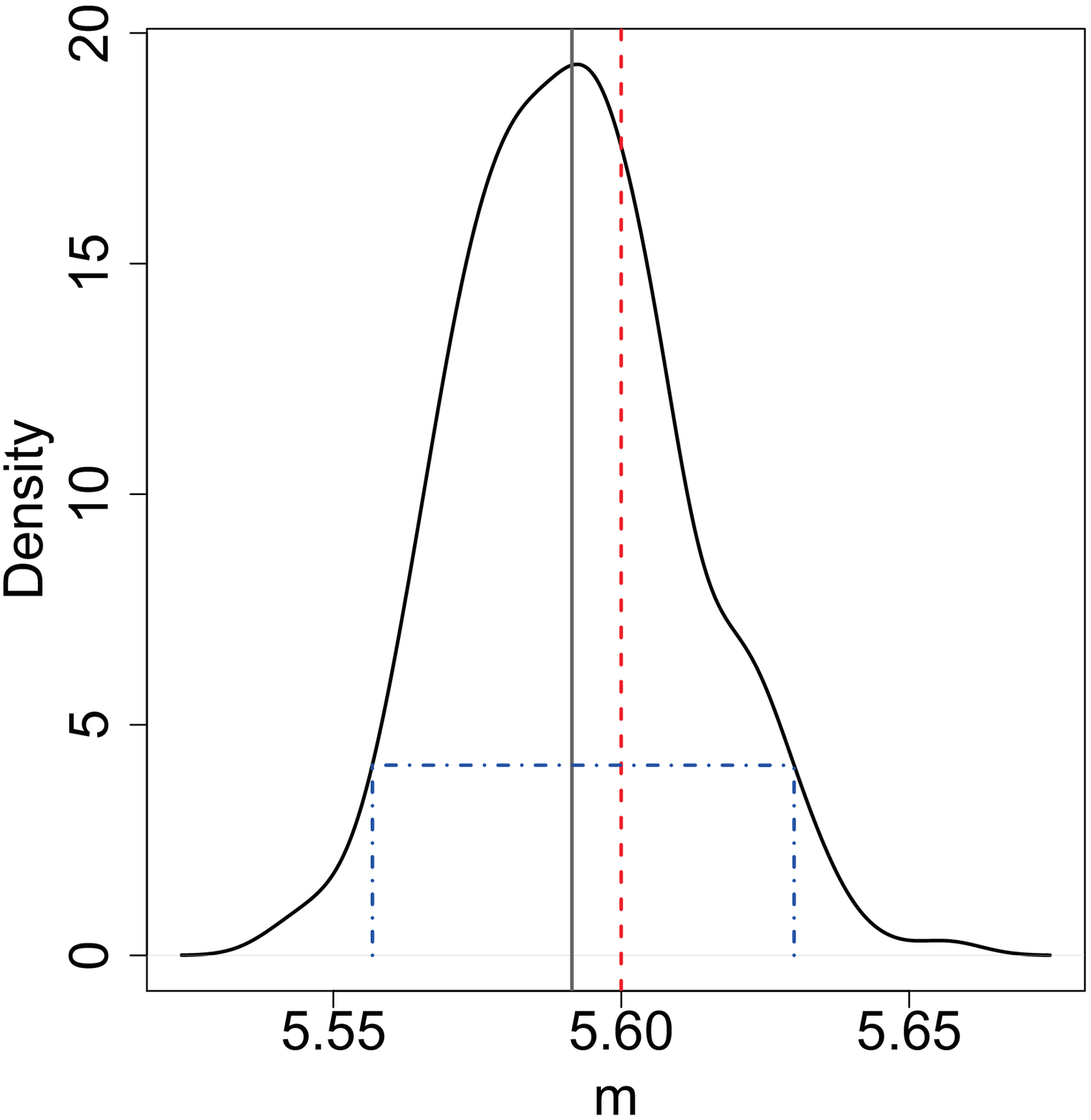} 
\includegraphics[width=0.32\textwidth]{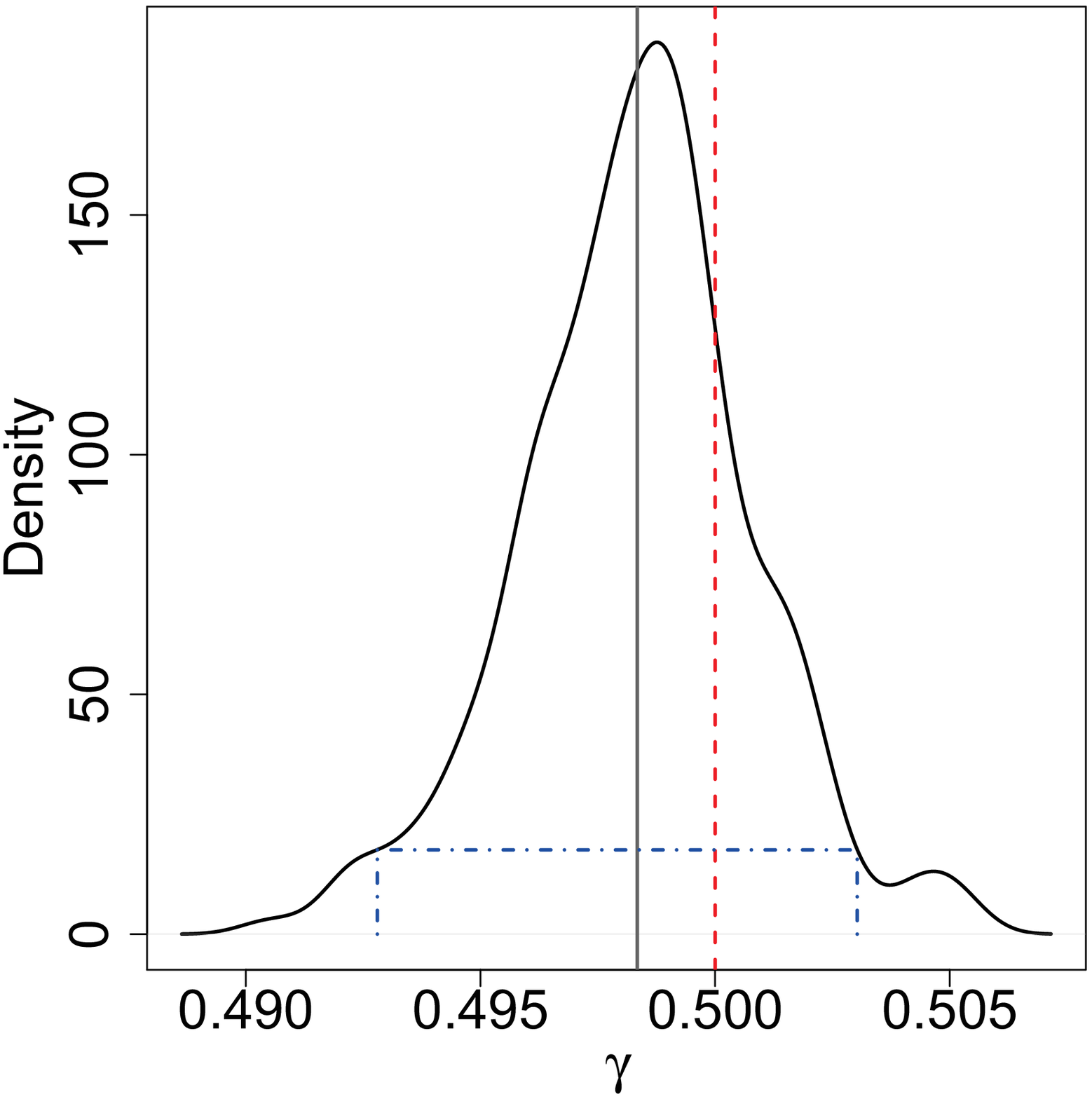}
\caption{Case 3. Estimates of the posterior distributions via the ABC algorithm with the local linear regression adjustment with $\widetilde{\kappa}_n=9$. Left: Contour plot of the estimates of the joint density $\pi(m,\gamma\mid\widetilde{\kappa}_n,\widetilde{\mathcal{Z}}_n^{obs})$, together with the curve $\tau m = 2.88$. The red point corresponds to the true values of the parameters and the grey point corresponds to the sample means. Centre: Estimate of  $\pi(m\mid\widetilde{\kappa}_n,\widetilde{\mathcal{Z}}_n^{obs})$. Right: Estimate of  $\pi(\gamma\mid\widetilde{\kappa}_n,\widetilde{\mathcal{Z}}_n^{obs})$.  Red dashed vertical lines represent the true value of the parameter, grey solid vertical lines are the sample means, and blue dashed-dotted vertical lines correspond to 95\%  HPD intervals.}\label{fig:c3mgama-second-step}
\end{figure}

\begin{figure}[H]
\centering\includegraphics[width=0.32\textwidth]{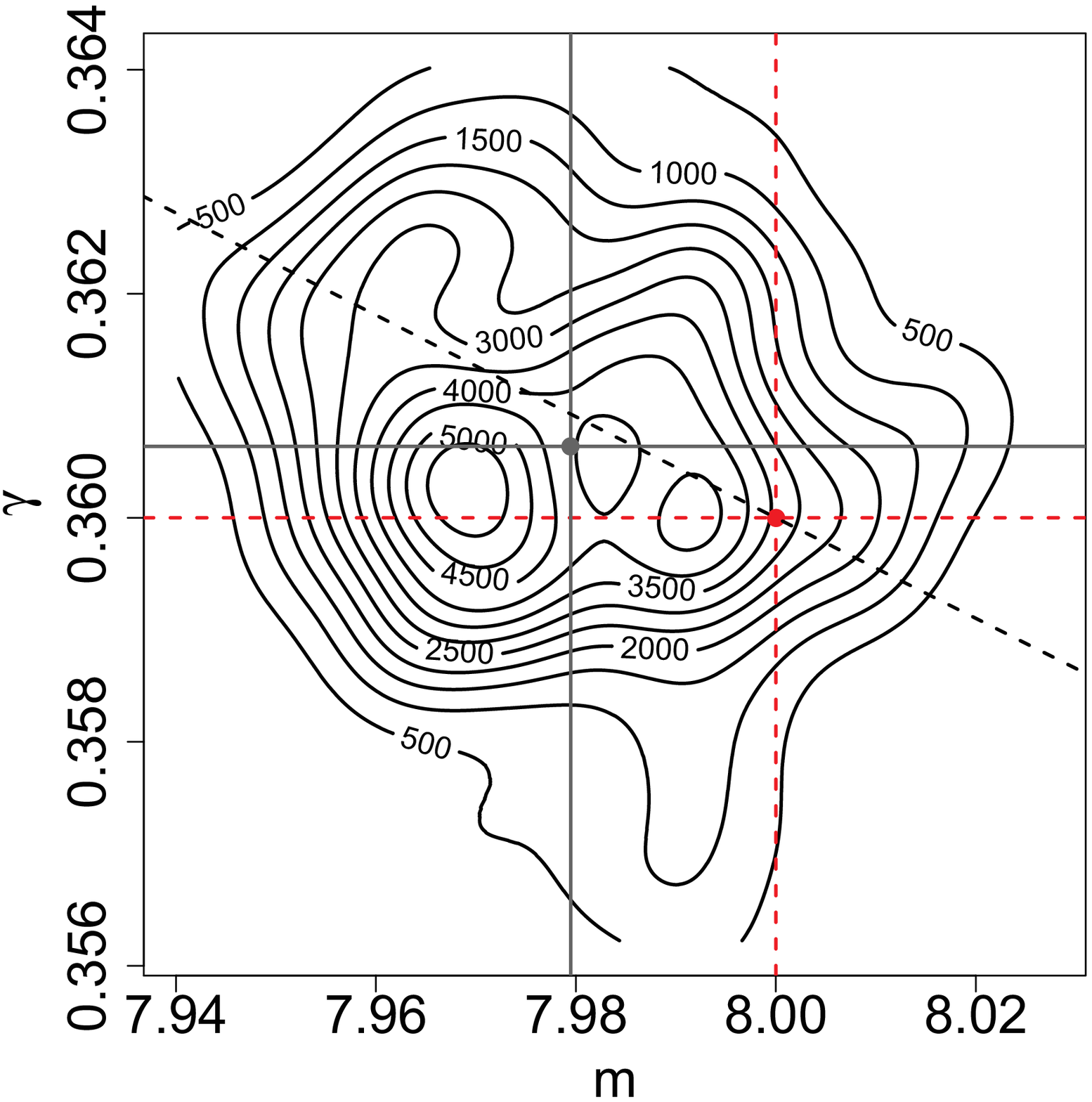}
\includegraphics[width=0.32\textwidth]{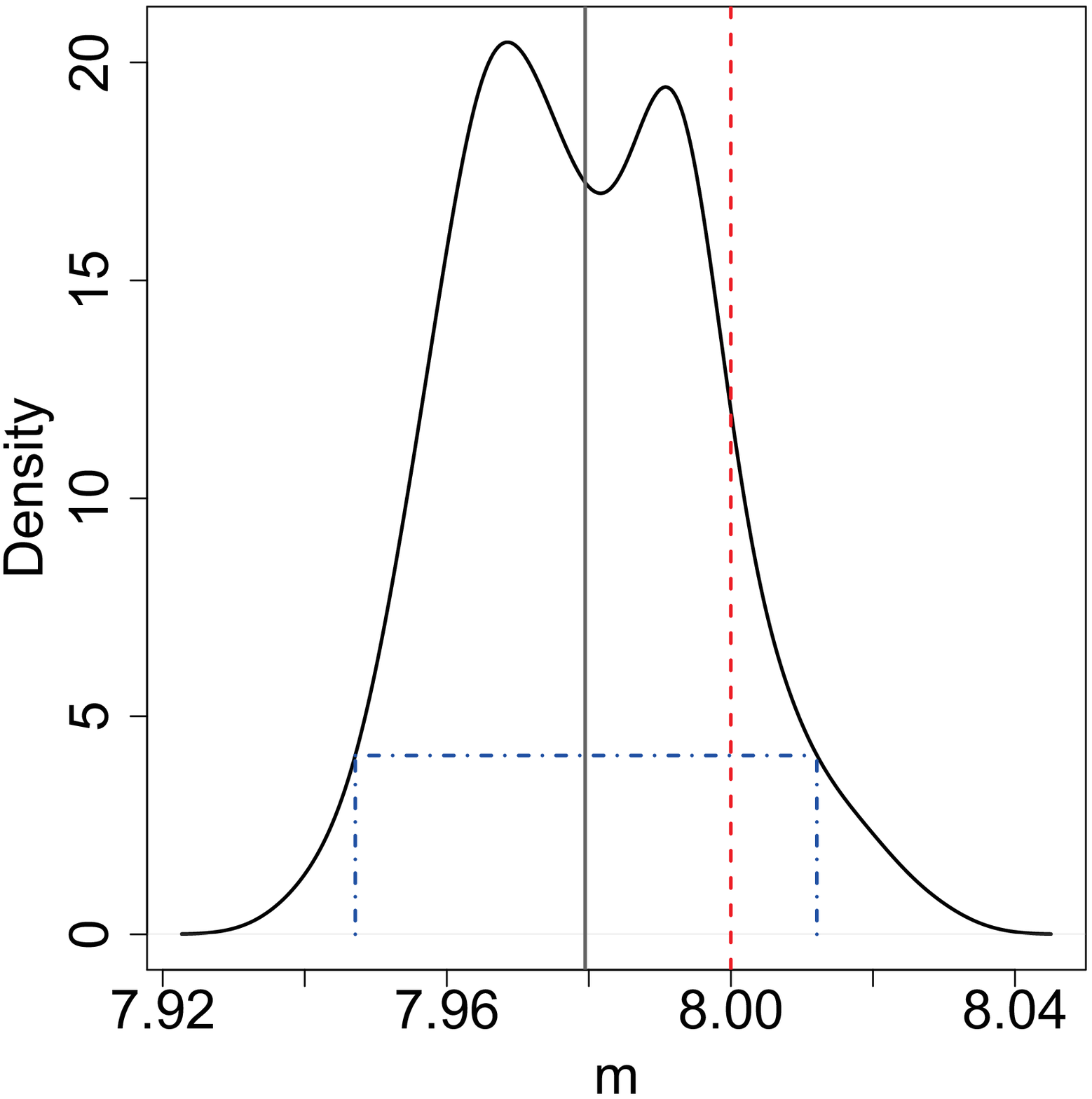} 
\includegraphics[width=0.32\textwidth]{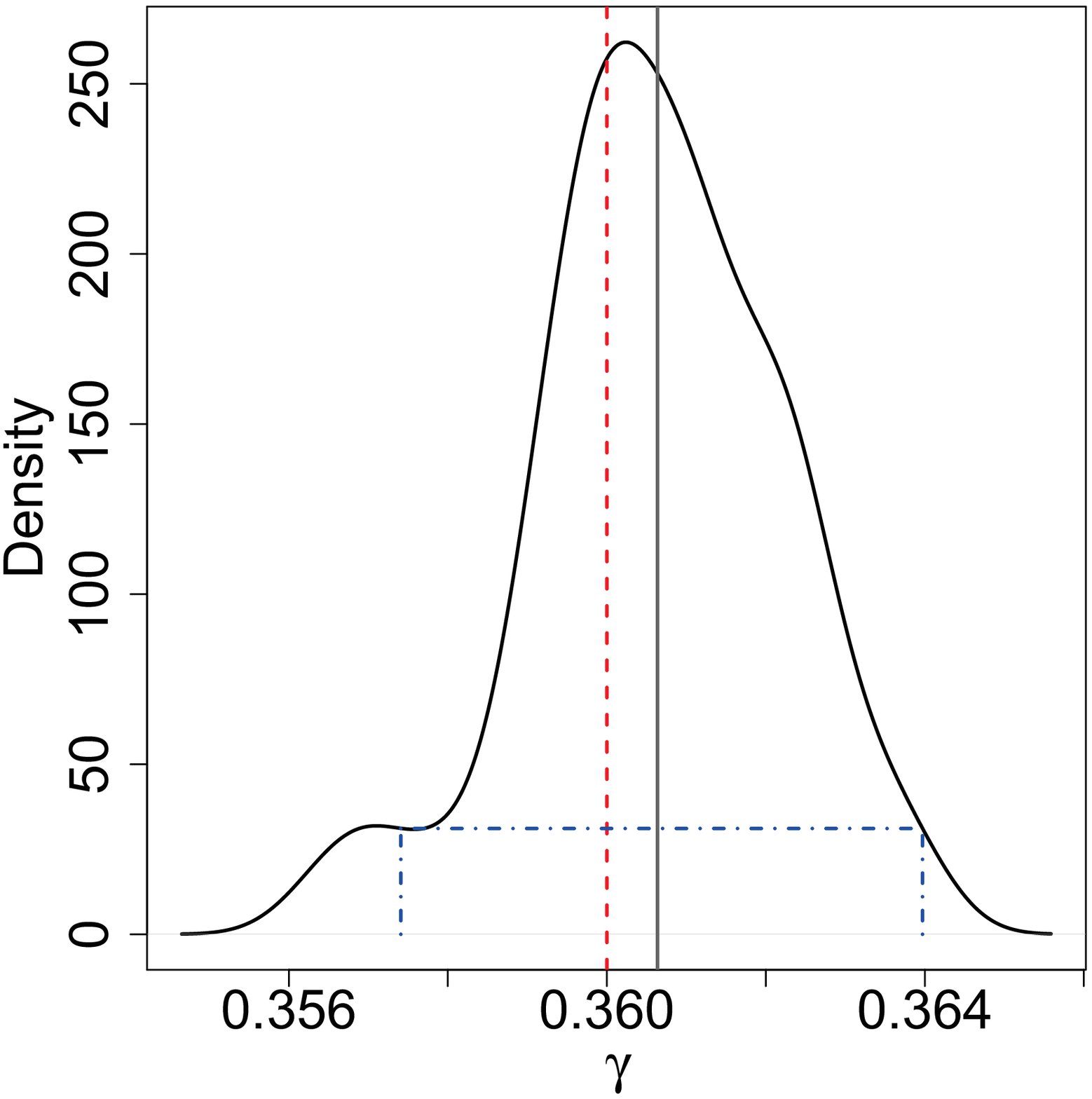}
\caption{Case 4. Estimates of the posterior distributions via the ABC algorithm with the local linear regression adjustment with $\widetilde{\kappa}_n=12$. Left: Contour plot of the estimates of the joint density $\pi(m,\gamma\mid\widetilde{\kappa}_n,\widetilde{\mathcal{Z}}_n^{obs})$, together with the curve $\tau m = 2.88$. The red point corresponds to the true values of the parameters and the grey point corresponds to the sample means. Centre: Estimate of  $\pi(m\mid\widetilde{\kappa}_n,\widetilde{\mathcal{Z}}_n^{obs})$. Right: Estimate of  $\pi(\gamma\mid\widetilde{\kappa}_n,\widetilde{\mathcal{Z}}_n^{obs})$.  Red dashed vertical lines represent the true value of the parameter, grey solid vertical lines are the sample means, and blue dashed-dotted vertical lines correspond to 95\%  HPD intervals.}\label{fig:c4mgama-second-step}
\end{figure}

We continued with the second step of our methodology by performing the tolerance-rejection algorithm and the post-processing method with the summary statistic to draw samples from distributions that approximate the posteriors $\pi(m\mid\widetilde{\kappa}_n,\widetilde{\mathcal{Z}}_n^{obs})$ and $\pi(\gamma\mid\widetilde{\kappa}_n,\widetilde{\mathcal{Z}}_n^{obs})$ in each case. The estimates of the joint posterior $\pi(m,\gamma \mid\widetilde{\kappa}_n,\widetilde{\mathcal{Z}}_n^{obs})$ densities and their marginal posterior distributions for each case are displayed in Figures~\ref{fig:c1mgama-second-step}-\ref{fig:c4mgama-second-step}.  In all cases one can observe that the estimated densities obtained are centred around the true values and their spread is relatively small. These results indicate that the method retrieves the parameters of interest reasonably well, which is a key property to predict the evolution  of the population.%of parameters of interest of the model such as the offspring mean $m$ and the control parameter $\gamma$. To that end,}%In all cases one can observe that while the estimated densities obtained with both methods are centred around the true values, the post-processing reduces dramatically the spread of the distributions. These results indicate that the method retrieves the parameters of interest reasonably well, which is a key property to predict the evolution  of the population.%of parameters of interest of the model such as the offspring mean $m$ and the control parameter $\gamma$. To that end,}

Besides the four particular examples presented above, in the second part of this subsection,  we analyse in more detail the accuracy of the methodology to estimate the posterior distributions for $\kappa$ when the support of the reproduction law is finite. %, running 100 times the proposed algorithm for each case. 
Specifically, for each of the previous four models, we simulated the first 10 generations of 100 processes starting with one individual for each of the cases (i.e., 100 different observed samples), and we ran the ABC SMC algorithm for model choice algorithm with each of these observed samples. To this aim, the same number of iterations, prior distributions,  tuning parameter as above are set, but we considered simulated pools of 16000, 80000 and 800000 of non-extinct CBPs at the corresponding iterations and fixed as the tolerance levels
$\epsilon_1$ , $\epsilon_2$ and $\epsilon_3$ the quantiles of orders 0.0125, 0.0025, and 0.00025, respectively, of the
sample of the distances between the simulated and the observed processes.   As a result, for each of the 100 observed paths we obtained a sample of  size 200 drawn from  the posterior of $\kappa$, and we computed the Bayesian point estimate $\widetilde{\kappa}_n$. Thus, we got a sample of size 100 of estimates, $\tilde{\kappa}_{n,1},\ldots,\tilde{\kappa}_{n,100}$,  for each model/case. The corresponding relative frequencies of the values of $\kappa$ are provided in Table \ref{tab:freq}.  We recall that the Cases 1 and 2 have the same offspring mean and control parameter, but the offspring distribution in Case 2 is concentrated in greater values than in Case 1 (see Table \ref{tab:binomialprob}).  Our results indicate that the algorithm proposed is  able to distinguish and to identify both cases reasonably well, as was reported above in the study developed above for each particular case. 
%probability with the same mean, being concentrated on the biggest values of it in Case 1 and on the smallest ones in Case 2 (see Table \ref{tab:binomialprob}). The control parameter is the same in both cases. The algorithm proposed is  able to distinguish and to identify reasonable well both cases. 
We also observe that the skewness of the reproduction law has some impact on the shape of the probability distribution of the Bayesian point estimator of $\kappa$, $\widetilde{\kappa}_n$. Indeed, the first offspring distribution is left-skewed, and the method tends to overestimate the value of $\kappa$, while the second one is right-skewed and the method tends to underestimate it.
%Due to the skewness that the offspring distributions show it justifies that the method tends to overestimate and underestimate  the respective values of $\kappa$. 
In particular, in the Case 1, the choices 5 and 6 cover the  86\% of the  values of the sample, where 5 has a relative frequency of 48\%. In the Case 2, the choices 6, 7 and 8 cover the 72\%,  where 7 has a relative frequency of 30\%; notice in this case that the cumulative probabilities for the values 6, 7, and 8, are 0.97, 0.994, and 0.999, respectively. Regarding the Cases 3 and 4, we remark that both of them have different offspring means and control parameters,  and the methodology is  able to discriminate satisfactorily between both.  Precisely, the choices 9 and 10 represent the 78\% of the values of the sample in the Case 3 whereas 11 and 12 correspond to the 84\% in the Case~4. It is also important to highlight that the range of selected values of $\widetilde{\kappa}_n$ for all the cases are different  (see Table~\ref{tab:freq}),  and consequently, the performance of the method enables us to  estimate adequately the support of the offspring distributions. 
%, being the value of 5 reaches in the 50\%. Respect to Cases 3 and 4, both have different offspring means and control parameters, being the proposed method again able to discriminate satisfactory between both, choosing 9 and 10, in Case 3, and 12 and 13, in Case 4, in the 72\% of the repeated runs in both cases. 

%The results are similar to those reported above 
\begin{table}[H]
\centering{\begin{tabular}{|c|c|c|c|c|}
%\cline{2-5}
%\multicolumn{1}{c}{ }& \multicolumn{4}{|c|}{Cases}\\
\hline\hline
$\kappa$ & Case 1 & Case 2 & Case 3 & Case 4  \\
\hline
2 &  &  &  &  \\
3 &  &  &  &   \\
4 &  $0.08$& $0.03$ &  &  \\
5 & $0.48$ &$0.16$  &  &  \\
6 & $0.38$ &$0.21$  &  &  \\
7 &  $0.06$&$0.30$  &  &  \\
8 &  &$0.21$  &$0.13$  &  \\
9 &  &$0.09$  &$0.34$  &  \\
10 &  &  & $0.44$ &$0.01$  \\
11&  &  & $0.08$ &$0.31$  \\
12 &  &  &  $0.01$&  $0.53$\\
13 &  &  &  & $0.14$ \\
14 &  &  &  & $0.01$ \\
15 &  &  &  &  \\
\hline\hline
\end{tabular}}
\caption{ {Relative frequencies of the values of $\kappa$ in the sample $\tilde{\kappa}_{n,1},\ldots,\tilde{\kappa}_{n,100}$ for each model.}} \label{tab:freq}
 %\caption{The frequency of reaching the nearest integer function of the posterior mean over 100 replicates are shown.} \label{tab:freq}
\end{table}

%\begin{figure}[H]
%\centering\includegraphics[width=0.35\textwidth]{./graphs/1_m_boxplot_q_corrected2}\hspace*{0.04\textwidth}
%\includegraphics[width=0.35\textwidth]{./graphs/1_gamma_boxplot_q_corrected2} 
%\includegraphics[width=0.35\textwidth]{./graphs/2_m_boxplot_q_corrected2}\hspace{0.04\textwidth}\includegraphics[width=0.35\textwidth]{./graphs/2_gamma_boxplot_q_corrected2}
%\caption{Box-plots of estimated posterior means  for $m$ and $\gamma$  based on 100 repeated runs of the rejection-tolerance ABC with post processing  with the selected $\kappa$ in the ABC SMC  model choice algorithm.  {First row: Case 1. Second row: Case 2.}}\label{fig:estimate}
%\end{figure}

\begin{figure}[H]
\centering\includegraphics[width=0.39\textwidth]{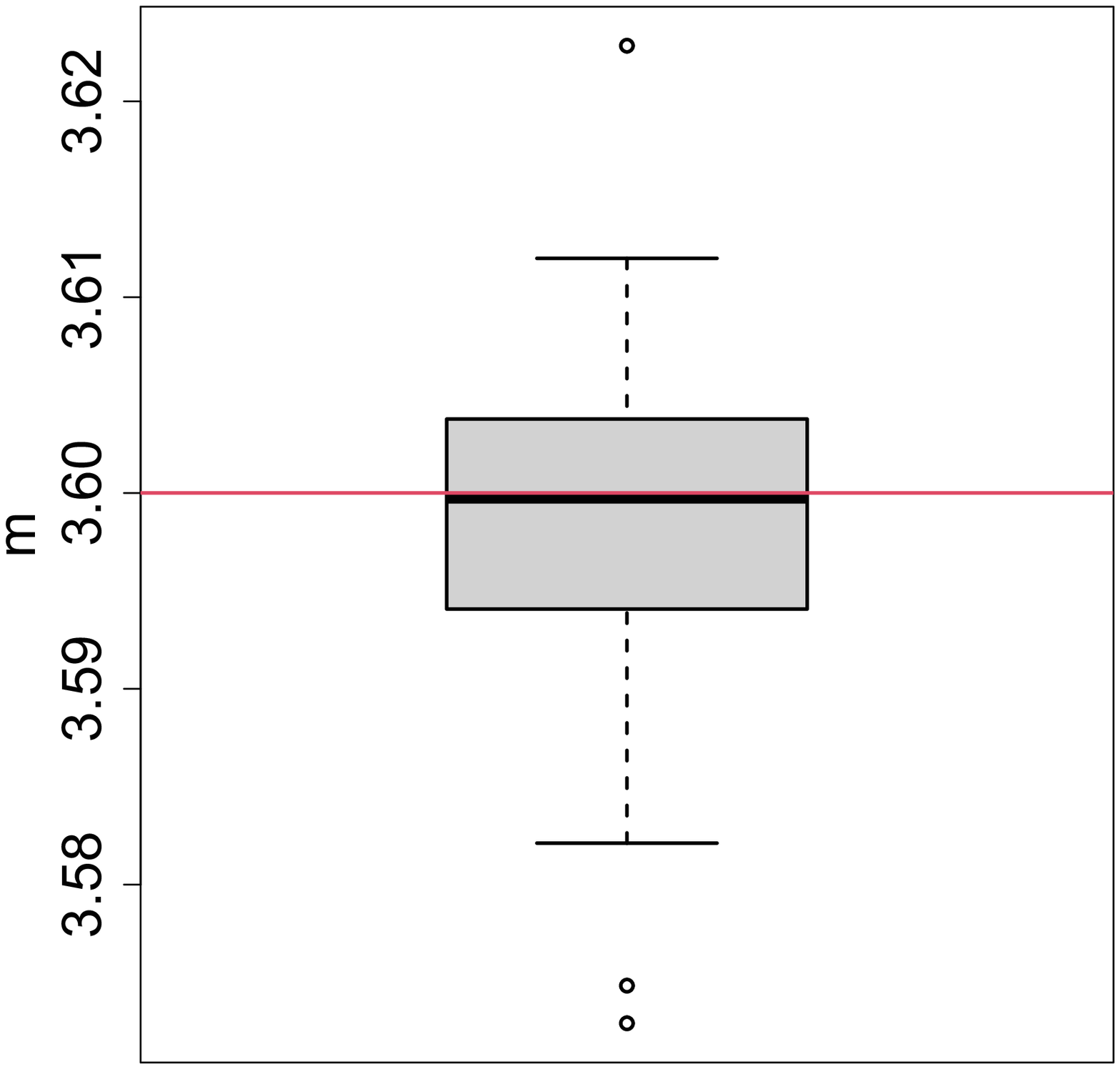}\hspace*{0.1\textwidth}
\includegraphics[width=0.39\textwidth]{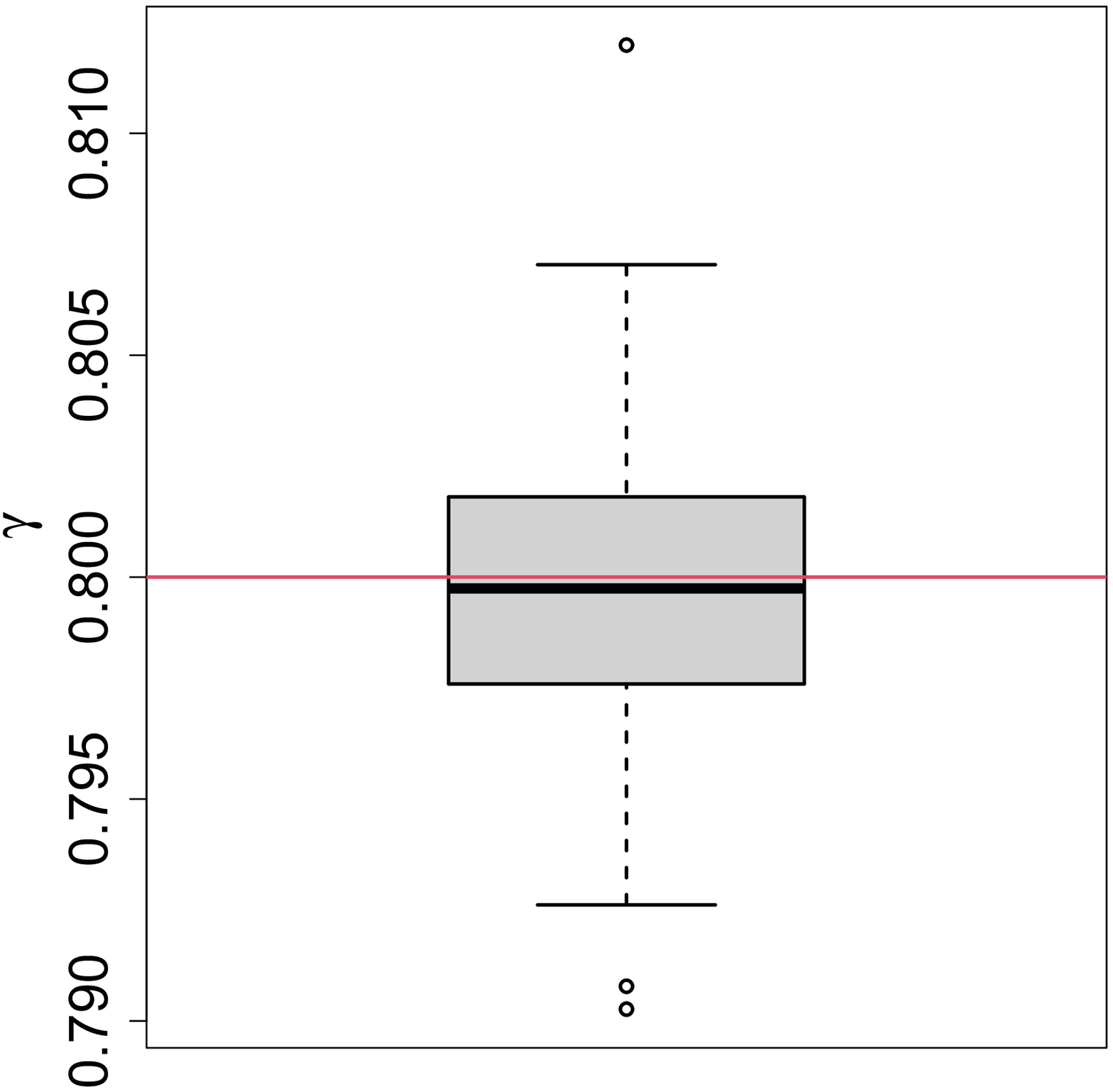} 
\caption{Case 1. Box-plots of Bayes point estimates of $m$ and $\gamma$, $\tilde{m}_{n,1},\ldots,\tilde{m}_{n,100}$ and $\tilde{\gamma}_{n,1},\ldots,\tilde{\gamma}_{n,100}$, based on the samples of 100 simulated processes. Horizontal red line corresponds to the true value of the parameter.}\label{fig:estimate1}
\end{figure}

\begin{figure}[H]
\centering\includegraphics[width=0.39\textwidth]{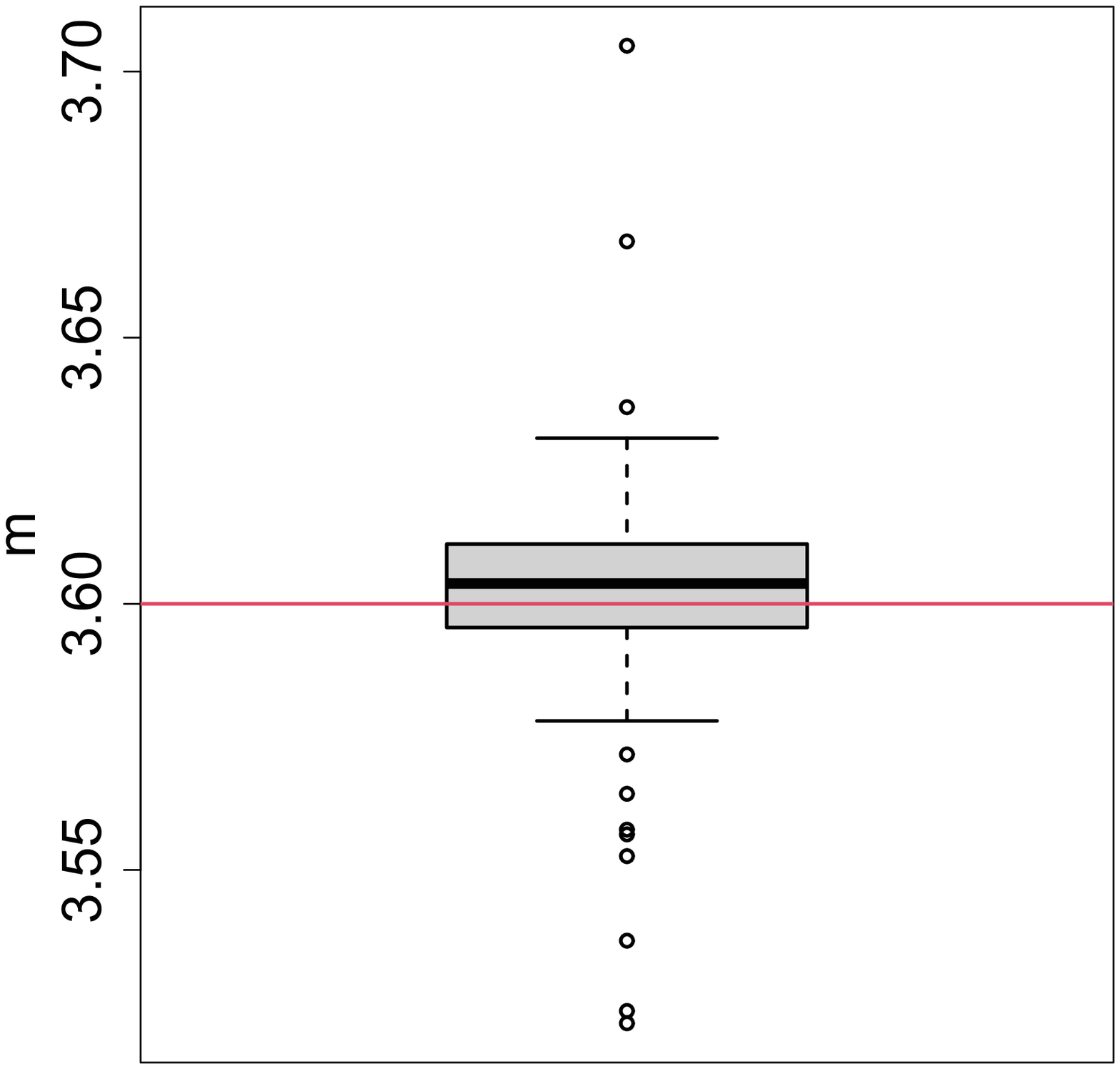}\hspace{0.1\textwidth}\includegraphics[width=0.39\textwidth]{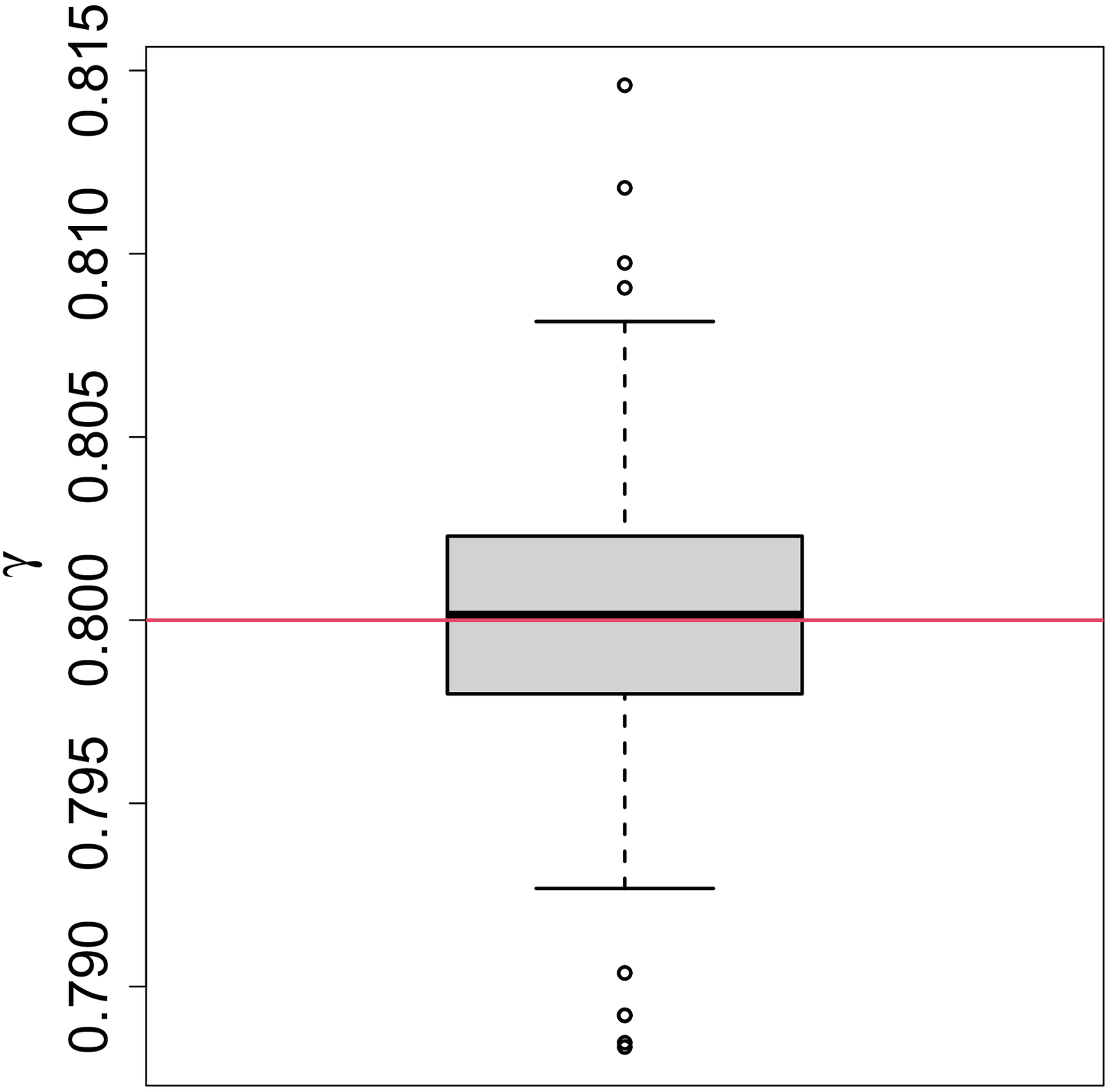}
\caption{Case 2.  Box-plots of Bayes point estimates of $m$ and $\gamma$, $\tilde{m}_{n,1},\ldots,\tilde{m}_{n,100}$ and $\tilde{\gamma}_{n,1},\ldots,\tilde{\gamma}_{n,100}$, based on the samples of 100 simulated processes. Horizontal red line corresponds to the true value of the parameter.}\label{fig:estimate2}
\end{figure}

\begin{figure}[H]
\centering
\includegraphics[width=0.39\textwidth]{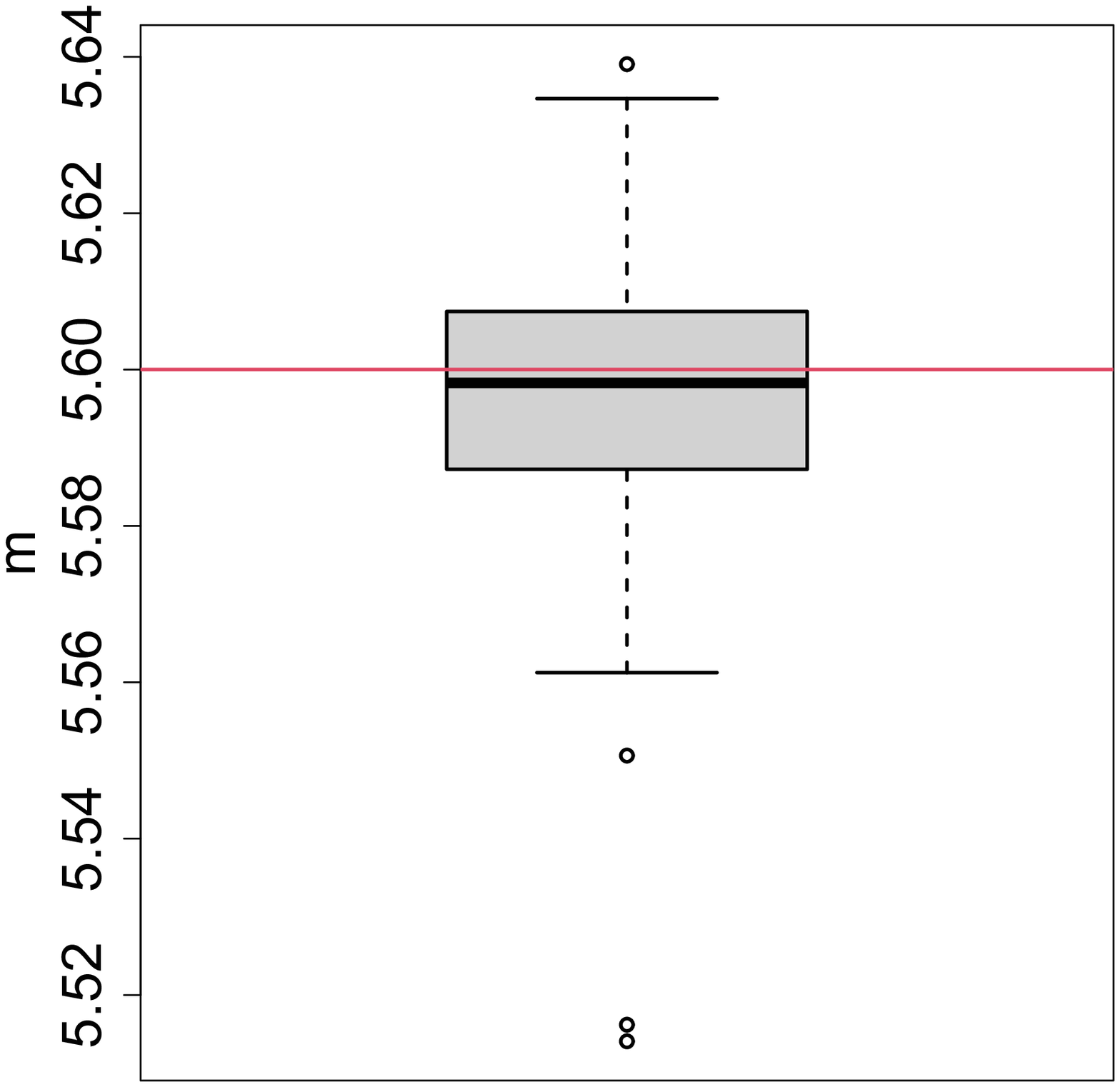}\hspace{0.1\textwidth}\includegraphics[width=0.39\textwidth]{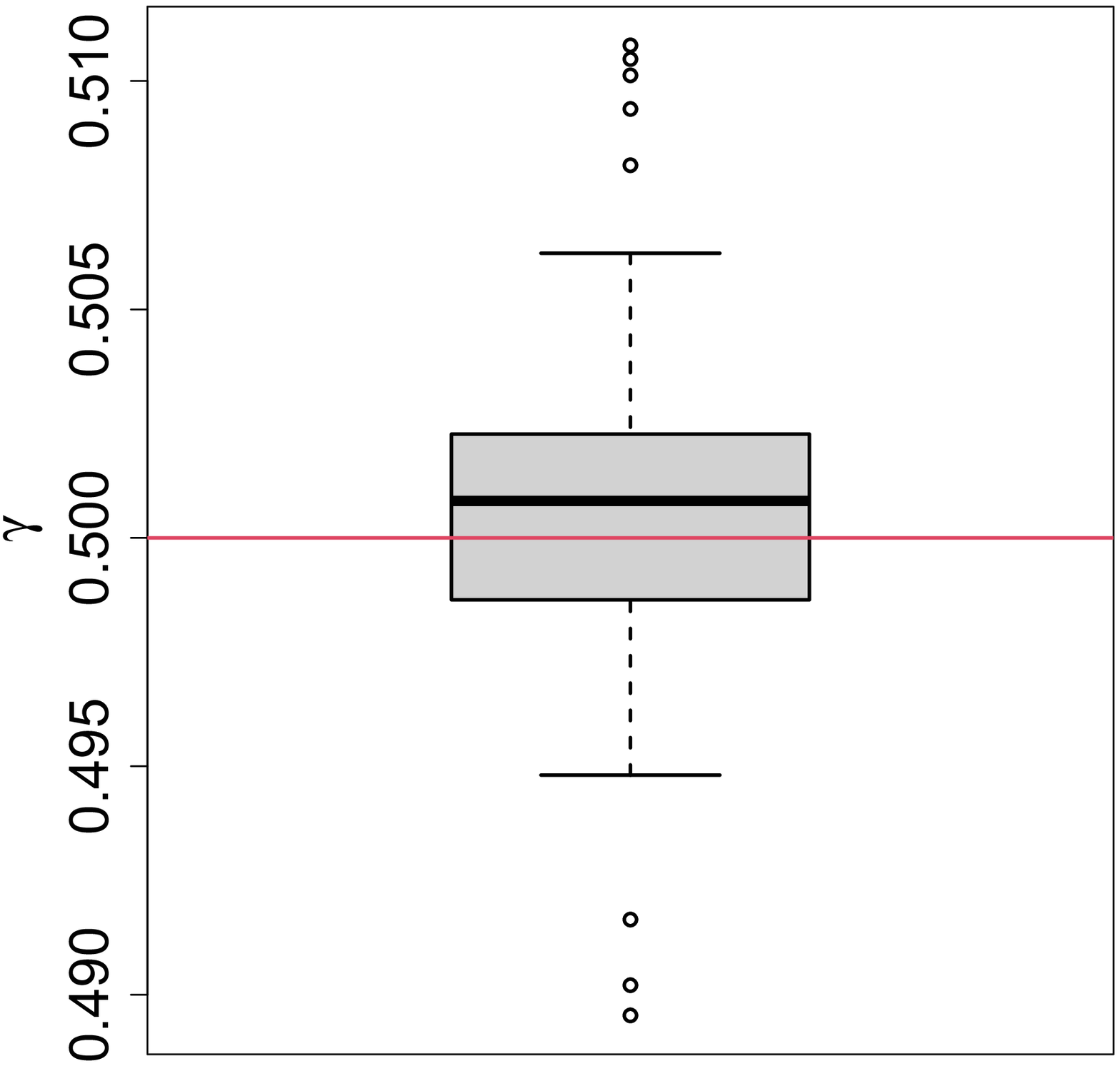}
\caption{Case 3.  Box-plots of Bayes point estimates of $m$ and $\gamma$, $\tilde{m}_{n,1},\ldots,\tilde{m}_{n,100}$ and $\tilde{\gamma}_{n,1},\ldots,\tilde{\gamma}_{n,100}$, based on the samples of 100 simulated processes. Horizontal red line corresponds to the true value of the parameter.}\label{fig:estimate3}
\end{figure}

\begin{figure}[H]
\centering
\includegraphics[width=0.39\textwidth]{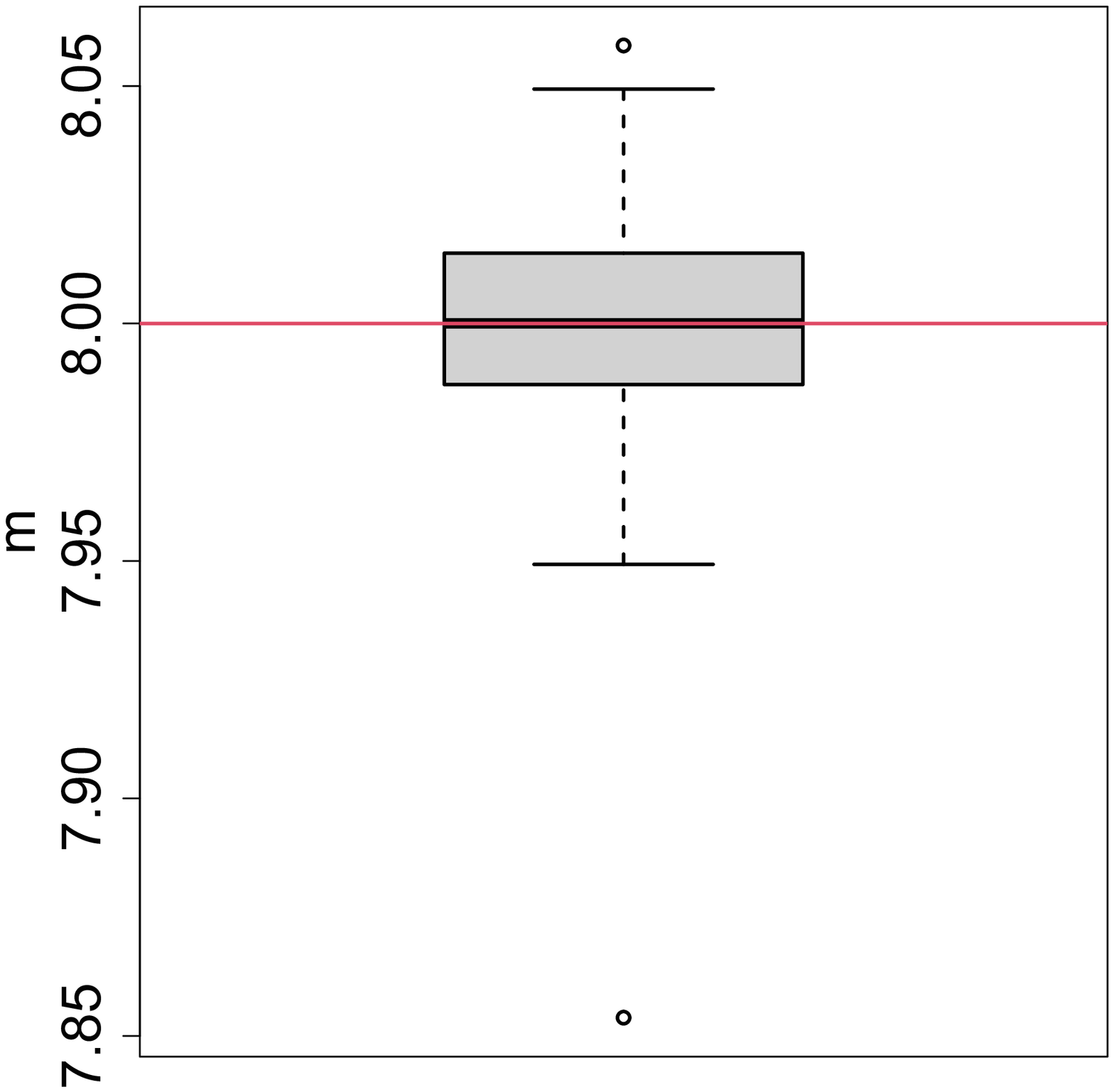}\hspace{0.1\textwidth}\includegraphics[width=0.39\textwidth]{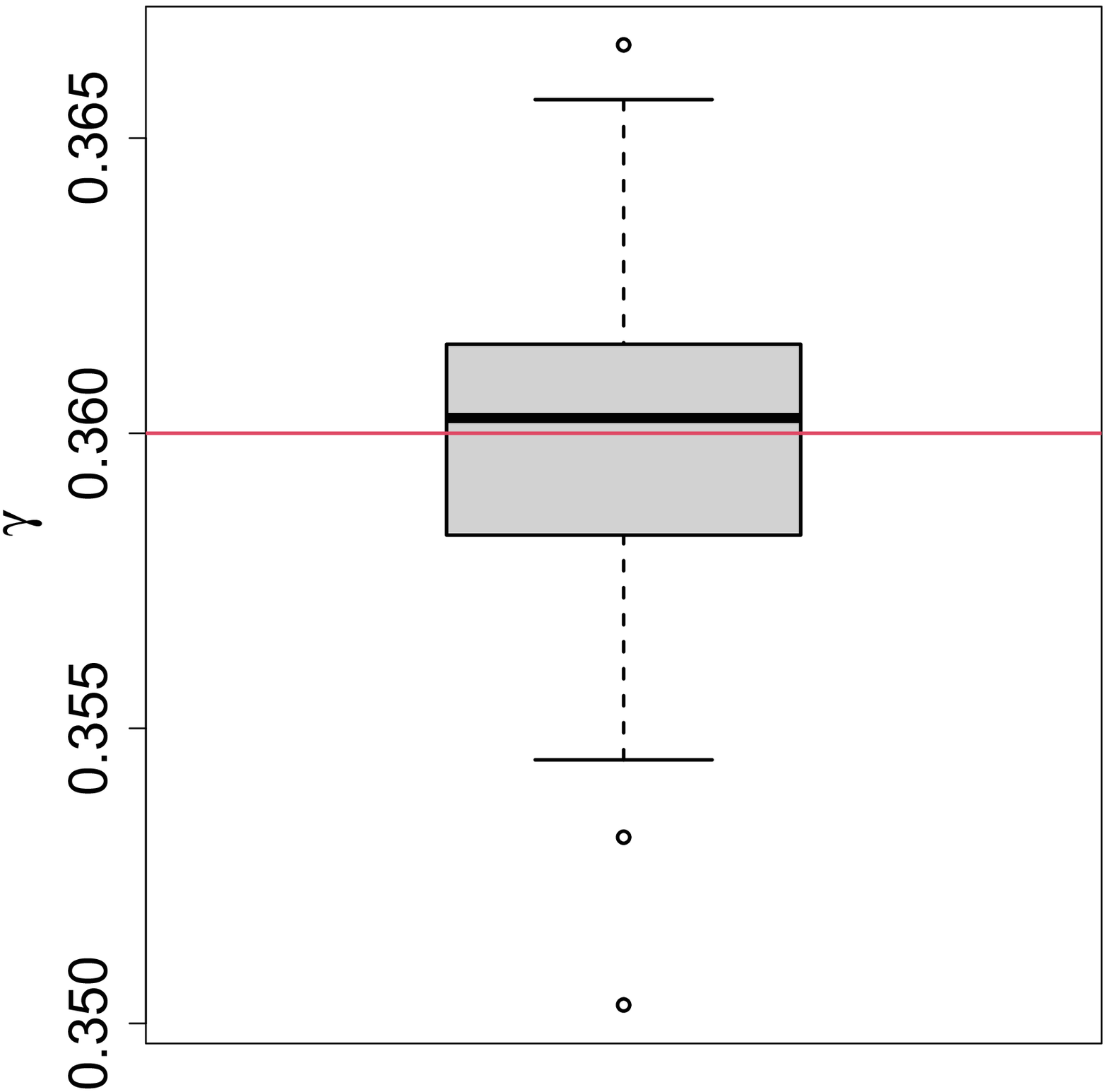}
\caption{Case 4.  Box-plots of Bayes point estimates of $m$ and $\gamma$, $\tilde{m}_{n,1},\ldots,\tilde{m}_{n,100}$ and $\tilde{\gamma}_{n,1},\ldots,\tilde{\gamma}_{n,100}$, based on the samples of 100 simulated processes. Horizontal red line corresponds to the true value of the parameter.}\label{fig:estimate4}
\end{figure}

Next, for each observed path we obtained a sample of the ABC approximation of the posterior distributions of  $\pi(m|\tilde{\kappa}_n,\widetilde{\mathcal{Z}}_n^{obs})$ and $\pi(\gamma|\tilde{\kappa}_n,\widetilde{\mathcal{Z}}_n^{obs})$ and we took the means of these samples as Bayesian point estimates of $m$ and $\gamma$, $\tilde{m}_{n,1},\ldots,\tilde{m}_{n,100}$ and $\tilde{\gamma}_{n,1},\ldots,\tilde{\gamma}_{n,100}$. The box-plots of these estimates are given in Figures \ref{fig:estimate1}, \ref{fig:estimate2}, \ref{fig:estimate3}, and \ref{fig:estimate4} in Cases 1, 2, 3, and 4, respectively. These show that the sample of each posterior distribution is centred around the true value of each parameter and their dispersion is not considerable. Thus, they lead to accurate estimates of the posterior of the parameters.

%\begin{figure}[H]
%\centering
%\includegraphics[width=0.35\textwidth]{./graphs/3_m_boxplot_q_corrected2}\hspace{0.04\textwidth}\includegraphics[width=0.35\textwidth]{./graphs/3_gamma_boxplot_q_corrected2}
%\includegraphics[width=0.35\textwidth]{./graphs/4_m_boxplot_q_corrected2}\hspace{0.04\textwidth}\includegraphics[width=0.35\textwidth]{./graphs/4_gamma_boxplot_q_corrected2}
%\caption{Box-plots of estimated posterior means  for $m$ and $\gamma$  based on 100 repeated runs of the rejection-tolerance ABC with post processing  with the selected $\kappa$ in the ABC SMC  model choice algorithm.  {First row: Case 3. Second row: Case 4.}}\label{fig:estimate1}
%\end{figure}

\subsection{Example 2}\label{subsec:example-infinite-sup}

We continue our simulation study with one of the examples in \cite{art-ABC-summary}. The considered CBP starts with $Z_0=1$ individual, the offspring distribution is a geometric distribution with parameter $q=0.4$ and the control variables $\phi_n(j)$ {follow} a binomial distribution with parameters $\xi(j)$ and $\gamma=0.75$, where the function $\xi(\cdot)$ was introduced in the previous example. The offspring mean and variance are $m=1.5$ and $\sigma^2=3.75$, the control means are $\varepsilon(j,\gamma)=\gamma \xi(j)=0.75 \xi(j)$, $j\in\N_0$, $\tau=\gamma=0.75$, and the threshold parameter is $\tau m=1.125$. Thus, taking into account the value of this last parameter, the CBP is supercritical. The simulated path and the observed sample $\widetilde{Z}_n^{obs}$  of the first 30 generations of such a process are presented in Table~\ref{tab:sup:sim-data} in Appendix. Note that the offspring distribution has infinite support.

\begin{wrapfigure}[22]{r}{0.5\textwidth}
  \centering \includegraphics[width=0.48\textwidth]{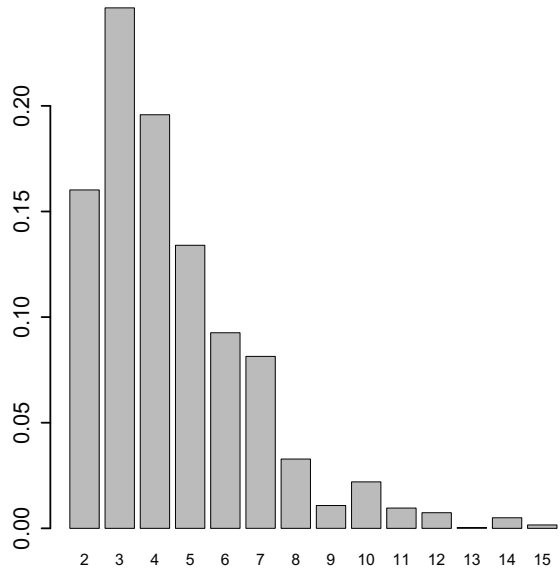}
  \caption{Estimate of the posterior distribution of {$\kappa$} in ABC SMC algorithm for model choice in Example \ref{subsec:example-infinite-sup}.}\label{fig:kappa}
\end{wrapfigure}

In Subsection 4.1 of  \cite{art-ABC-summary} we provided some inferential results obtained by using ABC algorithms  under the hypothesis of a parametric offspring distribution. Recall that this latter implies  that we assumed that we knew the parametric family of probability distribution to which the offspring distribution belonged, but the value of the parameter was unknown.  We now deal with the estimation of the posterior distributions of the stable parameters of the model as the offspring mean and the control parameter in a different framework.  To that end, throughout this example, we understand the \emph{maximum offspring capacity per individual}  as a number $\kappa$ such that the probability that an individual gives birth to more than $\kappa$ offspring is \emph{sufficiently} small, i.e., we look for a \emph{realistic} upper limit for the offspring capacity of the majority of the individuals of the population. Our goal is to estimate the posterior distribution of the maximum offspring capacity per individual with the aim of identifying    the stable parameters of the model properly. Thus, we assume that we can propose a reasonable upper bound, $K_{max}$, of this maximum offspring capacity in view of the knowledge of the population that we are modeling, as discussed in Section~\ref{sec:methods}.
We make use of the observed sample to estimate the joint  posterior distributions of the mean offspring and control parameter by assuming that our only knowledge on the offspring distribution is $K_{max}$, and the fact that the control laws for a population size $j$ are binomial distributions with parameters $\xi(j)$ and $\gamma$, with $\gamma\in (0,1)$ unknown.

%We make use again of this path to estimate the posterior distribution of the factors of the threshold parameter by assuming that our only knowledge on the offspring and control distribution is an upper bound for the maximum offspring per individual, and the fact that the control laws for a population size $j$ are binomial distributions with parameters $\xi(j)$ and $\gamma$, with $\gamma\in (0,1)$ unknown.

We implemented the ABC SBC  algorithm for model choice described in Section~\ref{sec:methods} by setting the same number of iterations, prior distributions, pools of non-extinct simulated processes, tolerance levels and tuning parameter as in Example 1. We therefore obtained a sample of length 5000 at each iteration. The resulting barplot of the sample obtained from the estimate of the posterior distribution $\pi(\kappa\mid\widetilde{\mathcal{Z}}_n^{obs})$ is shown in Figure \ref{fig:kappa}. The  closest integer to the  sample mean of the posterior distribution of $\kappa$ is 5, and then we propose $\widetilde{\kappa}_n=5$. We note that the probability that the true offspring distribution, a geometric distribution of parameter 0.4, is less than or equal to 5 is 0.9533. Consequently, the choice of 5 as the maximum number of offspring per individual is appropriate to explain the evolution of our data reasonably well. 

%\begin{figure}[H]
%\includegraphics[width=0.45\textwidth]{./graphs/evol-eps-converted-to}\hspace{0.04\textwidth}
%\centering
%\includegraphics[width=0.48\textwidth]{./graphs/k_barplot}
%  \caption{Estimate of the posterior distribution of {$\kappa$} in ABC SMC algorithm for model choice in Example \ref{subsec:example-infinite-sup}.}\label{fig:kappa}
%\end{figure}

\begin{figure}[H]
\includegraphics[width=0.32\textwidth]{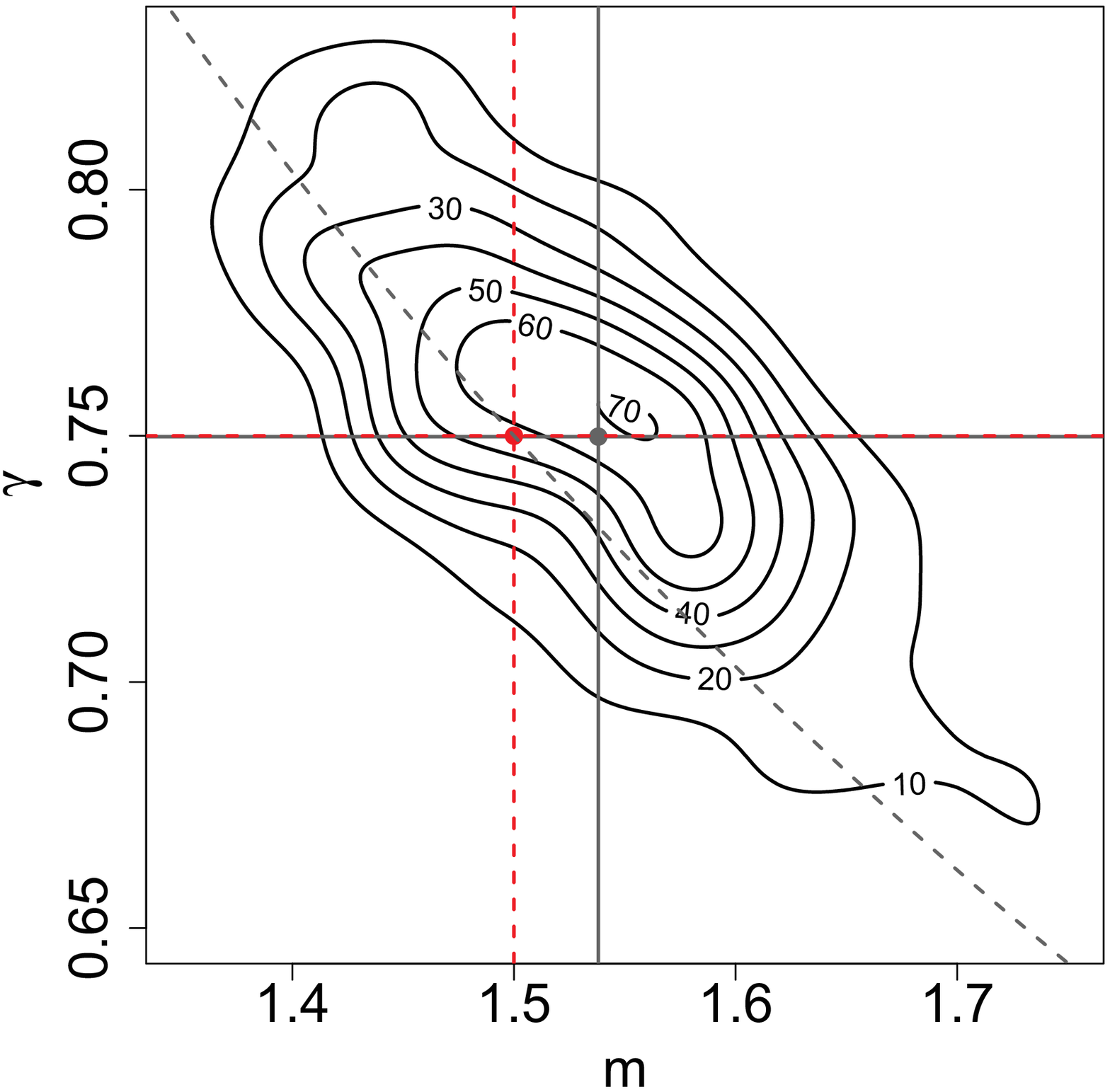}
\includegraphics[width=0.32\textwidth]{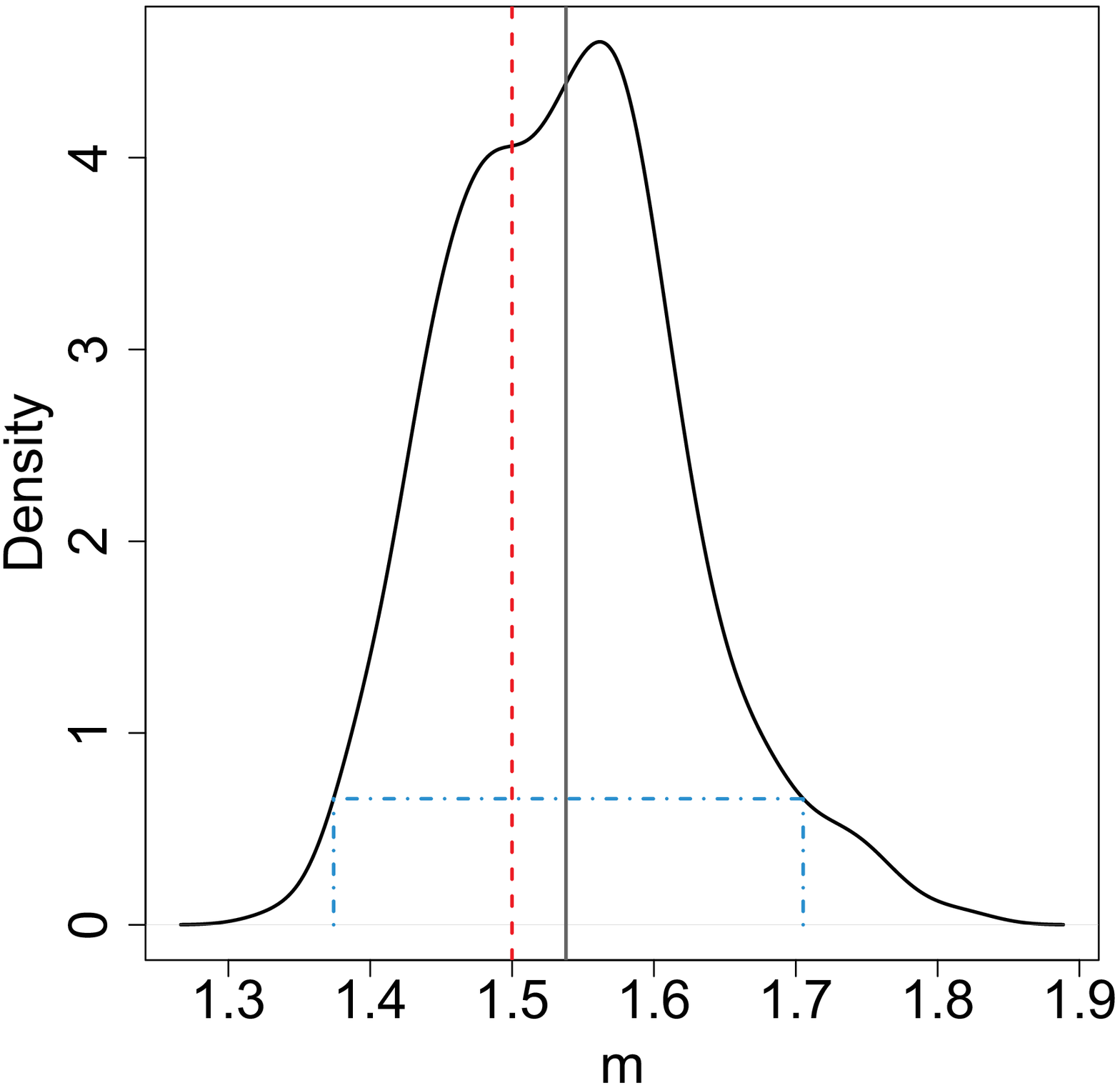} 
\includegraphics[width=0.32\textwidth]{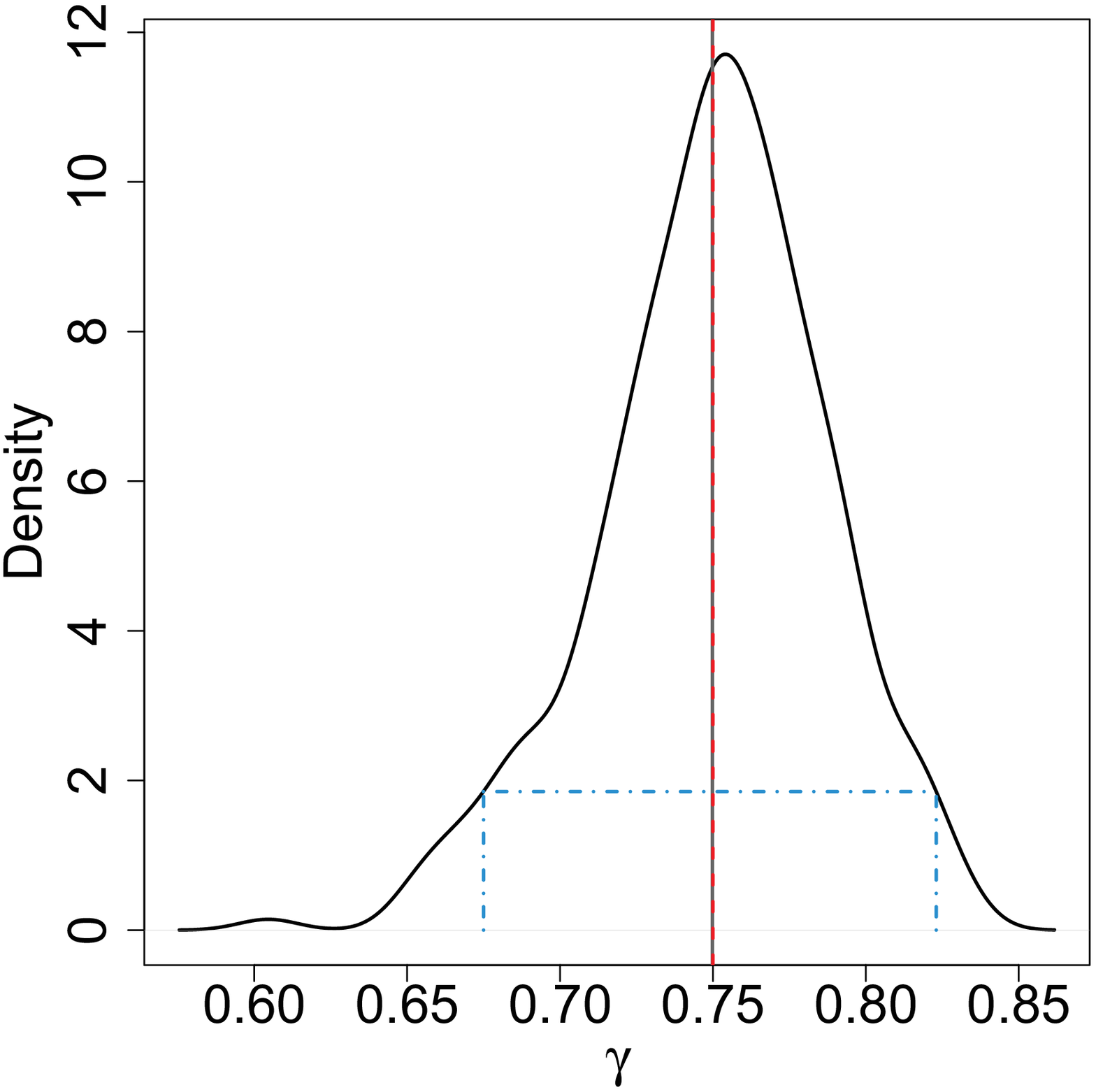}
\caption{Estimates of the posterior distributions via the ABC algorithm with the local linear regression adjustment with $\widetilde{\kappa}_n=5$. Left: Contour plot of the estimate of the joint  density $\pi(m,\gamma\mid\widetilde{\kappa}_n,\widetilde{\mathcal{Z}}_n^{obs})$ with the curve $\tau m = 1.125$. The red point indicates the true value of the parameters and the grey one represents the sample means.  Centre: Estimate of $\pi(m\mid\widetilde{\kappa}_n,\widetilde{\mathcal{Z}}_n^{obs})$. Right: Estimate of  $\pi(\gamma\mid\widetilde{\kappa}_n,\widetilde{\mathcal{Z}}_n^{obs})$. Red dashed vertical lines represent the true value of the parameter, grey solid vertical lines represent the sample means, and blue dashed-dotted vertical lines correspond to 95\% HPD intervals. }\label{fig:mgama-second-step}
\end{figure}

Next, we considered the marginal samples corresponding to  $\widetilde{\kappa}_n=5$ and applied the rejection condition  in the tolerance-rejection ABC algorithm and a local linear regression adjustment making use of the summary statistic  {in}  \eqref{eq:summary-statistic}, as described in Subsection \ref{subsec:second-part}. The results are plotted in Figures \ref{fig:mgama-second-step}. %and \ref{fig:contour-second-step}. 
Precisely, %in Figure \ref{fig:mgama-second-step} 
we represented the estimated posterior densities of $\pi(m\mid\widetilde{\kappa}_n,\widetilde{\mathcal{Z}}_n^{obs})$ and $\pi(\gamma\mid\widetilde{\kappa}_n,\widetilde{\mathcal{Z}}_n^{obs})$ and %. In Figure \ref{fig:contour-second-step}, 
 the contour plot %and the selected values of the estimate 
of the estimated joint posterior density of $\pi(m,\gamma\mid\widetilde{\kappa}_n,\widetilde{\mathcal{Z}}_n^{obs})$  together with the curve $\tau m=1.125$ (recall that in this case $\tau=\gamma$). This figure illustrates the correlation between $m$ and $\gamma$ given the observed sample. The results show that the proposed ABC algorithm estimates of the posterior densities are quite  accurate. It is worthy to point out that the implementation of the proposed methodology  is computationally simple, and provides a useful approach to make inference on the parameters of interest in a scenario that requires very little information about the true offspring law. This latter is a great advantage versus the previous methodology considered in \cite{art-ABC-summary} that assumed the knowledge of the parametric offspring  family to which the true offspring law belonged. 

%\begin{figure}[H]
%\includegraphics[width=0.48\textwidth]{./graphs/contour_m_gamma_K_max_5_2step} \hspace*{0.04\textwidth} \includegraphics[width=0.48\textwidth]{./graphs/m_gamma_K_max_5_2step}
%\caption{\blue{En ambos graficos las lineas punteadas las pondría completas, no solo hasta los puntos y les aumentaria el grosor, apenas se ven. Pondría los puntos coloreados en otro estilo y mas grandes para que se vean mejor, sobre todo en el segundo grafico. Merece la pena poner los dos graficos?} Left: Contour plot of the estimate of the joint posterior  density $\pi(m,\gamma\mid\widetilde{\kappa}_n,\widetilde{\mathcal{Z}}_n^{obs})$ with the curve $\tau m = 1.125$. Right: Sample drawn from the estimate of the joint posterior distribution of $(m,\gamma)$ obtained by the ABC algorithm with the local linear regression adjustment.  The coloured points indicate the true value of the parameters (red) and the sample mean obtained via the ABC rejection algorithm (blue),  with $\widetilde{\kappa}_n=5$.}\label{fig:contour-second-step}
%\end{figure}

\vspace{0.35cm}

\section{Real data examples}\label{sec:example_real}

In this section our aim is to apply the described methodology to real datasets that represent the logistic growth of  populations. These kinds of populations are characterized by an initial approximately exponential growth of the number of individuals till  they reach an equilibrium value around which they fluctuate. This equilibrium value, denoted as $K_e$, mainly depends on the maximum population size supported by the environment. We refer to the latter value as the carrying capacity of the population, denoted as $K$ (see \cite{Br-2012}). %We understood that this value can not be exceeded by the population size (some authors call carrying capacity what we call equilibrium value). 
Population-size dependent branching processes (PSDBPs) are often used to model these kind of data (see, for instance, \cite{KSVHJ-2011}, \cite{Hognas-2019}, or \cite{Braunsteins-Hautphenne-Minuesa-2021} and references therein). The PSDBP is a modification of a BGW process. Briefly, the assumption of identical offspring distribution for all the individuals in the BGW process is replaced with the assumption of offspring distributions in each generation which depend on the population sizes. In particular, in order to fit logistic growth data the reproduction laws depend on the current population size, the carrying capacity and on some other parameters. However, the existence of a carrying capacity does not necessarily imply that the reproductive capacity of an individual changes along generations, but rather the  probability that an individual successfully becomes a progenitor. Consequently, we propose a CBP to model population logistic growths by considering control laws defined by binomial distributions with a success probability depending on the current population size, $z$,  the carrying capacity, $K$,   and the offspring mean, $m$. We refer to $z/K$ as \emph{density}. More precisely, the random variable $\phi_0(z)$ is distributed following a binomial distribution of size $z$ and success probability given by a function $s(m,z,K)$. We consider that the process begins with a much smaller initial number of individuals $Z_0$ than $K$ and $m>1$. Under this consideration we have $E[Z_{n+1}\mid Z_n=z]= m z s(m,z,K)$. 
Although the probabilistic evolution of the described CBP with binomial control can be represented  equivalently as a PSDBP, from a practical view point the structure of a CBP makes easier to interpret the parameters involved.  %Although the probabilistic evolution of the described CBP with binomial control can be represented  equivalently as a PSDBP, from a practical view point the structure of a CBP \blue{is easier to interpret wins in interpretation of the parameters involved.  

Different functions $s(m,z,K)$ can be defined to introduce a density-dependent growth  inspired by deterministic models. Given their practical relevance we highlight the following ones and the corresponding deterministic models on which the functions $s(m,z,K)$ are based:
\begin{align*}
\begin{array}{rclcl}
s^V(m,z,K)&=&(1-z/K),&\qquad& \quad  \text{Verhulst logistic equation},\\ 
s^{L,\theta}(m,z,K)&=&m^{-(z/K)^\theta},&\quad \theta>0,&\quad  \text{$\theta$-logistic model},\\
s^{H,\beta}(m,z,K)&=&(1+(m-1)z/K)^{-\beta},&\quad \beta>0,&\quad\text{Hassell model},\\
s^{G}(m,z,K)&=&m^{-\log(z+1)/\log(K+1)},&\qquad& \quad \text{Gompertz model}.
\end{array}
\end{align*}

In particular, $\theta=1$ for the second function yields the Ricker model while $\beta=1$ in the third function gives us the Beverton-Holt model. 
%For instance,  $p^1(m,z,K)=(1-z/K)$, based on Verhulst logistic equation, $p^{2,\theta}(m,z,K)=m^{-(z/K)^\theta}$,  $\theta>0$, based on the $\theta$ logistic model ($\theta=1$ is Ricker model),   $p^{3}(m,z,K)=(1+(m-1)z/K)^{-\beta}$,  $\beta>0,$ based  Hassel model ($\beta=1$ is Beverton-Holt model), or $p^{4}(m,z,K)=m^{-(\log(z+1)/(\log(K+1))}$,   based on Gompertz model. 
We notice that, as is reasonable, a high value of density implies  a low probability of being progenitor in all the models.
The equilibrium value $K_e$ can be determined by solving the equation $E[Z_{n+1}\mid Z_n=z]=z$.
 The respective equilibrium values are $K_e^V=(1-m^{-1})K$, $K_e^{L,\theta}=K$, $K_e^{H, \beta}=K(m^{1/\beta}-1)/(m-1)$, and $K_e^{G}=K$.

%\blue{Now, we can apply the methodology described in %Section 2 by taking as control parameter $\gamma=K$. We %have tackled the estimation in two real datasets: seal %data and yeast data. } 

With the aim of making inference on the offspring mean and the equilibrium value for logistic growth data we implemented the ABC SBC algorithm for model choice and estimation of the parameters in Section \ref{sec:methods} by considering the binomial control distributions introduced above, with the control parameter $\gamma=K$. We tackled the estimation in two real datasets:  yeast data and seal data. %  Except the prior distributions, 
We set the same number of iterations, pools of non-extinct simulated processes, tolerance levels and tuning parameter as in the previous simulated examples. The details on the prior distributions are given below for each dataset.  %\blue{For this purpose, we} adapted the methodology by dropping the observation of  the  number  of  progenitors in the last generation, and what is even more interesting in presence of missing data (event that occurs in the second dataset). \blue{LA ULTIMA FRASE NO LA ACABO DE VER}

\begin{wrapfigure}[21]{r}{0.5\textwidth}
\vspace*{-0.85cm}
\centering\includegraphics[width=0.48\textwidth]{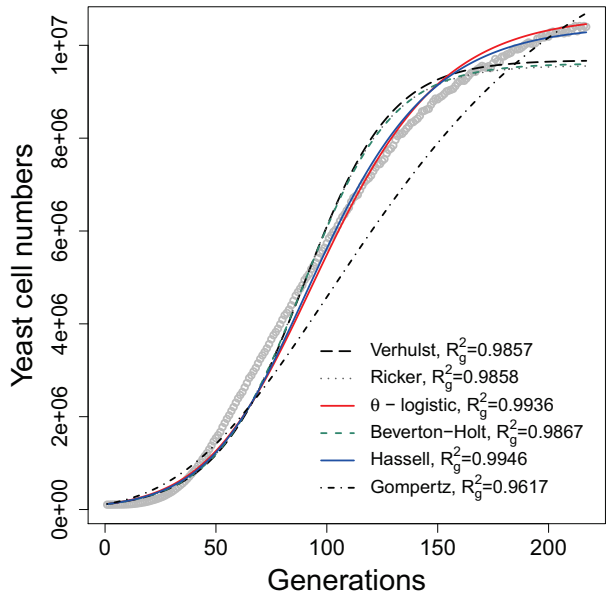}
\caption{Fitted logistic (expected values) curves together with the observed values (grey dots). For the $\theta$-logistic model $\theta=0.55$ and for the Hassell model $\beta=0.05$.}\label{fig:fitted_yeast}
\end{wrapfigure}

\subsection{Yeast dataset}

\begin{figure}[H]
\centering\includegraphics[width=0.32\textwidth]{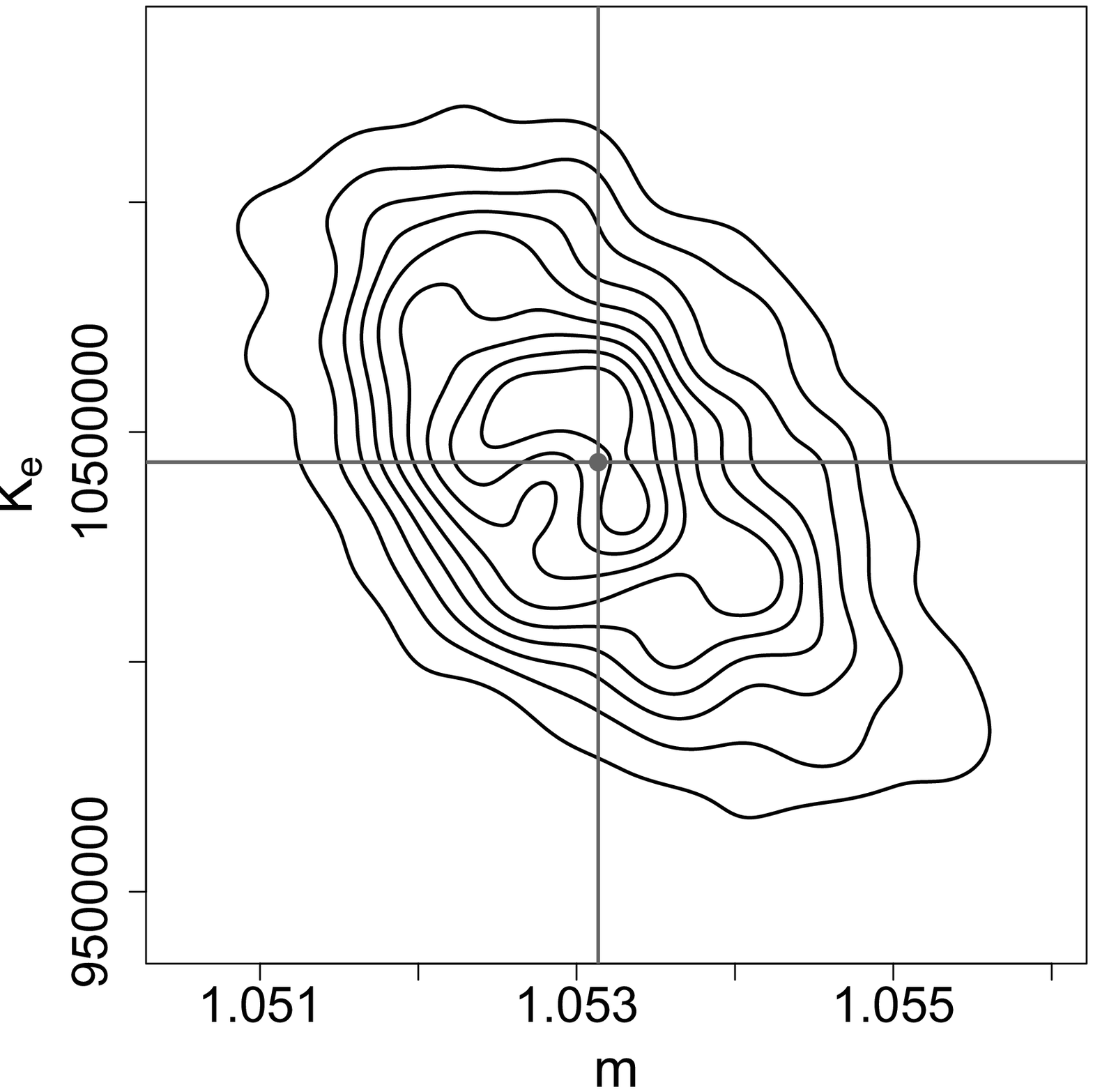}\includegraphics[width=0.32\textwidth]{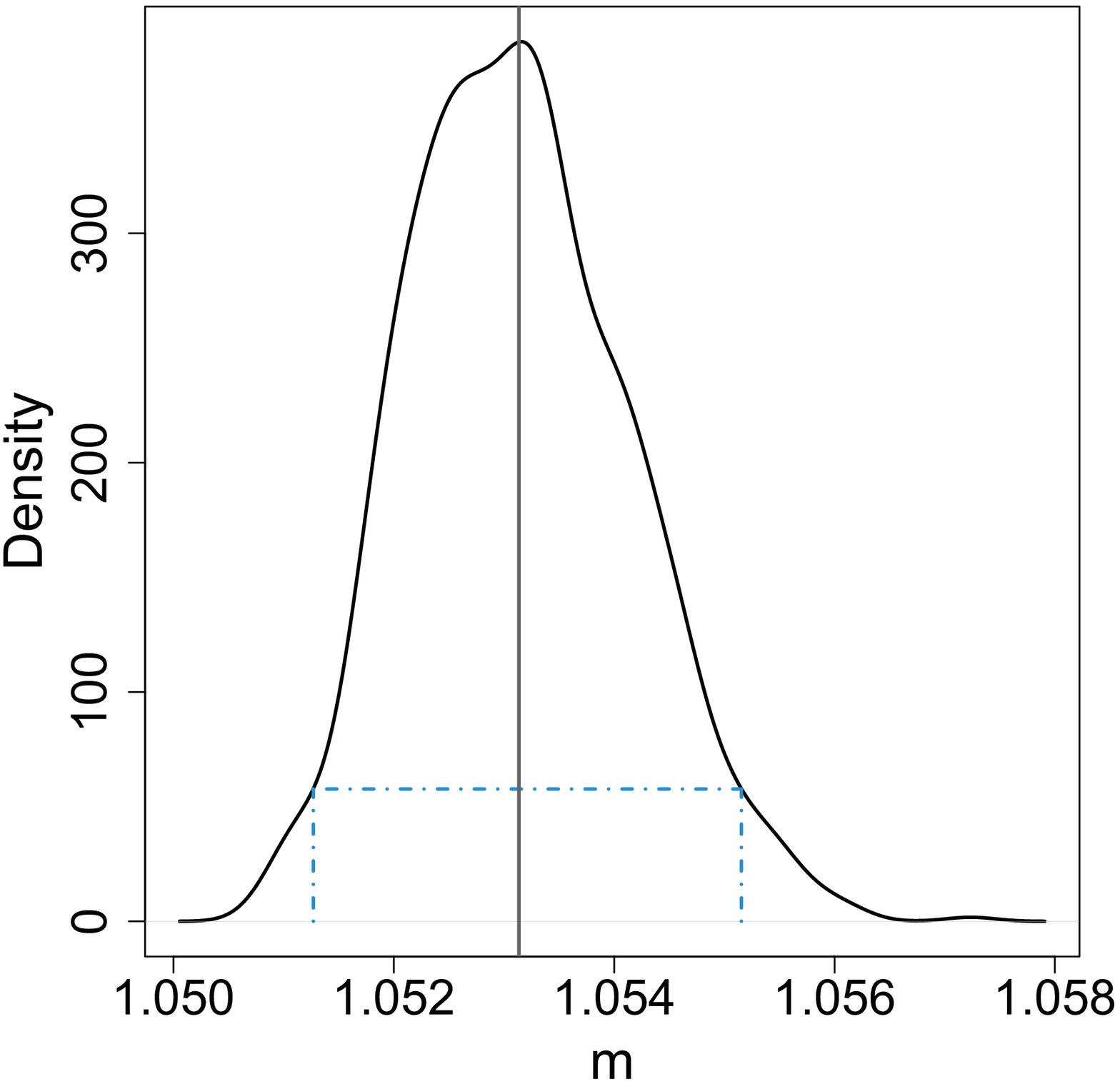}\includegraphics[width=0.32\textwidth]{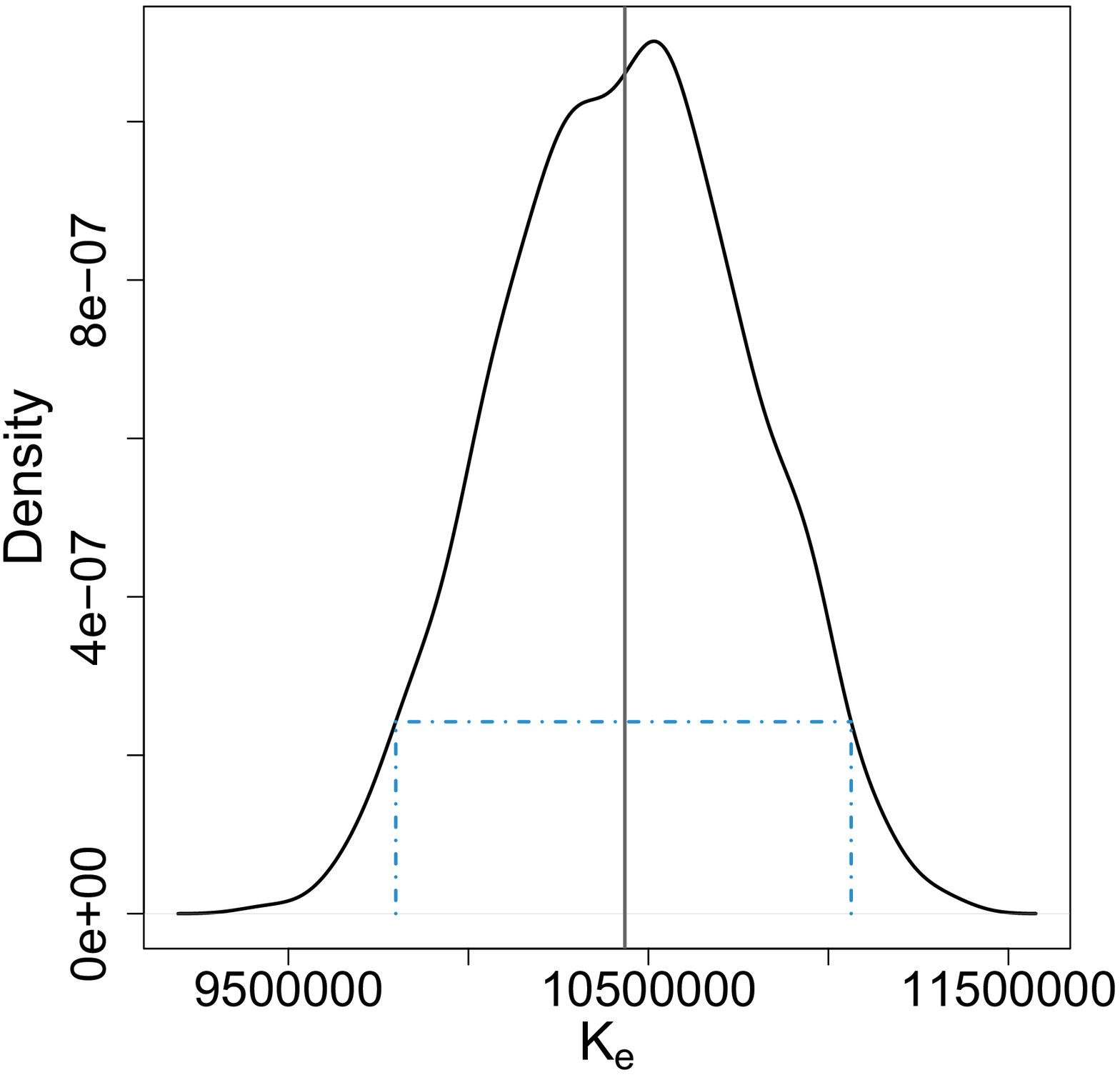}
\caption{Estimates of the posterior distributions via the ABC algorithm and the local linear regression adjustment with the Hassell model with $\beta=0.05$. Left: Contour plot of the estimate of the joint  density $\pi(m,K_e\mid\widetilde{\kappa}_n,\widetilde{\mathcal{Z}}_n^{obs})$. The grey point represents the sample means. Centre: Estimate of $\pi(m,\mid\widetilde{\kappa}_n,\widetilde{\mathcal{Z}}_n^{obs})$. Right: Estimate of $\pi(K_e,\mid\widetilde{\kappa}_n,\widetilde{\mathcal{Z}}_n^{obs})$. Grey solid lines are the sample means and blue dashed-dotted vertical lines correspond to 95\% HPD intervals.  }\label{fig: posterior_m_k_yeast}
\end{figure}

The yeast dataset was already studied  in \cite{weizhang2019} (see Figure~1~(a) in this paper% and Figure \ref{fig:fitted_yeast}, grey colour, in the current paper
) and it collects the yeast cell numbers in a replicate by colony scan-o-matic  from 0 and 72 hours of growth at 20 min intervals. These data are plotted in grey in Figure~\ref{fig:fitted_yeast} below. Note the high dimension of the data and that given the nature of the data, the observed sample is only given by the total size of each generation. To perform the algorithm, we set $K_{max}=6$ and the prior distribution for $\gamma=K$ as an uniform distribution on $(1\cdot 10^7, 1.1\cdot 10^7)$ interval. We note that a yeast cell might reproduce more than once in 20 minutes and $\kappa$ therefore represents the maximum number of yeast cells produced by a cell in this period of time. To choose the best choice of the $\theta$-logistic and Hassell models, we ran the algorithm for a grid of values of the $\theta$ and $\beta$ parameters and selected the corresponding models which provide the best adjustments. We based our decision on $R^2_g$, the fraction of variance in the growth data explained by the different logistic regression models, which is the adjustment measure considered in \cite{weizhang2019}. %We run the algorithm by achieving a grid of the $\theta$ and $\beta$ parameters in the $\theta$-logistic and Hassell models, respectively.   
In Figure  \ref{fig:fitted_yeast} we plotted %the best adjustments taking into account the fraction of variance in the growth data explained by the different logistic regression models, $R^2_g$ (adjustment measure considered in \cite{weizhang2019}).
a point estimates of the  expected values of each generation size given by the different logistic regression models and provide the fraction of variance explained by each of them. The maximum value of  $R^2_g$ is provided by Hassell logistic growth model with $\beta=0.05$, $R^2_g=0.9946$. It is worthy to point out that this latter value is similar to the one obtained in the study developed in  \cite{weizhang2019}. For this model, we also estimated the joint posterior distribution of the offspring mean, $m$, and the equilibrium value, $K_e$, and the corresponding marginal distributions in Figure~\ref{fig: posterior_m_k_yeast}. 

%\begin{figure}[H]
%\centering\includegraphics[width=0.48\textwidth]{yeast_fitted_yeast}
%\caption{Fitted logistic (expected values) curves together with the observed values (grey dots). For the $\theta$-logistic model $\theta=0.55$ and for the Hassell model $\beta=0.05$.}\label{fig:fitted_yeast}
%\end{figure}

\vspace*{-0.85cm}

%Notice that for any of the definition of $p(m,z,K)$ we can adapt the methodology described in Section 2 by considering the control parameter $\gamma=K$. We have estimated the posterior distribution of $m$  and the $K_e$ for two real data: seal data and yeast data. 

\begin{wrapfigure}[21]{r}{0.5\textwidth}
\vspace*{-1.7cm}
\centering\includegraphics[width=0.48\textwidth]{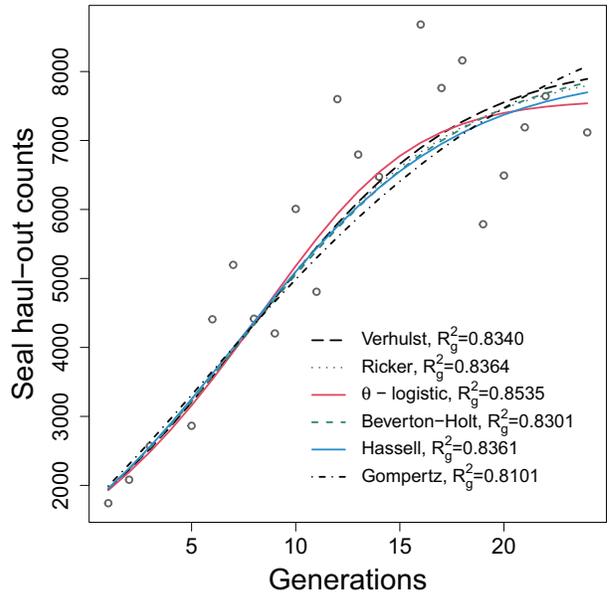}
\caption{Fitted logistic (expected values) curves together with the observed values (grey dots). For the $\theta$-logistic model $\theta=2$ and for the Hassell model $\beta=1.25$.}\label{fig:fitted_seals}
\end{wrapfigure}

\subsection{Seal dataset}

\begin{figure}[H]
\centering\includegraphics[width=0.32\textwidth]{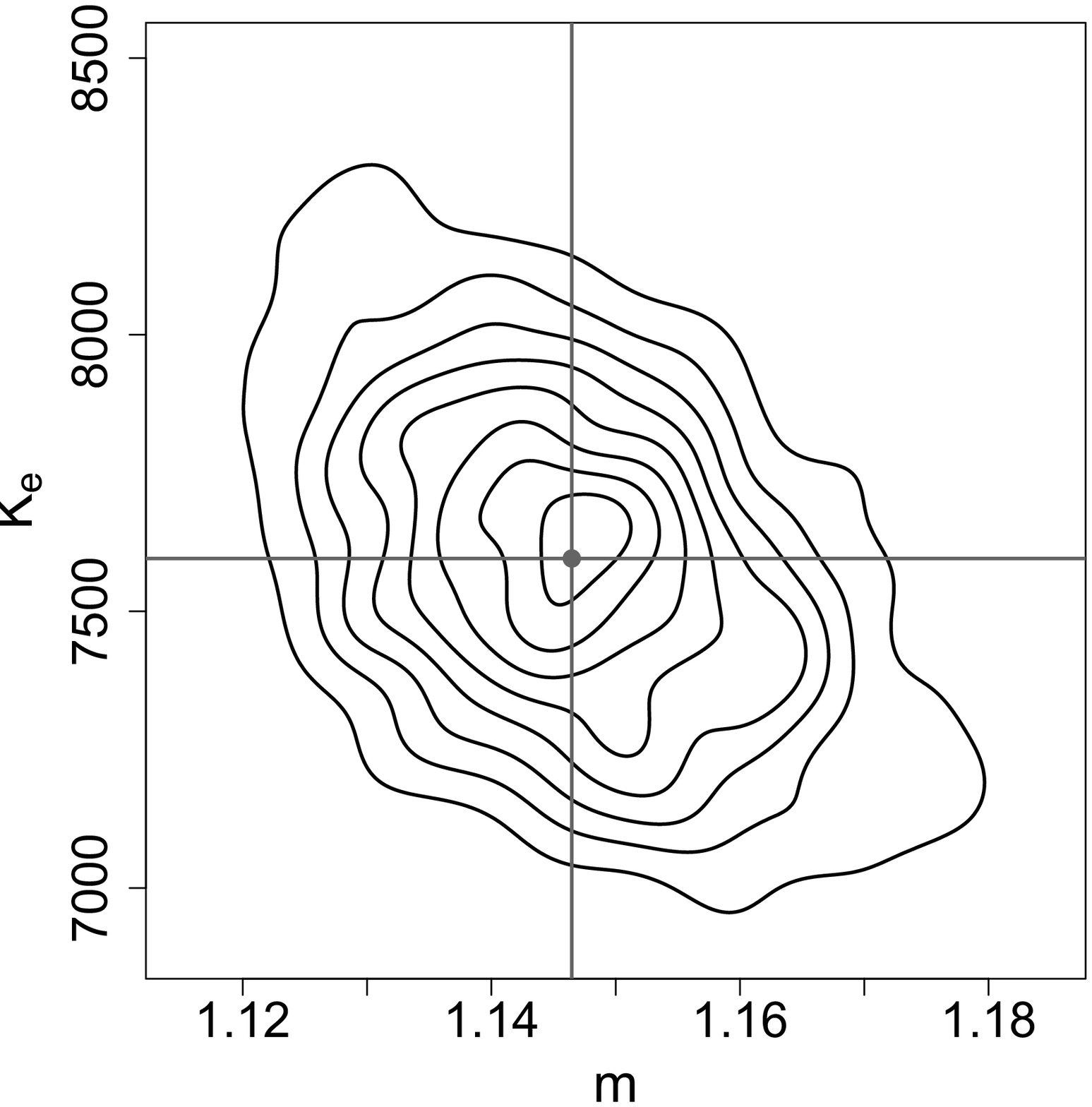}\includegraphics[width=0.32\textwidth]{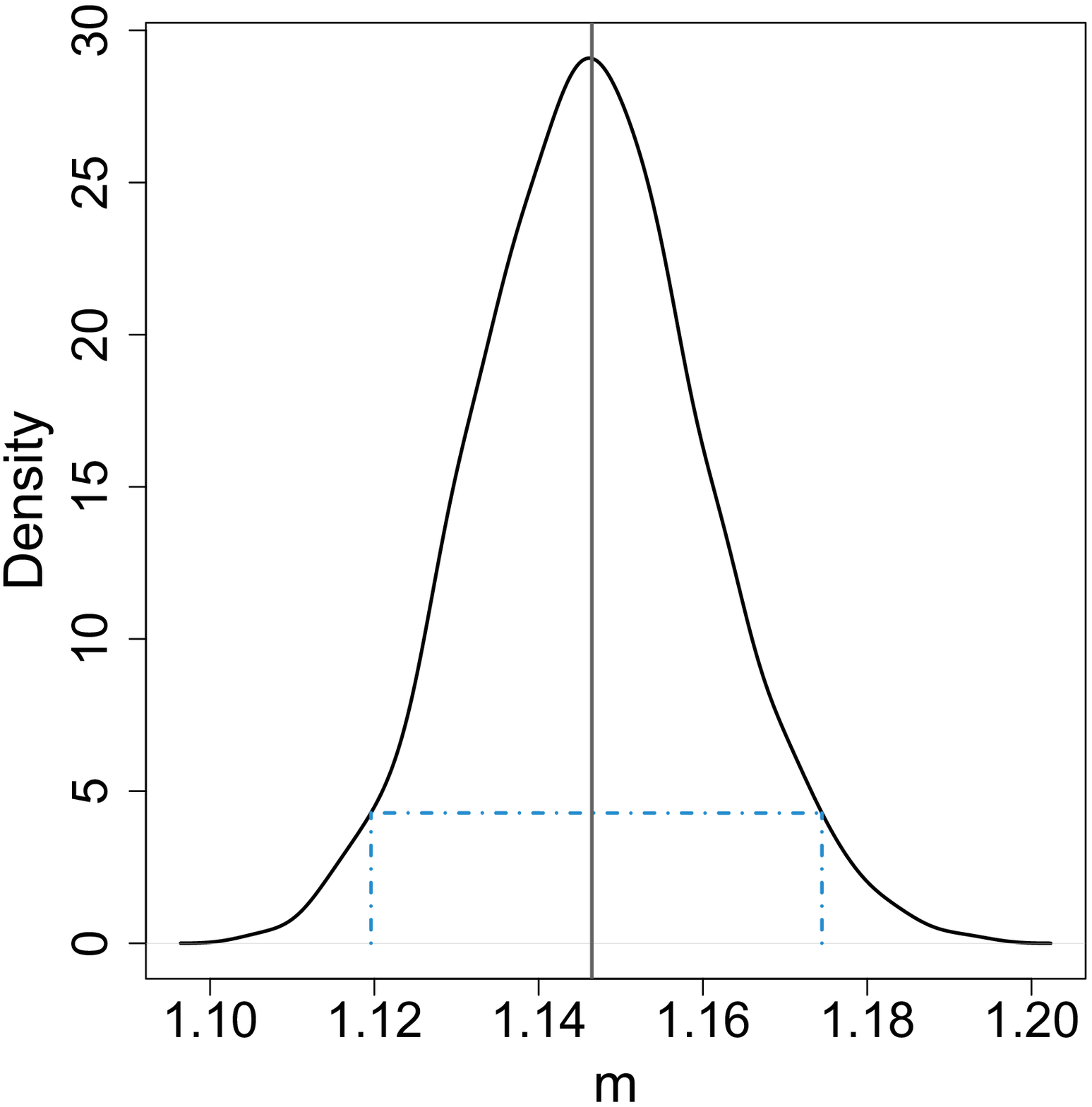}
\includegraphics[width=0.32\textwidth]{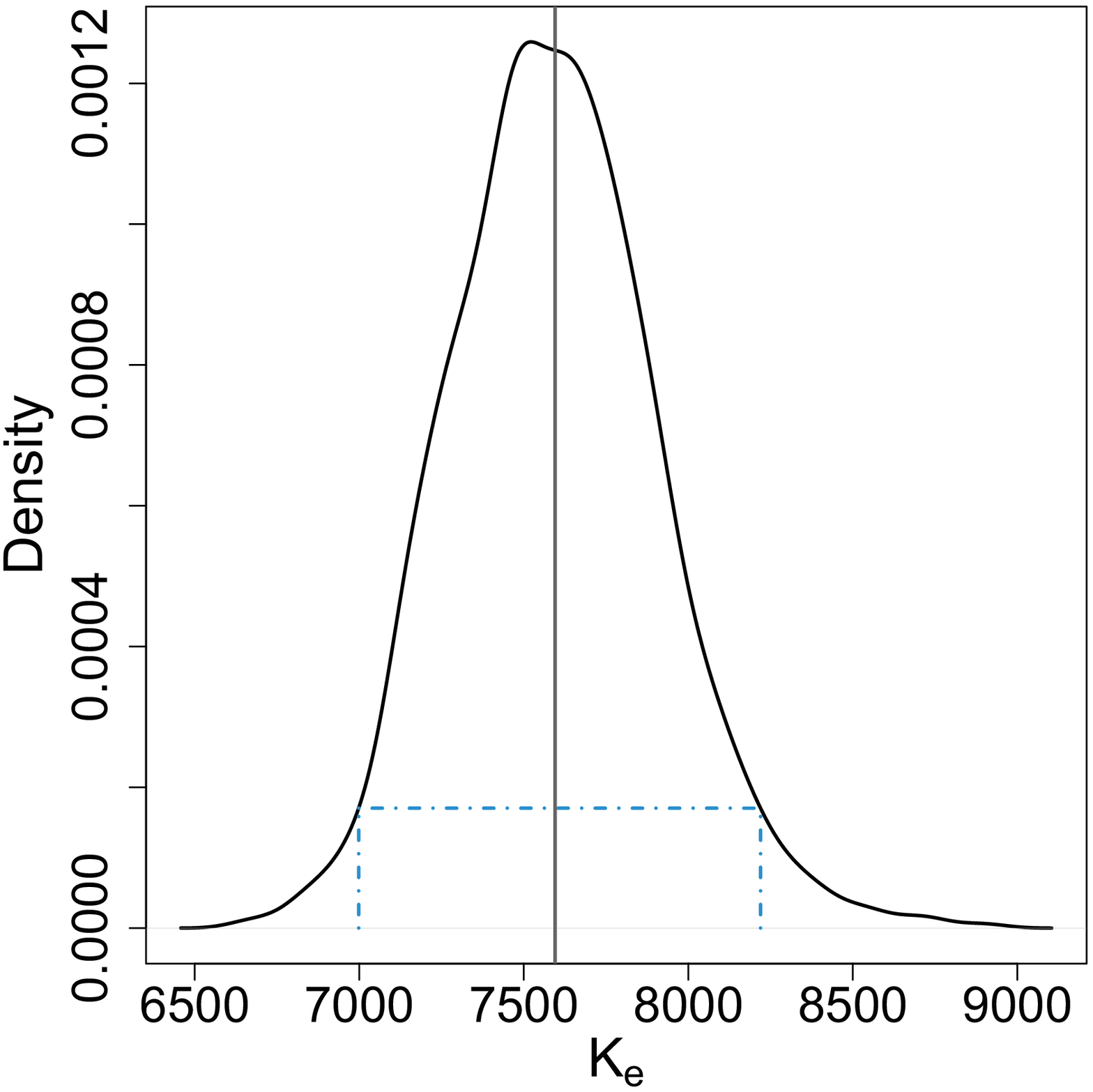}
\caption{Estimates of the posterior distributions via the ABC algorithm and the local linear regression adjustment with the $\theta$-logistic model $\theta=2$. Left: Contour plot of the estimate of the joint  density $\pi(m,K_e\mid\widetilde{\kappa}_n,\widetilde{\mathcal{Z}}_n^{obs})$. The grey point represents the sample means. Centre: Estimate of $\pi(m,\mid\widetilde{\kappa}_n,\widetilde{\mathcal{Z}}_n^{obs})$. Right: Estimate of $\pi(K_e,\mid\widetilde{\kappa}_n,\widetilde{\mathcal{Z}}_n^{obs})$. Grey solid lines are the sample means and blue dashed-dotted vertical lines correspond to 95\% HPD intervals.}\label{fig: posterior_m_k_seals}
%\caption{Estimates of the posterior distributions via the ABC algorithm with the local linear regression adjustment  for $m$ (left) and $K_e$ (right) for . Blue dashed-dotted vertical lines correspond to HPD intervals.  }\label{fig: posterior_m_k_seals}
\end{figure}

The seal dataset collects the average annual harbor seal haul-out counts in the coastal estuarine environment of Washington State, USA, from 1975 to 1999. These are provided in Table \ref{tab:seals} in the Appendix (see \cite{Jeffries2003} for further details on this dataset) and represented in Figure \ref{fig:fitted_seals}. It is worthy to point out that these data show missing values and a greater dispersion than yeast data. In this case we use the same value of $K_{max}$ as in the yeast data example and  for prior distribution of $\gamma=K$ we set a uniform distribution on the interval $(5000, 10000)$. Based on the values of $R_g^2$ the best adjustment is provided by $\theta$-logistic model with $\theta=2$. For this model the estimated joint posterior distribution of the offspring mean, $m$, and the equilibrium value, $K_e$, and the corresponding marginal distributions, which are plotted in Figure~\ref{fig: posterior_m_k_seals}.

%\begin{figure}[H]
%\centering\includegraphics[width=0.48\textwidth]{seals_fitted_seals}
%\caption{Fitted logistic (expected values) curves together with the observed values (grey dots). For the $\theta$-logistic model $\theta=2$ and for the Hassell model $\beta=1.25$.}\label{fig:fitted_seals}
%\end{figure}

%\blue{The second dataset} collect the yeast cell numbers in a replicate by colony scan-o-matic  from 0 and 72 h of growth at 20 min intervals, already studied  in \cite{weizhang2019}, see Figure 1 (a). 

%%%%%%%%%%%%%%%%%%%%%%%%%%%%%%%%%%%%%%%%%%%%%%%%%%%%%%%%%%%%%%%%%%%%%%%%%%%%%%%%%%%%%%%%%%%%%%%%%%%%
\section{Concluding remarks}\label{sec:conclusion}
%%%%%%%%%%%%%%%%%%%%%%%%%%%%%%%%%%%%%%%%%%%%%%%%%%%%%%%%%%%%%%%%%%%%%%%%%%%%%%%%%%%%%%%%%%%%%%%%%%%%
We dealt with the Bayesian estimation of the main parameters of a CBP in a general context. Precisely, we assumed a parametric framework for the control laws and that the only information about the offspring distribution was an upper bound for the maximum number of offspring per individual. The two main goals in this setting were to estimate the posterior distribution of the maximum number of offspring per individual, $\kappa$, and to estimate the posterior distribution of other parameters such as the offspring mean and control parameter based on the Bayes point estimate of $\kappa$ under the quadratic loss function. To that end, we considered the sample defined by the population sizes in all the generations and the number of progenitors in the last generation.

The methodology that we proposed consists of two steps. In the first one, we used the parameter $\kappa$ as a model index and applied a SMC ABC algorithm for model choice with the raw data to draw a sample from the estimate of the posterior distribution of $\kappa$. From this sample, we also proposed the sample mean as point estimate of the value of $\kappa$. In the second step, given this point estimate and the samples obtained in the last iteration of the method in the previous step, we made use of an  ABC algorithm together with a local linear regression adjustment to draw samples from the estimates of the posterior distribution of the parameters of interest related to the offspring and control distribution. In this stage, we introduced an appropriate summary statistic to identify the parameters of the model. 

Our empirical results support the suitability of the methodology proposed. First,  via several simulated examples, we showed that SMC ABC algorithm for model choice with the raw data enables us to obtain a sample of the posterior distribution of $\kappa$ relatively easily and identify the main parameters of the reproduction and control laws through the second stage of the algorithm with the summary statistic. Indeed, the resulting posterior distributions are centred around the true value of the parameters. Second, turning to the simulation study in  \cite{art-ABC-summary} we applied the method to estimate the posterior distribution of the offspring mean and control parameter when the support of the offspring distribution is  infinite. In this setting, as indicated above, the parameter $\kappa$ is now interpreted as such a quantity satisfying that probability that an individual has at most $\kappa$ offspring is large enough, that is a \emph{realistic} upper bound for the reproduction capacity of the majority of the individuals. Again, the results obtained are quite satisfactory even in  this  miss-specified model framework. 

We also used our methodology to estimate the posterior distribution of the offspring mean and the equilibrium value for two real datasets that present logistic growth of populations. 
To the best of our knowledge, this was the first time that CBPs were used as models for populations whose evolution is conditioned by the existence of a maximum capacity of the environment in which evolve. We highlight that the methodology is quite flexible and works reasonably well even with missing  values, as happens in seal dataset, and with high value data, as happens in both examples - mainly in the yeast one. In both datasets the adjusted models  fit the observed data quite well, providing suitable estimates of the parameters of interest.

We finally remark that in situations where the knowledge on the reproduction law is limited, the computational simplicity of the methodology makes it an appropriate way to generate samples of the the estimate of the posterior distributions of the target parameters. This represents a clear progress compared to previous works in this setting such as \cite{chap-proceedings-2016}, \cite{art-ABC-summary}, \cite{art-Dposterior} and \cite{drovandi-Pettit-al-2016}, even when working with high value data.

\vspace{0.5cm}

\appendix

\section*{Appendix}

\subsection*{Simulated examples. Example 1.}
The data  of the particular simulated examples in  Subsection~\ref{subsec:example-finite-sup} developed in the first part  are  % {plotted} in Figure \ref{fig:sup:path} and
provided in Table \ref{tab:sup:sim-data-ex2}.

\subsection*{Simulated examples. Example 2.}

The data of the simulated example in  Subsection~\ref{subsec:example-infinite-sup}, previously analysed  in \cite{art-ABC-summary}, are  % {plotted} in Figure \ref{fig:sup:path} and
provided in Table \ref{tab:sup:sim-data}. Recall that for the simulated CBP, which starts with $Z_0=1$ individual, the reproduction law  is a geometric distribution with parameter $q=0.4$, and for each $k\in\N_0$, the probability distribution of the control variable $\phi_n(k)$ is a binomial distribution with parameters $\xi(k)$ and $\gamma=0.75$, with $\xi(k)= k+\lfloor\log (k)\rfloor$, for each $k\in\N$ and $\xi(0)=0$.

\begin{table}[H]
\centering
\begin{tabular}{|c|c|c|c|c|}
  \cline{2-5}\cline{2-5}
\multicolumn{1}{ c|}{ } & Case 1 & Case 2 & Case 3 & Case 4 \\ 
  \hline
    $Z_0$ & 1  & 1  & 1  & 1  \\ 
  $Z_1$  & 4  & 5 & 7  & 10  \\ 
  $Z_2$  & 12  & 21  & 31  & 53  \\ 
  $Z_3$ & 30  & 59  & 82  & 131  \\ 
  $Z_4$  & 84  & 168  & 237  & 372  \\ 
  $Z_5$  & 249  & 467  & 617  & 1045  \\ 
 $Z_6$  & 728  & 1242  & 1637  & 3085  \\ 
  $Z_7$  & 2148  & 3614  & 4328  & 8539  \\ 
  $Z_8$  & 6165  & 10282  & 12368  & 24730  \\ 
 $Z_9$  & 17883  & 29600  & 34593  & 69854  \\ 
 $Z_{10}$  & 51412  & 85501  & 96321  & 202339  \\ 
  $\phi_{9}(Z_{9})$  &  14281 & 23668  & 17238  & 25309  \\ 
   \hline\hline
\end{tabular}
\caption{Observed samples for the particular cases studied in Example 1.}\label{tab:sup:sim-data-ex2}
\end{table}

% {Creo que es mejor no utilizar el mismo grafico que en el otro articulo, puedo generarlo de nuevo con colores para que no sea exactamente igual, o incluso representar unicamente lo que es nuestra muestra.}

%\begin{figure}[H]
%\centering\includegraphics[width=0.45\textwidth]{./graphs/evol-eps-converted-to}
%\hspace{0.04\textwidth}
%\includegraphics[width=0.45\textwidth]{./graphs/k_barplot}
 % \caption{Temporal evolution of the number of individuals (solid line) and progenitors (dashed line) of the observed sample.}
%\label{fig:sup:path} %Right: Barplot of selected {$\kappa$} in ABC SMC model choice algorithm.}\label{fig:sup:path}
%\end{figure}

\begin{table}[H]
\centering
\begin{tabular}{|c|c|c|c|c|c|c|c|c|c|c|c|c|c|c|c|c|}
\hline\hline
 $n$ & 0& 1 & 2&3&4&5&6&7&8&9&10&11&12&13&14&15 \\
\hline
$Z_n$ & \textbf{1}&\textbf{4}&\textbf{6}&\textbf{4}&\textbf{11}&\textbf{6}&\textbf{9}&\textbf{19}&\textbf{26}&\textbf{14 }&\textbf{10}&\textbf{11}&\textbf{9}&\textbf{12}&\textbf{14}&\textbf{15}\\ 
$\phi_n(Z_n)$ &1  &3  &5&3&10&7&7&13&19&9&9&9&7&8 &12&12\\
\hline
\multicolumn{17}{ c }{ }\\
\cline{1-16}
 $n$ & 16& 17 & 18&19&20&21&22&23&24&25&26&27&28&29&30& \multicolumn{1}{ c }{ }\\
\cline{1-16}
$Z_n$ &\textbf{9}&\textbf{3}&\textbf{6}&\textbf{13}&\textbf{17}&\textbf{23}&\textbf{35}&\textbf{58}&\textbf{75}&\textbf{73 }&\textbf{103}&\textbf{107}&\textbf{141}&\textbf{166}&\textbf{216}&\multicolumn{1}{ c }{ }\\ 
$\phi_n(Z_n)$ &5  &3  &7&13&15&18&32&46&61&51&78&83&100 &\textbf{131}&& \multicolumn{1}{ c }{ }\\
\cline{1-16}\cline{1-16}
\end{tabular}
\caption{Simulated data and observed sample $\widetilde{\mathcal{Z}}_{30}^{obs}$ in bold.}\label{tab:sup:sim-data}
\end{table}

\subsection*{Real data examples. Seal dataset.}

Table \ref{tab:seals} gathers the average annual harbor seal haul-out counts in the coastal estuarine environment of Washington State from 1975 to 1999. These data were previously provided and analysed in \cite{Jeffries2003} (see Table 1 in the aforementioned paper).
 %The average annual harbor seal haul-out counts in the coastal estuarine environment of Washington, USA, from 1975 to 1999 are provided in Table \ref{tab:seals} (source: \cite{Jeffries2003}).

\begin{table}[H]
\centering
\begin{tabular}{|c|c|c|c|c|c|c|c|c|c|c|c|c|}
\hline\hline
 Year & 1975& 1976 & 1977&1978&1979&1980&1981&1982&1983&1984&1985&1986 \\
\hline 
$Z_n$ & 1694&1742&2082&2570&NA&2864&4408&5197&4416&4203&6008&4807
\\

\hline\hline
 Year & 1987& 1988 & 1989&1990&1991&1992&1993&1994&1995&1996&1997&1998 \\
\hline 
$Z_n$ & 7600&6796&6475&NA&8681&7761&8161&5786&6492&7191&7643&NA
\\ 
\hline\hline
 Year & 1999&  & &&&&&&&&& \\
\hline 
$Z_n$ &7117 &&&&&&&&&&& \\\hline
\end{tabular}
\caption{The average annual harbor seal haul-out counts.}\label{tab:seals}
\end{table}

%c(7117)

%%%%%%%%%%%%%%%%%%%%%%%%%%%%%%%%%%%%%%%%%%%%%%%%%%%%%%%%%%%%%%%%%%%%%%%%%%%%%%%%%%%%%%%%%%%%%%%%%%%%%
\section*{Funding}
%%%%%%%%%%%%%%%%%%%%%%%%%%%%%%%%%%%%%%%%%%%%%%%%%%%%%%%%%%%%%%%%%%%%%%%%%%%%%%%%%%%%%%%%%%%%%%%%%%%%%
This research has been supported by  the Junta de Extremadura (grant GR18103), the Spanish State Research Agency (PID2019-108211GBI00/AEI/10.13039/501100011033) and the Fondo Europeo de Desarrollo Regional.

%%%%%%%%%%%%%%%%%%%%%%%%%% REFERENCES %%%%%%%%%%%%%%%%%%%%%%%%%%%%%%%%
%%\addtocontents{toc}{\protect\setcounter{tocdepth}{1}}
%\addcontentsline{toc}{section}{References}
%\bibliographystyle{plainnat}
%\bibliography{./bibtes_source}

%\subsection{Example 2}

\end{document}